\documentclass[a4paper,11pt]{article}
\pdfoutput=1 

\usepackage{jheppub} 

\usepackage{amsmath}
\usepackage{slashed}
\usepackage{graphicx}
\usepackage{caption}
\usepackage{subcaption}
\usepackage{amssymb}
\usepackage{mathrsfs}
\usepackage{bm}
\usepackage{xspace}
\usepackage{booktabs}
\usepackage{xcolor}
\usepackage{listings}
\usepackage{verbatim}
\usepackage{mathtools}
\usepackage{hyperref}
\usepackage{multicol}
\usepackage{multirow}

\newcommand{\eq}[1]{Eq.~\eqref{eq:#1}}
\newcommand{\eqs}[2]{Eqs.~\eqref{eq:#1} and \eqref{eq:#2}}
\renewcommand{\sec}[1]{Sec.~\ref{sec:#1}}
\newcommand{\secs}[2]{Secs.~\ref{sec:#1} and \ref{sec:#2}}
\newcommand{\app}[1]{App.~\ref{app:#1}} 
\newcommand{\fig}[1]{Fig.~\ref{fig:#1}}
\newcommand{\figs}[2]{Figs.~\ref{fig:#1} and \ref{fig:#2}}
\newcommand{\tab}[1]{Table~\ref{tab:#1}}
\newcommand{\be}{\begin{equation}}
\newcommand{\ee}{\end{equation}}

\newcommand{\ord}[1]{{\mathcal O}(#1)}
\newcommand{\ORd}[1]{{\mathcal O}\Bigl(#1\Bigr)}
\newcommand{\ORD}[1]{{\mathcal O}\biggl(#1\biggr)}
\newcommand{\mae}[3]{\langle#1\rvert#2\rvert#3\rangle}

\newcommand{\td}[2]{\frac{\text{d}#1}{\text{d}#2}}

\newcommand{\nn}{\nonumber}

\newcommand{\img}{\mathrm{i}}

\newcommand{\tr}{\mathrm{tr}}

\newcommand{\al}{\alpha}

\newcommand{\ga}{\gamma}
\newcommand{\Ga}{\Gamma}
\newcommand{\de}{\delta}
\newcommand{\De}{\Delta}
\newcommand{\eps}{\epsilon}

\newcommand{\si}{\sigma}
\newcommand{\w}{\omega}

\newcommand{\cB}{\mathcal{B}}

\newcommand{\cS}{\mathcal{S}}
\newcommand{\cC}{\mathcal{C}}
\newcommand{\cO}{\mathcal{O}}
\newcommand{\cP}{\mathcal{P}}

\newcommand{\bA}{\mathbb A}
\newcommand{\bB}{\mathbb B}

\newcommand{\bD}{\mathbb D}

\newcommand{\bN}{\mathbb N}

\newcommand{\bS}{\mathbb S}
\newcommand{\bT}{\mathbb T}

\newcommand{\bZ}{\mathbb Z}

\def\disc{{\mathrm{Disc\,}}}

\newcommand{\LO}{\mathrm{LO}}
\newcommand{\NLO}{\mathrm{NLO}}

\allowdisplaybreaks[2]

\title{Resurgence analysis of the Adler function at $\ord{1/N_f^2}$}
\author[1,2,3]{Eric Laenen,}
\author[1,2]{Coenraad Marinissen,}
\author[1] {Marcel Vonk}

\affiliation[1]{Institute of Physics, University of Amsterdam, Science Park 904, 1098 XH Amsterdam, The Netherlands}
\affiliation[2]{Nikhef, Theory Group, Science Park 105, 1098 XG, Amsterdam, The Netherlands}
\affiliation[3]{Institute for Theoretical Physics, Utrecht University, Leuvenlaan 4, 3584 CE Utrecht, The Netherlands}

\emailAdd{e.l.m.p.laenen@uva.nl}
\emailAdd{c.b.marinissen@uva.nl}
\emailAdd{m.l.vonk@uva.nl}

\abstract{
\noindent
We compute non-perturbative contributions to the Adler function, the derivative of the vacuum polarization function in gauge theory, using resurgence methods and Borel-summed gauge field propagators. At 2-loop, to order $1/N_f$, we construct the full 2-parameter transseries and perform the sum over the non-perturbative sectors. We then introduce a convolution-based method to derive the transseries structure of product series, which can also be used to study higher orders in the expansion in $1/N_f$.  We compute 3-loop planar diagrams, at order $1/N_f^2$, and for each diagram study the asymptotic behavior and resulting non-perturbative information in the transseries.  A structure emerges that, from a resurgence point of view, is quite different from toy models hitherto studied. We study in particular the first and second non-perturbative sectors, their relation to UV and IR renormalons, and how their presence influences the perturbative expansions in neighbouring sectors. Finally, finding that many non-perturbative sectors have asymptotic series, we derive relations among all of them, thus providing an interesting new perspective on the alien lattice for the Adler function.}

\begin{document}
\maketitle

\section{Introduction}
\label{sec:intro}

The perturbative description of gauge theory quantities in terms of a coupling constant is of central
importance to our understanding of such theories, and thereby of our
ability to use them for phenomenology. A full description of these quantities in fact
should include non-perturbative contributions, usually in the form of
power corrections, themselves multiplied by associated perturbative series in the coupling constant.
The direct calculation of such non-perturbative corrections is in many cases
very challenging, and often not even fully understood.

Access to non-perturbative sectors can be gained via the perturbative
series itself. Being an asymptotic series, the series sum must be properly defined.
The intrinsic ambiguities of such a definition take the form of
non-perturbative power corrections.  This interesting aspect of
perturbative series was first explored in \cite{PhysRevD.10.3235,
  Lautrup:1977hs} and in particular in \cite{tHooft:1977xjm}, where the
factorial growth of the perturbative series due to fermion
loop diagrams was discussed, and termed {\em renormalon divergence}. 
Excellent reviews on renormalons are \cite{Beneke:1998ui,Shifman:2013uka}.
In this paper, we shall explore such aspects for the Adler function. To learn more about its non-perturbative aspects, we apply the mathematical techniques of resurgence to this function.

The Adler function $D(Q^2)$ is defined as the logarithmic derivative with respect to $Q^2$ of the
vacuum polarization $\Pi(Q^2)$ in a gauge theory:
\begin{equation}
\label{eq:adlerintro}
D(Q^2) = 4\pi^2Q^2\td{\Pi(Q^2)}{Q^2}\,,
\end{equation}
where $\Pi(Q^2)$ is related to the correlator of two vector currents
$j^\mu=\bar\psi \gamma^\mu \psi$,
\begin{equation}
  \label{eq:12}
  (-\img) \int d^4x \, e^{-\img q x} \,\langle 0 \vert T \,\big\{ j^\mu(x)
  j^\nu(0)\big\} \vert 0 \rangle = (q^\mu q^\nu - \eta^{\mu\nu}q^2) \Pi(Q^2)\,, \qquad Q^2 = -q^2\,.
\end{equation}
The Adler function has long featured as an object to study  asymptotic behaviour
of perturbative series and renormalons, see e.g. \cite{Beneke:1992ch, Broadhurst:1992si, Neubert:1994vb}. This is because, as a
self-energy, it depends on only one scale, and because its
imaginary part, for timelike $Q^2$, is 
directly related to the $R$-ratio of the inclusive
$e^+ e^-$ cross section to hadrons and muons.

There has been much earlier work on renormalon contributions to the Adler function  \cite{Beneke:1992ch,Beneke:1993ee,Broadhurst:1992si,Broadhurst:1993ru,Neubert:1994vb,Vainshtein:1994ff,Ball:1995ni,Beneke:1995qq,Peris:1997dq}, as well as their connection to its operator production expansion \cite{Parisi:1978az,Mueller:1984vh}. The Adler function appears in more recent  renormalon studies \cite{Cvetic:2018qxs,Ayala:2021mwc,Ayala:2022cxo,Boito:2018rwt,Caprini:2020lff}, including a study whether field theories can have renormalons beyond the usual first Borel plane \cite{Cavalcanti:2020osb}. The Adler function has recently been studied using resurgence techniques \cite{Maiezza:2021mry,Maiezza:2021bed}, based on renormalization group considerations\footnote{Note that in these works, assumptions are used that reproduce the expectations by Parisi and 't Hooft \cite{Parisi:1978az, Parisi:1978bj, tHooft:1977xjm} that renormalons occur at half-integer multiples of $1/\beta_0$, the inverse of the leading beta function coefficient. It was found recently \cite{Marino:2021dzn} that in several examples, this pattern is in fact {\em not} present -- even though the breaking of the pattern only shows up at finite $N_f$ and not in a perturbative expansion in the reciprocal of that parameter.} \cite{Maiezza:2019dht,Bersini:2019axn}. 

In this paper we dive deeper into the method of resurgence, a set of mathematical techniques already introduced by J.~Écalle in the 1980s \cite{ecalle1985fonctions}. It has become a popular technique for the study of non-perturbative effects in quantum field theories and string theory over the past decade or so. Good reviews from a physics point of view are e.g.\ \cite{Aniceto:2018bis, Marino:2012zq, Dorigoni:2014hea}, as well as the review sections of \cite{Aniceto:2011nu}. For a nice mathematical introduction, see \cite{Sauzin:2014}. The application of resurgence to renormalon physics was studied e.g.\ in \cite{Shifman:2022xsa} and in a series of papers starting with \cite{Marino:2019eym} and summarized in \cite{Reis:2022tni}.

Resurgence provides tools to systematically decode non-perturbative information directly from the perturbative data. The techniques usually require a substantial number of perturbative terms (of order 10 at least, and for precision numerics of order 100), which explains why resurgence has not been widely applied to particle physics and phenomenology yet. Let us note that more broadly in quantum field theory, resurgence techniques have been introduced in e.g. \cite{Dunne:2012ae, Dunne:2013ada, Krichever:2020tgp, Borinsky:2022knn} and follow-up works by the same authors.  The time seems ripe for application to more phenomenologically relevant quantities, since for many observables in particle physics a  substantial number perturbative terms can be computed. 
The Adler function is a good starting point. In fact, we shall be able extract new non-perturbative results for the Adler function at $\ord{1/N_f}$ and $\ord{1/N^2_f}$, where $N_f$ is the number of flavours in the fermion loop diagram. In the resurgence literature the description of both perturbative and non-perturbative contributions is conveniently summarized in a  {\em transseries} (see e.g.\ \cite{edgar2010transseries} for an introduction to the concept) and here we will do the same.

To obtain our perturbative data, we compute two- and three-loop skeleton diagrams for the Adler function for gauge theory, using the gauge field propagator with fermion loop insertions summed in Borel space. For the two-loop case ($\ord{1/N_f}$) we find, using our resurgence techniques,the complete transseries of the Adler function due to renormalon contributions, including subleading effects, which reproduces and extends earlier results in \cite{Beneke:1992ch,Broadhurst:1992si,Neubert:1994vb,Maiezza:2021mry}.  Subsequently we develop a powerful convolution method that facilitates the calculation of renormalon contributions due to adding new fermion-loop-summed propagators. 

We test our findings in calculations of three-loop skeleton diagrams ($\ord{1/N^2_f}$), where we combine the decomposition into master integrals with the convolution method and very high order expansions in the Borel plane. We find asymptotic series in the non-perturbative sectors, and operators that establish relations between these sectors. Moreover, we find a new logarithmic type of non-perturbative power correction in the coupling constant plane. When we put all ingredients together, an intricate transseries structure emerges. Our results derive from thorough analytical and numerical analyses, which we describe extensively in the later sections. 

\bigskip
\noindent Before we describe the organization of the paper in more detail, we note that it addesses two communities: the high energy physics community interested in renormalons, and the mathematical physics community interested in resurgence. Our hope is to interest the former community about the latter topic and vice versa. For this reason, we have tried to make the paper relatively self-contained on both topics.

Thus, the paper starts with introductory sections on renormalons and  resurgence.  \sec{renormalons} covers the basics of renormalons, in particular as they occur in QED and QCD. After introducing the technique of Borel summation that is central to all of our studies, we explain the difference between UV and IR renormalons and see how they originate in the so-called bubble diagrams. \sec{resurgence} then contains an introduction to the topic of resurgence. We do not aim to be exhaustive, but introduce the concepts that are relevant for us, such as alien calculus, transseries, alien chains/lattices and large order relations and refer to the literature mentioned above for further details. We end the section by highlighting in which ways the resurgent structures appearing in this paper go beyond the simplest toy models and general lore one usually encounters.

We then turn to the Adler function. \sec{adlerLO} discusses how renormalons are usually studied for the Adler function: by introducing a single bubble chain in the Feynman diagrams. This section contains many known results, but we try to present these in a way that is most suitable for a resurgent analysis. We start with a brief introduction on the flavour expansion in QED/QCD, formalizing the bubble diagrams already discussed in \sec{renormalons} and showing a framework in which these bubble diagrams can be ordered. We give the exact expression of the Adler function  at leading order in the flavour expansion. Subsequently we perform the resurgence analysis of the Adler function at this order, showing how resurgence and large order relations can be used to extract all non-perturbative sectors in the corresponding transseries. We finish this section with a discussion on some of the more subtle issues encountered at this order.

Our aim is then to explore higher orders in the flavour expansion and investigate which aspects of the leading order analysis persist, and which ones may even show a richer structure. Before doing  computations at next to leading order, we discuss in \sec{convoInt}  the convolution integral that is an important ingredient in the calculation at higher orders. Building on toy models, we explore the resurgent structure that arises from these convolution integrals in the computations for Adler function Feynman diagrams with several bubble chains. While many of the mathematical results in this section may be familiar to the resurgence community,  our application and interpretation of these results sheds a new light.

In \sec{adlerNLO} we then compute a set of planar diagrams present at next to leading order in the flavour expansion of the Adler function. The degree of difficulty here varies strongly from diagram to diagram; we are able to obtain results for some diagrams as arbitrary order expansions and for one as an expansion to a certain finite order. Despite only having partial data, we can investigate much of the resurgent structure that occurs at order $1/N_f^2$. We discuss the non-perturbative sectors for individual diagrams as well as the alien lattice structure, which is considerably richer than at order $1/N_f$.

We summarize our findings in \sec{conclusion} and discuss some open questions. In four appendices, we provide background on the calculation of the diagrams and give details on the numerical methods used in the main text.

\section{Renormalons}
\label{sec:renormalons}
In gauge theories, and more generally in quantum physics, perturbatively expanded observables $F = \sum_n c_n \, \alpha^{n+1}$ often have coefficients that grow as $c_n \sim n!$. This causes the perturbative series to be asymptotic, at first decreasing with increasing order, but then succumbing to the factorial growth. In particular, the series has zero radius of convergence irrespective of the size of $\alpha$. The factorial growth of the coefficients often indicates the presence of effects that are non-analytic at $\al = 0$, generally of the form $e^{-A/\al}$. One well-known source of such effects is the occurrence of instantons in the theory. The occurrence of an instanton is associated to the fact that the number of Feynman diagrams that one can draw at order $\al^n$ itself grows like $n!$ \cite{Lipatov:1976ny, tHooft:1977xjm, LeGuillou:1990nq}.

However, there is a second type of effect that causes perturbative coefficients to grow as $c_n \sim n!$: {\em renormalons}. Renormalon divergences \cite{tHooft:1977xjm} occur when individual diagrams in the Feynman expansion at order $\al^n$ are of size $n!$. Like instantons, these effects lead to singularities of the Borel transform (to be introduced shortly), but in this case their cause is the large or small loop momentum behaviour of these particular diagrams \cite{Beneke:1998ui}. Often, these diagrams contain chains of fermion bubbles, and as we shall review below, this is also the case for the Adler function. 

To an asymptotic perturbative series, whether it comes from instantons or from renormalons, a 
finite value can nevertheless be assigned, even for non-zero $\alpha$. The simplest method to do so is by summing the series up to its smallest term. However, any assignment leaves ambiguities,
which will in fact be the main occupation of this paper.  Many quantum field theories exhibit
renormalon behavior, including gauge theories in four dimensions, and their associated ambiguities can act as a portal to non-perturbative information -- see e.g.\ \cite{Beneke:1998ui,Shifman:2013uka} for reviews.

\subsection{Borel summation}
\label{sec:borelSummation}
An alternative way of assigning a finite value to an asymptotic series
involves the Borel transform. Effectively, it amounts to dividing the $n$th coefficient by $n!$:
\begin{equation}
\label{eq:2.1}
F=\sum_nc_n\alpha^{n+1}
\quad\to\quad
\cB[F](t)=\sum_n \frac{c_n}{n!} t^n,   
\end{equation}
where $t$ is the Borel transformation parameter conjugate to $\alpha$,
and where we have included an extra power of $\alpha$ on the left for
later convenience. For asymptotic series whose coefficients grow factorially, the additional factorial in the denominator leads to a finite radius of convergence. In this case
one may endeavour to sum the Borel tranformed series, and after that
invert the transform to obtain a ``Borel-summed" expression for the
original series.  To make the inversion precise one traditionally defines the Borel
sum as follows
\begin{equation}
\label{eq:borelResummation}
\cS_0[F](\alpha)
= \int_0^\infty dt\, \cB[F](t)\, e^{-\frac{t}{\alpha}}.
\end{equation}
Instead of integrating along the real positive line, one can instead integrate in a different direction $\theta$ in the complex $t$-plane for which the integral converges. 
This leads to a Borel sum $\cS_\theta[F]$ where the integration in \eq{borelResummation} is now from $0$ to $e^{\img\theta}\infty$. We will need such more general Borel transforms in what follows.

It might appear as if the process of Borel transforming and then Borel summing is a tautology, since the integral reinstates the factorial growth, e.g.
\begin{equation}
\int_0^\infty dt\, t^n\, e^{-\frac{t}{\alpha}}
= \Ga(n+1)\alpha^{n+1}.
\end{equation}
However, one can now first perform the infinite sum $\cB[F]$, 
analytically continue in the complex $t$ plane and then perform the integral in
\eq{borelResummation}, thus effectively interchanging integration and
summation. One can show that this leads to a finite function $\cS_0[F]$
\cite{Borel:1899}.  

\begin{figure}
    \centering
    \includegraphics[width=.6\textwidth]{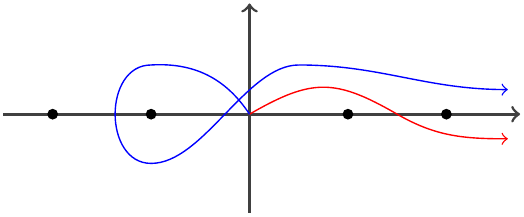}
    \caption{Deformations of the Borel integration contour.}
    \label{fig:lasso}
\end{figure}

The Borel sum $\cB[F]$ can still have singularities in the
complex $t$ plane, which are in fact related to the factorial
growth. A simple example illustrates this point. In \eq{2.1} let us
choose $c_n=n!$. In that case we have
\begin{equation}
    \label{eq:irexample}
    c_n = n!  \,, \quad \to \quad \cB[F] = \frac{1}{1-t}\,,
\end{equation}
leading to a pole at $t=1$. As another example, consider
\begin{equation}
\label{eq:4}
c_n = (n-1)!\,,
\quad\to\quad    
\cB[F](t)=\sum_{n=1}^\infty \frac1n t^n = -\log(1-t)\,,
\end{equation}
which leads to a branch cut running from 1 to $\infty$.
In the case that these singularities lie on the integration contour
from $0$ to $\infty$ in \eq{borelResummation} this leads to an
ambiguity, as one can deform the contour such
that one integrates either slightly above the singularity under an angle $\theta=0^+$, or below the singularity under an angle $\theta=0^-$, see the red curve in \fig{lasso}. 
Deciding to integrate either above or below for the example \eq{irexample} e.g.\ leads to an ambiguity
\begin{equation}
\label{eq:3}
  \cS_{0^+}[F]-\cS_{0^-}[F] = 2\pi\img\, e^{-\frac1\alpha}.
\end{equation}
Other singularities in the complex $t$ plane will influence the
integral in \eq{borelResummation}; the contour may be chosen in many ways, for example as the
blue curve in \fig{lasso}, adding sensitivity to singularities on the
negative Borel axis to the Borel-summed result. Notice the
non-perturbative aspect of the ambiguities like in \eqref{eq:3}.

For QED and QCD the type of (renormalon)
singularity in the Borel plane depends on the loop-momentum regions, specifically the
ultraviolet (UV) and infrared (IR) ones. We review this in the next subsection.

\subsection{UV and IR renormalons}
\label{sec:uvirrenormalons}

\begin{figure}
    \centering
    \includegraphics[width=.6\textwidth]{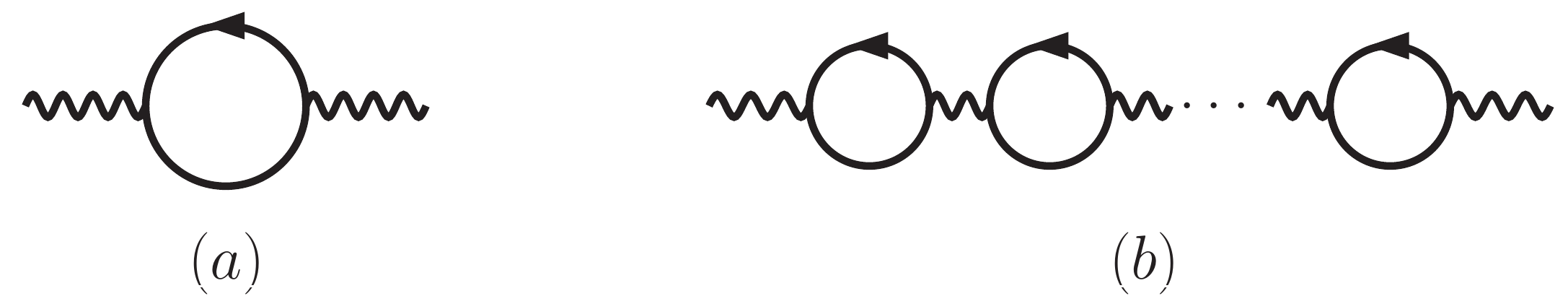}
    \caption{(a) One fermion bubble, (b) chain of fermion bubbles.}
    \label{fig:vacpol}
\end{figure}

A class of diagrams that by itself leads to factorial coefficient
growth, and is typical for renormalons, are bubble-chain diagrams, in which $n$ fermionic self energies
are inserted in the photon or gluon propagator, see
\fig{vacpol}. To see the occurrence of the factorial growth, consider its constituent element, the
one-loop vacuum-polarization graph (``bubble'') in \fig{vacpol}(a). Its expression reads
\begin{equation}
    \label{eq:vacpolone}
    \Pi_{\mu\nu} = (k_\mu k_\nu- k^2 \eta_{\mu\nu}) \pi(k,\mu)
\end{equation}
with
\begin{equation}
    \label{eq:vacpolint}
    \pi(k,\mu) =\frac{-2\mathrm{i} N_f e^2 \mu^{2\epsilon}}{(2\pi)^n} \left( \frac{2-n}{1-n} \right) \int d^nq \frac{1}{(q+\tfrac{1}{2}k)^2(q-\tfrac{1}{2}k)^2}\,.
\end{equation}
This integral can be readily carried out using a Feynman parameter leading to 
\begin{align}
    \label{eq:vacpolint2}
    \pi(k,\mu) = &\frac{2N_f e^2}{(4\pi)^2}\left(\frac{4\pi \mu^2}{-k^2} \right)^\epsilon \frac{(2-2\epsilon) \Gamma(1+\epsilon)\Gamma(1-\epsilon)^2}{\epsilon\, (3-2\epsilon) \Gamma(2-2\epsilon)}\nonumber \\
    =& \frac{\alpha N_f}{3\pi}\left[\frac{1}{\epsilon} -\gamma_E + \log 4\pi - \log\left(\frac{-k^2}{\mu^2} \right) + \frac{5}{3} \right] + \cO(\epsilon)\,.
\end{align}
To remove the ultraviolet divergence in \eqref{eq:vacpolint2} one
includes an $\overline{\mathrm{MS}}$ counterterm, which leads to 
\begin{equation}
        \label{eq:vacpolint3}
    \pi(k,\mu) = -\alpha \beta_{0f}
    \log\left(\frac{-k^2 e^{-5/3}}{\mu^2}
      \right) \,,
    \end{equation}
    where $\beta_{0f} = T \, N_f /3\pi$, with $T=1$ for QED and
    $T=1/2$ for QCD.
Bubble-chain diagrams produce factorial growth at leading power in
$N_f$, i.e. the $n$-bubble contribution is proportional to $(\alpha N_f)^n$.

\begin{figure}
    \centering
    \includegraphics[width=.6\textwidth]{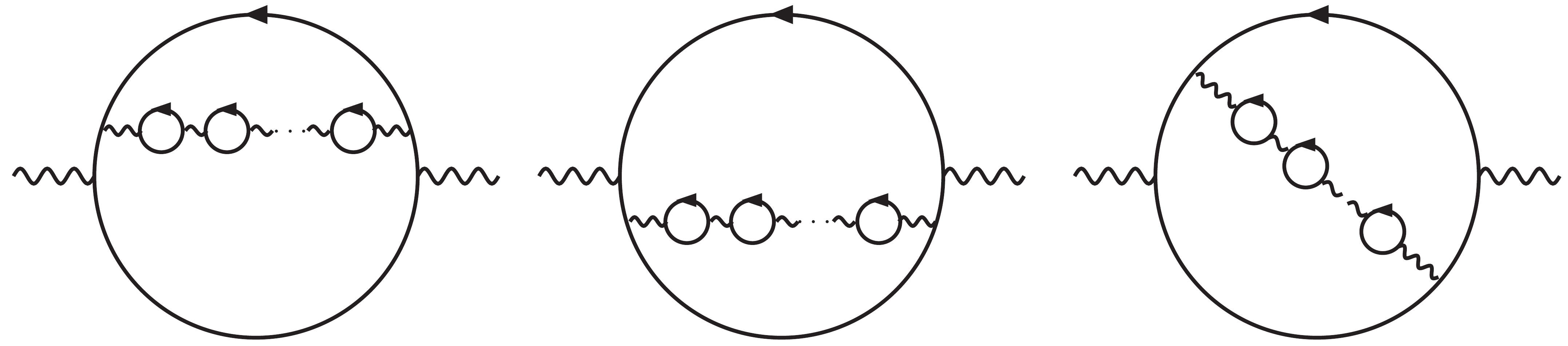}
    \caption{Bubble chain inserted into Adler function. }
    \label{fig:vacpol2}
  \end{figure}

For the Adler function, we consider the diagrams in \fig{vacpol2}, where a fermion bubble chain with $n$ bubbles 
is inserted. 
The diagrams, summed over the number of bubbles, yield the expression
  \begin{equation}
    \label{eq:1}
    D = \al\int_0^\infty \frac{d\hat{k}^2}{\hat{k}^2} \, F(\hat{k}^2)
    \sum_n \left[ \alpha \beta_{0f}
    \log\left(\frac{\hat{k}^2Q^2 e^{-5/3}}{\mu^2}
      \right) \right]^n\,,
  \end{equation}
where $\hat{k}^2 = -k^2/Q^2$, and $F$ represents the rest of the
diagram (see \cite{Neubert:1994vb}). We can examine the $n$-dependence of
the integral separately for small  and large $\hat{k}^2$ (the
separation being defined by the argument
in the logarithm in \eqref{eq:1} being unity), using the small and
large $\hat{k}^2$ dependence of $F$. Up to subleading corrections the result \cite{Neubert:1994vb, Beneke:1998ui}  reads
\begin{equation}
  \label{eq:2}
  D = \frac{C_F}{\pi} \sum_{n=0}^\infty \alpha_s^{n+1}
\Big[\frac{3}{4}\left(\frac{Q^2 e^{-5/3}}{\mu^2}
      \right)^{-2} \left(\frac{-\beta_{0f}}{2} \right)^2\,n!  +
      \frac{1}{3} \frac{Q^2 e^{-5/3}}{\mu^2} \beta_{0f}^n \, n!\,
      \left(n+\frac{11}{6}\right)\Big]\,,
\end{equation}
where the first part is due to the IR behavior and the second to the
UV behavior of the integrand in \eqref{eq:1}. 

To see what this implies for the Borel transform of the
Adler function, we use \eqref{eq:2.1}. 
For the IR case one finds  ($u\equiv-\beta_{0f} t$)
\begin{equation}
  \label{eq:11}
     \cB[D]_{\mathrm{IR}} (u) \sim  \frac{1}{u-2}\,, \qquad     \cB[D]_{\mathrm{UV}} (u) \sim   \frac{1}{(u+1)^2}\,,
   \end{equation}
   where we only exhibit the leading $u$-poles and suppress prefactors.
Thus IR renormalon poles for QED  will lie on the negative
Borel parameter axis, and vice versa for the UV renormalon
poles. As we will discuss in section \ref{sec:flavour}, for QCD one
changes $\beta_{0f} \to \beta_{0} = -(11-2N_f/3)/(4\pi)$, which also
implies a change of sign for $u$, so that IR (UV) renormalon poles are
on the positive (negative) real $u$ axis.

As shown in \eqref{eq:3} and \eqref{eq:borelResummation}, the poles on
the positive $u$-axis imply ambiguities in the Borel-resummed
perturbative series. In fact, they imply non-perturbative corrections;
the QCD version of the ambiguity in equation \eqref{eq:3}, due
the pole at $u=2$ in \eqref{eq:11}, is proportional to
\begin{equation}
  \label{eq:10}
e^{\frac{2}{\beta_0\alpha}} \sim \left(\frac{\Lambda}{Q}\right)^4\,,
\end{equation}
with $\Lambda$ the QCD scale parameter, and $Q$ the
 scale of the Adler function. Non-perturbative effects in QCD are
 proportional to this typically very small ratio, and are therefore
 often also referred to as power corrections. Our goal in this paper, then, is
 to learn more about power corrections for the Adler function, using
 the methods of resurgence.

\section{Resurgence}
\label{sec:resurgence}
\noindent
In the previous section we saw that factorial growth in the perturbative
series of a quantity leads to ambiguities that have the form of non-perturbative
corrections expressed in terms of $e^{-\frac{1}{\al}}$. This indicates that correction terms must be added to the
series expansion to remove such ambiguities, and as a bonus
non-perturbative information is included. {\em Resurgence}, an idea which originated in \cite{ecalle1985fonctions}, is a way to do
this systematically; the word refers to non-perturbative sectors
resurging through analysis of asymptotic behaviour of other sectors.
The perturbative expansion enhanced by such sectors is known as the
\emph{transseries} (see \cite{edgar2010transseries} for an
introduction), which will feature extensively in what follows.

The goal of this section is twofold. First, we introduce notation and definitions needed for the resurgence analysis of the Adler function in later sections. 
Secondly, for those who have not yet encountered the concepts and
methods of resurgence, this section can be read as a concise
introduction to the topic. Of course, resurgence is a broad topic, and
we refer the interested reader to \cite{Aniceto:2018bis, Marino:2012zq, Dorigoni:2014hea, Sauzin:2014} or the review sections of \cite{Aniceto:2011nu} for a more extensive exposition. Below, we will mostly adhere to the notation and presentation of \cite{Aniceto:2018bis}.

\subsection{From Borel summation to alien calculus}
\label{sec:alienCalculus}
\noindent
In physics, singularities and divergences are usually regarded as
troublesome. However, in a resurgence context, all the
information about an asymptotic series and the associated
non-perturbative data is encoded in the singularity
structure of its Borel transform. Before discussing how to
systematically extract this information, we need to introduce some
concepts and terminology.

Let us again consider a formal power series and its Borel transform,
\be
F(\al) = \sum_{n=0}^\infty f_n \al^{n+1}, 
\qquad 
\cB[F](t) = \sum_{n=0}^\infty \frac{f_n}{\Ga(n+1)} t^n.
\label{eq:pertBorel}
\ee
A function whose expansion gives a formal power series $F(\al)$ is said to be a {\em simple resurgent function} if the Borel transform $\cB[F](t)$ has only simple poles or logarithmic branch cuts as singularities. That is, near each singularity, say at $t = \w$, we have that
\begin{equation}\label{eq:defSimpleResurgentFunction}
\cB[F](t) 
= \frac{a}{2\pi \img\, (t-\w)} + \Psi(t-\w)\frac{\log(t-\w)}{2\pi \img} + \Phi(t-\w),
\end{equation}
for some $a\in\mathbb{C}$ and where the functions $\Psi,\Phi$ are
analytic around the origin. The Borel transform can also contain other
singularities, e.g.\ double or higher order poles. Although this will
be the case for the Adler function, most of our discussion in this
section is focused on the simplest case of a single pole and a $\log$-branch
cut that are commonly considered in the literature. In \sec{generalizations}, we then give the necessary details to
extend the resurgence analysis needed for the discussions of the Adler
function in \secs{adlerLO}{adlerNLO}.

To obtain a better understanding of these Borel
singularities we introduce {\em alien calculus}. The
fundamental object in alien calculus is the linear differential
operator $\Delta_\w$, the {\em alien derivative}, which acts on simple
resurgent functions. Being a derivative, it
satisfies the Leibniz rule when acting on a product of simple
resurgent functions. Furthermore, for those functions it has a rather
simple expression: by rewriting $\Psi(t)$ in
\eq{defSimpleResurgentFunction} as the Borel transform of a resurgent
function $G(\al)$, i.e.\ $\cB[G](t) = \Psi(t)$, the alien derivative
at a singular point $\w$ is
\begin{equation}\label{eq:formalDefAlienDerivative}
\Delta_\w F(\al) = a + G(\al).
\end{equation}
In fact, this operation is defined both `before' and `after' Borel summation: the alien derivative can also be viewed as a map relating the power series expansion of $F(\al)$ to that of $G(\al)$.
When $\w$ is not a singular point of $\cB[F](t)$, then $\Delta_\w F(\al) = 0$. 
In other words, the alien derivatives fully encode the singular properties of the Borel transforms of simple resurgent functions.

Next, let us consider a singular direction $\theta$ in the complex
plane of the Borel variable $t$ along which $\cB[F]$ has
singularities. Such a direction is known as a {\em Stokes line}, since as we shall see Stokes' phenomenon \cite{stokes_2009a, stokes_2009b} occurs there. Because of
the singularities on the integration path, the Borel summation
$\cS_\theta[F]$ is no longer well-defined. Of course, the
singularities are easily avoided by integrating slightly above or
below them, leading to summations that we denote by
$\cS_{\theta^+}[F]$ and $\cS_{\theta^-}[F]$. These different
summations lead to an ambiguity and one may ask how these two
distinct functions are related. They are in fact connected by the {\em
  Stokes automorphism} $\underline{\mathfrak{S}}_\theta$, or its
related {\em discontinuity} $\disc_\theta$.
These are defined as 
\begin{equation}\label{eq:defDiscontinuity}
\cS_{\theta^+} 
= \cS_{\theta^-}\circ \underline{\mathfrak{S}}_\theta 
=\cS_{\theta^-} \circ (1 - \disc_{\theta^-}).
\end{equation}
One can show (see e.g.\ \cite{Sauzin:2014}) that these operators can be expressed in terms of the alien derivatives via the exponential map
\begin{equation}\label{eq:StokesAutomorphism}
	\underline{\mathfrak{S}}_\theta
	= \exp \Bigg[\sum_{\{\w_\theta\}} e^{-\frac{\w_\theta}{\al}}\Delta_{\w_\theta}\Bigg],
\end{equation}
where the set $\{\w_\theta\}$ denotes all the singular points along
the $\theta$-direction. The main point now is that if we know the
Stokes automorphisms (or equivalently, the alien derivatives), then we
also know how to relate the different summations. Consequently, at least in principle a full
reconstruction anywhere in the complex $\alpha$ plane, including non-perturbative sectors, of the function $F(\alpha)$ is then possible.

For future purposes we define the {\em pointed alien derivative}, related to the ordinary alien derivative by
\begin{equation}
	\dot{\Delta}_\w \equiv e^{-\frac{\w}{\al}}\Delta_\w.
\end{equation} 
This turns out to be a convenient operator as it commutes with the usual derivative (see again \cite{Sauzin:2014} for details), i.e. 
\begin{equation}
\label{eq:commutatorDeltaDotPartial}
[\dot{\Delta}_\w,\partial_\al] = 0.
\end{equation}
This will momentarily be used to derive a bridge equation.

\subsection{Transseries and bridge equations}
\label{sec:bridgeEquation}
\noindent
From the relation between Stokes automorphisms and alien derivatives,
\eq{StokesAutomorphism}, we notice that the ambiguity arising in Borel
summation is non-perturbative in $\al$, being of order
$e^{-\frac\omega\al}$.
This implies that the non-perturbative solution we are trying to
construct must
contain such non-perturbative exponential contributions. 
As we will see in a moment, resurgence relations can be captured in a
universal way through transseries \cite{edgar2010transseries}. Transseries are
generalizations of perturbative series by the inclusion of terms with
non-perturbative (non-analytic) factors like
$e^{-\frac{\omega}{\al}}$. Factors of this type are called {\em transmonomials}.

Let us start by assuming that our resurgent function arises as a
solution to some (potentially non-linear) problem depending on a
single boundary condition, i.e.\ we consider the single parameter
transseries Ansatz
\begin{equation}\label{eq:transseries}
F(\al,\si) = \sum_{n=0}^\infty \si^n e^{-n\frac{A}{\al}} \Psi^{(n)}(\al)\,,
\qquad\text{with}\qquad
\Psi^{(n)}(\al) 
= \al^{\beta_n} \; \sum_{h=0}^\infty f_h^{(n)}\al^{h}
\end{equation}
where $\Psi^{(0)}$ is simply a perturbative series, as in \eq{pertBorel}, and 
$\Psi^{(n)}$, for $n\geq1$, are the non-perturbative contributions. The transseries parameter $\si$ counts the number of  $e^{-\frac{A}{\al}}$ factors \cite{Aniceto:2018bis} and
parameterizes different choices of the boundary condition. The
$\beta_n$ are called the {\em characteristic exponents}; we discuss their
role in resurgence equations in the end of \sec{generalizations}. In
the resurgence literature, $\Psi^{(n)}$ is often
called the {\em $n$-instanton sector}, even though in practice $n$ may
count solitons, renormalons, or some other non-perturbative physical
quantity. To avoid confusing readers with a high-energy background, we
shall call these quantities the {\em $n$-th non-perturbative sector} instead.

After introducing an intuitive pictorial representation of
non-perturbative transseries sectors and their interrelations in the
form of the {\em alien chain} in the next subsection, it will be
straightforward in \sec{generalizations} to generalize this one-parameter transseries to
multi-parameter transseries, by including further non-perturbative
monomials like $e^{+nA/\al}$, $\log^n(\al)$, etc.

We saw that the alien derivatives play an important role in the
construction of the complete non-perturbative solution to a
problem. It is however still unclear how to compute these derivatives
in a systematic way. This is done through the construction of the {\em
  bridge equations}, so named because they form a bridge between the
ordinary calculus of derivatives and alien calculus. 
Assume for the moment that $F(\al,\si)$ is the solution to some differential equation (in the variable $\al$). From \eq{commutatorDeltaDotPartial} we get that acting on this equation with $\dot{\Delta}_{\w}$ yields a new, linear differential equation to which $\dot{\Delta}_{\w} F(\al,\si)$ is a solution. At the same time, acting on the original equation with $\partial_\si$ shows that $\partial_\si F(\al,\si)$ is a solution to the {\em same} equation. 
As an example, consider the non-linear differential equation
\begin{equation}
\partial_\al F(\al,\si) 
= 6+ F(\al,\si)^3.
\end{equation}
Acting on this with $\dot{\Delta}_{\w}$ yields
\begin{equation}\label{eq:bridgeEqExample}
\partial_\al \dot{\Delta}_{\w} F(\al,\si) = 3 F(\al,\si)^2 \cdot \dot{\Delta}_{\w} F(\al,\si),
\end{equation}
which is a new, linear differential equation for $\dot{\Delta}_{\w} F(\al,\si)$. Similarly, acting with $\partial_\si$ leads to the same equation as \eq{bridgeEqExample}, with $\dot{\Delta}_{\w} F(\al,\si)$ replaced by $\partial_\si F(\al,\si)$.
Supposing the new linear differential equation is of first order (as is natural for a problem with a single boundary condition) we conclude that the two new solutions must be proportional to each other, i.e.
\begin{equation}\label{eq:bridgeEq}
\dot{\Delta}_{\w} F(\al,\si) = S_\w(\si) \partial_\si F(\al,\si),
\end{equation}
where $S_\w(\si)$ is a proportionality factor which still may depend
on $\si$. This relation is \'Ecalle's bridge equation \cite{ecalle1985fonctions}; it indeed
presents a bridge between the alien derivatives and the regular ones. By substituting $\dot{\Delta}_{\w} = e^{-\frac{\w}{\al}}\Delta_{\w}$ and expanding the transseries, the LHS equals
\begin{equation}\label{eq:bridgeEqLHS}
\dot{\Delta}_{\w} F(\al,\si) 
= \sum_{n=0}^\infty \si^n e^{-\frac{nA+\w}{\al}}
\Delta_{\w}\Psi^{(n)}(\al),
\end{equation}
while the RHS yields
\begin{equation}\label{eq:bridgeEqRHS}
S_\w(\si) \partial_\si F(\al,\si) 
= S_\w(\si)\sum_{n=0}^\infty n\, \si^{n-1} e^{-n\frac{A}{\al}}\Psi^{(n)}(\al).
\end{equation}
We need to match both sides term by term according to the powers of $\si$ and  $e^{-\frac{A}{\al}}$. 
To this end, \cite{Aniceto:2011nu} defined a notion of {\em degree} as
\begin{equation}
\deg\big(\si^n e^{m\frac{A}{\al}}\big) = n+m.
\end{equation}
Since the transseries \eq{transseries} has degree $\deg( F(\al,\si))=0$, it follows that the bridge equation can only contain nontrivial information at $\w=\ell A$, $\ell\in \mathbb{Z}$. Thus, for transseries of this type we expect singularities in the Borel plane at $\omega = A$ but also at integer multiples of $A$ -- something which we will also see for the Adler function. Note that here $\ell=0$ is excluded because the Borel transform is regular at the origin. Furthermore, $\dot{\Delta}_{\ell A} F(\al,\si) $ only contains positive powers of $\si$, so we can write the proportionality constant as a formal power series expansion
\begin{equation}
S_{\ell A}(\si) = \sum_{k=0}^\infty S_\ell^{(k)}\si^k.
\end{equation}
Taking the degree of both \eqs{bridgeEqLHS}{bridgeEqRHS} implies that $S_\ell^{(k)}$ is only nonzero at $k=1-\ell$, and therefore, writing $S_\ell^{(1-\ell)} \equiv S_\ell$, we have that
\begin{equation}
S_{\ell A}(\si) = S_\ell \,\si^{1-\ell}, \quad \ell\leq1.
\end{equation}
The bridge equation \eq{bridgeEq} now reads
\begin{equation}
\sum_{n=0}^\infty \si^n e^{-(n+\ell)\frac{A}{\al}}
\Delta_{\ell A}\Psi^{(n)}(\al)
= \sum_{n=0}^\infty S_\ell\, n\, \si^{n-\ell} e^{-n\frac{A}{\al}}\Psi^{(n)}(\al),
\end{equation}
or equivalently
\begin{equation}\label{eq:resurgenceEqs}
\Delta_{\ell A} \Psi^{(n)}(\al) = 
\begin{cases}
	0 & \ell > 1,\\
	 (n+\ell)S_\ell \Psi^{(n+\ell)} & \ell\leq 1, \quad \ell\neq0,
\end{cases}
\end{equation}
where we used that $\Psi^{(n)}= 0$ for all $n< 0$. 
With this expression, we can compute all alien derivatives as long as
the yet unknown constants $\{S_1,S_{-1},S_{-2},...\}$ are known. We
refer to these constants as {\em Stokes constants}. In general,
computing Stokes constants is a difficult task and depends on the
specific system that one tries to solve, and we shall not need to do so in this paper. 

Although the alien derivative has an involved definition, we see a
remarkable result in alien calculus: the final result for the alien
derivative is surprisingly simple, and works purely algebraically on
the building blocks of the transseries, \eq{transseries}.

For simple resurgent functions with a single parameter, we now have two equations for the alien derivative, i.e.\ \eqs{formalDefAlienDerivative}{resurgenceEqs}. Comparing the two yields
\begin{equation}\label{eq:borelSingularities}
\cB[\Psi^{(n)}](t+\ell A) 
= \frac{a}{2\pi \img\, t} + \mathsf{S}_{n\to n+\ell}\ \cB[\Psi^{(n+\ell)}](t)\frac{\log(t)}{2\pi \img} + \text{holomorphic}.
\end{equation}
That is, near the singularity $t=\ell A$ of $\cB[\Psi^{(n)}](t)$, we find the resurgence of the the $\Psi^{(n+\ell)}$ sector. 
In this expression, the so-called {\em Borel residues} $\mathsf{S}_{n\to n+\ell}$ are constants that can be expressed in terms of the Stokes constants via Eqs.\ \eqref{eq:defDiscontinuity}, \eqref{eq:StokesAutomorphism} and \eqref{eq:resurgenceEqs}. For example we find
\begin{equation}
\mathsf{S}_{n\to n+1}
= -(n+1)S_1,
\quad
\mathsf{S}_{n\to n+2} = -\frac12 (n+1)(n+2) S_1^2,
\quad\text{etc.}
\end{equation}
and similarly 
\begin{equation}
\mathsf{S}_{n\to n-2}
= -(n-1)S_{-1},
\quad
\mathsf{S}_{n\to n-2}
= -(n-2)\left(S_{-2} + \frac12(n-1)S_{-1}^2\right),
\quad\text{etc.}
\end{equation}
See \cite{Aniceto:2011nu} for closed-form expressions for the Borel residues.

\subsection{One-dimensional alien chain}
\label{sec:1Dalienchain}
\noindent
Instead of entering the world of alien calculus, we will follow the
more pedagogical picture of the {\em alien chain} developed in
\cite{Aniceto:2018bis} as it will help us build an intuitive language
in which resurgence equations can be better understood.  For example,
in the case of a single boundary condition, Stokes' automorphism 
\eq{StokesAutomorphism} can be fully computed using
\eq{resurgenceEqs} (see e.g. \cite{Aniceto:2011nu}), but once one
needs to generalize this to multi-parameter transseries, the equations
can become quite intricate. Instead, for practical situations, the
simple setup of the alien chain can be used and generalizations will
come naturally. (For multi-parameter transseries, we will introduce
alien {\em lattices} in \sec{generalizations}.)

From the point of view of the alien chain, the sectors $\Psi^{(n)}$ of the
transseries, \eq{transseries}, are viewed as nodes:
\begin{equation*}
\vcenter{\hbox{\includegraphics[width=.8\linewidth]{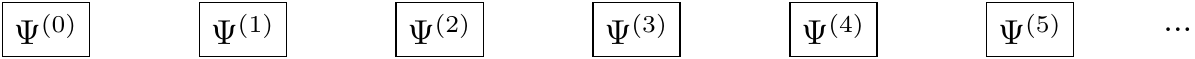}}}
\end{equation*}
Later these will be connected by alien derivatives to form a chain. We can then reinterpret the resurgence equation, \eq{resurgenceEqs},
as a set of allowed resurgence ``motions'' along this chain. That is,
there is only one type of forward motion (i.e.\ with $\ell$ positive)
because of the constraint that $\De_{\ell A}\Psi^{(n)} = 0$ for
$\ell>1$. However, from the fact that nonzero $\ell\leq1$ can give
nonvanishing alien derivatives, we see that there are multiple
backwards motions. Recall that (for real and positive $A$) to compute the Stokes
automorphism \eq{StokesAutomorphism} at $\theta = 0$ one only needs
alien derivatives with $\ell>0$, whereas the Stokes automorphism at
$\theta=\pi$ only requires $\ell<0$. Therefore, we never need to
consider combinations of forward and backward motions together.

As an example, there is only one forward path to go from $\Psi^{(1)}$ to $\Psi^{(4)}$ by repeatedly acting with $\De_A$:
\begin{equation*}
\includegraphics[width=.5\linewidth]{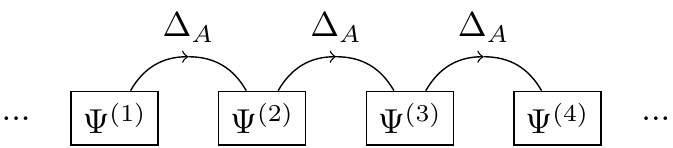}   
\end{equation*}
However, there are multiple
backwards motions to go from $\Psi^{(4)}$ to $\Psi^{(1)}$ using different
combinations of $\De_{-A}$, $\De_{-2A}$ and $\De_{-3A}$:
\begin{equation*}
\includegraphics[width=.5\linewidth]{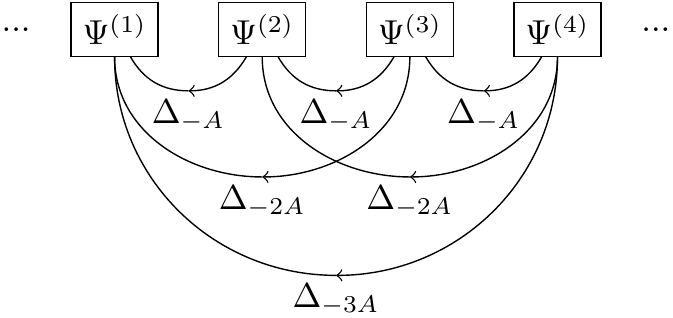}   
\end{equation*}
Before we can compute Stokes automorphism using allowed motions on the
alien chain, we need to introduce some further terminology and set up
some computational rules.  We denote a step $\cS_{n\to m}$ as a single link connecting two nodes
$n$ and $m$ on the chain and a path $\cP$ as a combination of
consecutive steps. The length $\ell(\cP)$ of a path is
then defined as the number of steps composing the path:
\begin{equation}
    \ell(\cP) = \#\{\cS_{n\to m}\in\cP\}\,.
\end{equation}
Looking at the proportionallity factor on the RHS of
\eq{resurgenceEqs}, we see that it is natural to define the weight
$w$ of a step $\cS_{n\to m}$ in terms of the Stokes constants $S_k$ as
\begin{equation}\label{eq:weightStep}
w(\cS_{n\to m}) = mS_{m-n}\,.
\end{equation}
The weight of a path is then simply the product of the weights of the
steps that make up the path
\begin{equation}
w(\cP) = \prod_{\cS_i\in\cP}w(\cS_i)\,.
\end{equation}
Finally, it is convenient to define
 a ``statistical factor'' \cite{Aniceto:2018bis} linking two nodes as
\begin{equation}\label{eq:statisticalFactor}
\text{SF}_{(n\to m)}
=\sum_{\cP(n\to m)}\frac{w(\cP)}{\ell(\cP)!}\,,
\end{equation}
where the sum is over all allowed paths linking nodes $n$ and $m$.

Equipped with these definitions, let us now see how they turn up in the
calculation of Stokes automorphisms, through a specific
example. Reviewing \eqs{StokesAutomorphism}{resurgenceEqs}, we notice
that the actual form of Stokes automorphism depends on the object it
acts on. Focusing on the example of the transseries sector $\Psi^{(3)}$, it follows from
the allowed motions on the alien chain that this sector has
singularities both in the $\theta=0$ and $\theta=\pi$ directions. As
the only allowed forward motions consist of repeated actions of
$\De_A$, we see that the Stokes automoprhism in the $\theta=0$
direction takes a simple form:
\begin{align}
\underline{\mathfrak{S}}_0\Psi^{(3)} 
&= \exp\Big(e^{-\frac{A}{\alpha}}\De_A\Big)\Psi^{(3)}\nn\\
&= \Big[1+e^{-\frac{A}{\alpha}}\De_A+\frac{1}{2!}\Big(e^{-\frac{A}{\alpha}}\De_A\Big)^2
+\frac{1}{3!}\Big(e^{-\frac{A}{\alpha}}\De_A\Big)^3+\ldots\Big]\Psi^{(3)}\nn\\
&= \Psi^{(3)} +  4S_1e^{-\frac{A}{\alpha}}\Psi^{(4)} + \frac{20}{2!}S_1^2e^{-2\frac{A}{\alpha}}\Psi^{(5)}
+ \frac{120}{3!}S_1^3e^{-3\frac{A}{\alpha}}\Psi^{(6)}+\ldots
\end{align}
In the $\theta=\pi$ direction, the Stokes automorphism does not take
such a simple form, as there are multiple allowed backward motions on
the alien chain. Luckily, when acting on $\Psi^{(3)}$ the possible set of
backward paths is finite, and we obtain
\begin{align}
\underline{\mathfrak{S}}_\pi\Psi^{(3)} 
&= \exp\bigg(\sum_{\ell=1}^3e^{\ell\frac{A}{\alpha}}\De_{-\ell A}\bigg)\Psi^{(3)}\nn\\
&= \bigg[1+\sum_{\ell=1}^3e^{\ell\frac{A}{\alpha}}\De_{-\ell A}
+\frac{1}{2!}\bigg(\sum_{\ell=1}^3e^{\ell\frac{A}{\alpha}}\De_{-\ell A}\bigg)^2
+\frac{1}{3!}\bigg(\sum_{\ell=1}^3e^{\ell\frac{A}{\alpha}}\De_{-\ell A}\bigg)^3+\ldots\bigg]\Psi^{(3)}\nn\\
&= \Psi^{(3)} + 2S_{-1}e^{\frac{A}{\al}}\Psi^{(2)} + \big(S_{-2}+S_{-1}^2\big)e^{2\frac{A}{\al}}\Psi^{(1)}.
\end{align}
Having computed these actions explicitly,  let us now
translate these results to the terminology we introduced above. We see
that under the action of $\underline{\mathfrak{S}}_0$ on $\Psi^{(3)}$ we
obtain an infinite sum of higher sectors $\Psi^{(n\geq3)}$: the nodes on
the alien chain that can be reached by forward motions. The
coefficients of the terms containing these sectors can be expressed in
terms of statistical factors \eq{statisticalFactor}. For example going from $\Psi^{(3)}\to\Psi^{(4)}$, we
see that there is only a single path, of length $\ell=1$ and weight
$w=4S_1$. Furthermore, we have to
include a non-perturbative factor $e^{-\frac{A}{\al}}$. Likewise, the
path to go from $\Psi^{(3)}\to\Psi^{(5)}$ has length $\ell=2$ with weight
$w=20S_1^2$. In order to get the correct coefficient of the $\Psi^{(5)}$
term, we have to multiply by $\frac{1}{2!}$ and a non-perturbative
term, i.e. SF$_{(3\to5)}e^{-2\frac{A}{\al}}$ in total. Similarly, the
coefficient in front of $\Psi^{(6)}$ is SF$_{(3\to6)}e^{-3\frac{A}{\al}}$
etc. Adding up all different terms for all possible paths gives the
full action of $\underline{\mathfrak{S}}_0$ on $\Psi^{(3)}$.  

For the full action of $\underline{\mathfrak{S}}_\pi$, we need to
consider the allowed backward motions. Again, there is only a single
path of length $\ell=1$ to go from $\Psi^{(3)}\to\Psi^{(2)}$, leading to a
statistical factor SF$=2S_{-1}$ and non-perturbative term
$e^{\frac{A}{\al}}$. To go from $\Psi^{(3)}\to\Psi^{(1)}$ however, we have two
allowed paths. One has $\ell=1$ with $w=S_{-2}$, and the other path
has $\ell=2$ with $w=2S_{-1}^2$, so SF$_{(3\to1)}=S_{-2}+S_{-1}^2$. In
both cases we need to multiply by the non-perturbative factor
$e^{2\frac{A}{\al}}$. Adding all terms for all possible paths again
gives the action of the Stokes automorphism.

To summarize what we have learned: $\underline{\mathfrak{S}}_0\Psi^{(n)}$
(resp. $\underline{\mathfrak{S}}_\pi\Psi^{(n)}$) is a sum over all forward
(backward) paths linking nodes to the right (left) of $\Psi^{(n)}$,
i.e.\ the terms in this sum can be written as
\begin{equation}\label{eq:StokesAutoInTermsOfAllowedMotions}
\underline{\mathfrak{S}}_0\Psi^{(n)}
= \Psi^{(n)}+\sum_{m>n}\text{SF}_{(n\to m)}e^{-(m-n)\frac{A}{\al}}\Psi^{(m)} ,
\end{equation}
and likewise for $\underline{\mathfrak{S}}_\pi\Psi^{(n)}$, where the only
difference is that one should sum over $m<n$.

\subsection{Large order behaviour and asymptotics}
\label{sec:asymptotics}
\begin{figure}
  \centering
  \includegraphics[width=0.4\linewidth]{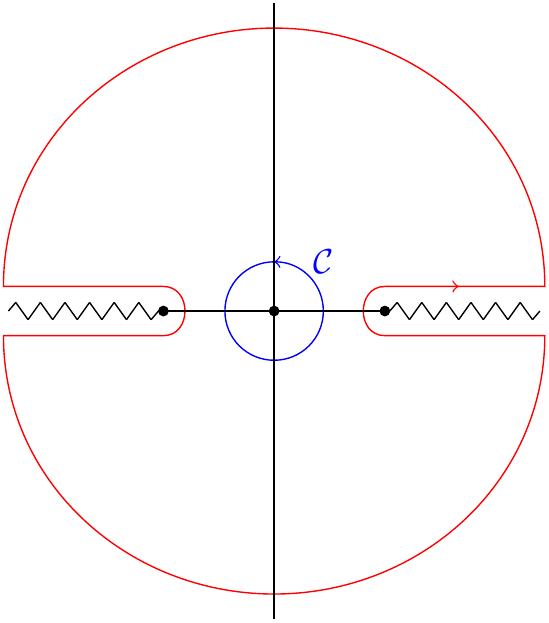}
  \caption{In blue, (the inner contour) we show the contour $\cC$ of the first line of the
    Cauchy integral, \eq{Cauchy}. In red (the outer contour) the deformed contour
    of the second line in that equation is shown. The zigzag lines on the right and
    left denote the rays where $\disc_0$ and $\disc_\pi$ occur.}
  \label{fig:deformedCauchyContour}
\end{figure}

\noindent
With the knowledge of the previous sections, we can now return to our
main goal: the understanding of asymptotic behaviour of
perturbative series in QFT and its relation to
non-perturbative sectors. In fact, the resurgent structure is even
more general and can be used to relate the asymptotic series of
all non-perturbative sectors to each other. To see this, we
apply Cauchy's theorem 
\begin{align}\label{eq:Cauchy}
  f(\al)
  &= \oint_{\cC}\frac{dy}{2\pi\img}\frac{f(y)}{y-\al}\nn\\
  &= -\int_{0}^{\infty} \frac{dy}{2\pi\img}\frac{\disc_0 f(y)}{y-\al}
    -\int_{0}^{-\infty} \frac{dy}{2\pi\img}\frac{\disc_\pi f(y)}{y-\al} +\oint_{(\infty)}\frac{dy}{2\pi\img}\frac{f(y)}{y-\al},
\end{align}
where we assumed discontinuities only in the $\theta=0$ and $\theta=\pi$
directions. See \fig{deformedCauchyContour} for the deformation of the
contour to go from the first to the second line in \eq{Cauchy}. Notice that we deformed the contours even further, such that the first two integrals in \eq{Cauchy} start from $0$.

For example, we can apply this to the perturbative sector
$\Psi^{(0)}(\al)$. The discontinuities of this sector are
easily computed using the rules from the previous subsection:
\begin{equation}
  \disc_0 \Psi^{(0)} = (1 - \underline{\mathfrak{S}}_0)\Psi^{(0)} = -\sum_{\ell=1}^\infty S_1^\ell e^{-\ell\frac{A}{\al}}\Psi^{(\ell)},
  \qquad\text{and}\qquad
  \disc_\pi \Psi^{(0)} = 0.
\end{equation}
In many cases (including the ones in this paper), the integral around
infinity vanishes, yielding
\begin{equation}\label{eq:largeOrderWithIntegral}
  \sum_{k=0}^\infty f_k^{(0)}\al^{k}
  = \sum_{\ell=1}^\infty \frac{S_1^\ell}{2\pi\img} \int_{0}^\infty dy  
  \frac{e^{-\ell\frac{A}{y}}y^{\beta_\ell}}{y-\al}\sum_{h=0}^\infty f_h^{(\ell)}y^h.
\end{equation}
By expanding around $\al=0$, we can match equal powers in $\al$
on both sides of this equation and perform the $y$-integrations, after
which we arrive at a remarkable connection between the perturbative
and non-perturbative expansions
\begin{align}
  f_k^{(0)}
  &\sim\sum_{\ell=1}^\infty\frac{S_1^\ell}{2\pi\img}\sum_{h=0}^\infty  f_h^{(\ell)}\frac{\Ga(k-h-\beta_\ell)}{(\ell A)^{k-h-\beta_\ell}}.
    \label{eq:largeOrderRelation}
\end{align}
Here, we used a $\sim$ symbol instead of an equals sign to indicate that this relation only describes the behavior as $k \to \infty$. In particular, we have exchanged the sum over $h$ and the integral in \eq{largeOrderWithIntegral}, which is only allowed formally -- leading e.g. to the fact that $\Gamma(k-h-\beta_\ell)$ may be ill-defined when $h$ is large compared to $k$.

In words: \eq{largeOrderRelation} states that asymptotic behaviour of the perturbative
coefficients $f_k^{(0)}$ is completely determined by the
non-perturbative expansion coefficients $f_h^{(\ell)}$. In
\secs{adlerLO}{adlerNLO} we shall explain in detail how one can unpack
this equation systematically, and decode from the perturbative
coefficients the non-perturbative ones. 

We can repeat this derivation to obtain resurgent large order
relations for other non-perturbative sectors as well. The key
ingredient is to rewrite Cauchy's theorem in terms of the Stokes
discontinuities, so that from the rules discussed in the previous
subsection, one can write down large order relations by looking at
allowed motions on the alien chain. Let us note that of course all of this is only possible if the non-perturbative sectors indeed have an asymptotic expansion (as opposed to a convergent or even finite one) -- we shall see later that this is not always the case for the Adler function. 

To be explicit, we substitute
\eq{StokesAutoInTermsOfAllowedMotions} into Cauchy's theorem to get
\begin{equation}\label{eq:Cauchy3}
  \Psi^{(n)}(\al)
  = \sum_{\ell>n} \frac{\text{SF}_{(n\to \ell)}}{2\pi\img}\int_{0}^\infty dy  
  \frac{e^{-(\ell-n)\frac{A}{y}}}{y-\al}\Psi^{(\ell)}(y)
  +\sum_{\ell< n} \frac{\text{SF}_{(n\to \ell)}}{2\pi\img}\int_{0}^{-\infty} dy  
  \frac{e^{-(\ell-n)\frac{A}{y}}}{y-\al}\Psi^{(\ell)}(y).
\end{equation}
Again expanding around $\al=0$, matching equal powers in $\al$ and
performing the $y$-integrals, one finds
\begin{equation}\label{eq:1DlargeOrderEquation}
  f_k^{(n)}
  \sim \sum_{\ell\neq n} \frac{\text{SF}_{(n\to \ell)}}{2\pi\img}\chi_{(n\to\ell)}(k),
\end{equation}
where it is convenient to define 
the large order factor\footnote{Note that we define these factors
  slightly different from \cite{Aniceto:2018bis}, as we include an
  explicit factor of $\Gamma(k)/A^k$.}
\begin{equation}\label{eq:1DlargeOrderFactor}
  \chi_{(n\to\ell)}(k)
  =\sum_{h=0}^\infty  f_h^{(\ell)}
  \frac{\Ga(k+\beta_n-h-\beta_\ell)}{((\ell-n) A)^{k+\beta_n-h-\beta_\ell}}.
\end{equation}
Thus, \eqs{1DlargeOrderEquation}{1DlargeOrderFactor} show how, using the alien chain formalism, the
asymptotic behaviour of expansion coefficients in non-perturbative sectors encodes all
expansion coefficients in other non-perturbative sectors.

\subsection{Generalizations and extensions}
\label{sec:generalizations}
\noindent
There are many generalizations of the concepts and constructions we
have seen so far, but in this work we only need two of them.  First, we
have worked thus far with examples with a single boundary condition, but we
will also need the concept of {\em multi-parameter transseries}.
Second, we need to consider transseries with terms that include
logarithmic factors. In this section, we  follow the
exposition in \cite{Aniceto:2011nu,Aniceto:2018bis}. We finish 
 with a short discussion on the types of singularities in the
Borel plane that are different than those discussed so far. 

As we will see, for the Adler function it is not enough to capture all
its
non-perturbative contributions with a single exponential transmonomial
$e^{-\frac{A}{\al}}$. Instead, we need to allow for more such
exponentials, $e^{-\frac{A_i}{\al}}$. In fact, two
exponents seem to suffice for the Adler function at the order we are
interested in, but we shall be somewhat  more
general. Thus, by writing $\bm{A}=(A_1,...,A_k)$, a typical
Ansatz for a $k$-parameter transseries solution to a non-linear
problem is
\begin{equation}\label{eq:multiParameterTransseries}
  F(\al,\bm{\si})
  = \sum_{\bm{n}\in\mathbb{N}^k_0}\bm{\si}^{\bm{n}}e^{-\frac{\bm{n}\cdot \bm{A}}{\al}}\Psi^{(\bm{n})}(\al),
  \qquad\text{with}\qquad
  \Psi^{(\bm{n})}(\al) = \al^{\beta_{\bm{n}}}\sum_{h=0}^\infty f_h^{(\bm{n})}\al^h.
\end{equation}
Here we used the notation
$\bm{\si}^{\bm{n}}=\prod_{i=1}^k \si_i^{n_i}$. Understanding the
resurgence properties of such a transseries is again best understood
in terms of the alien derivatives $\De_\w$.  For the one-parameter
transseries, a key ingredient in the calculation of alien derivatives
was the bridge equation \eq{bridgeEq}. In the case of the
multi-parameter transseries, the bridge equation usually takes the
form
\begin{equation}\label{eq:multiParameterBridgeEq}
  \Delta_{\bm{\ell}\cdot \bm{A}} \Psi^{(\bm{n})}(\al) = 
  \begin{cases}
    \bm{S}_{\bm{\ell}}\cdot (\bm{n}+\bm{\ell}) \Psi^{(\bm{n}+\bm{\ell})}
    & \quad\ell_i\leq \de_{ij}, \quad \bm{\ell}\neq\bm{0},\\
    0 & \quad\text{elsewhere},
  \end{cases}
\end{equation}
where for each combination $\bm{\ell}\cdot \bm{A}$ we now need a
whole vector of Stokes parameters
$S_{\bm{\ell}}=(S_{\bm{\ell}}^{(1)},...,S_{\bm{\ell}}^{(k)})$.  This
equation can be derived by generalizing the steps we took in
\sec{bridgeEquation} to the case of a multi-parameter transseries.
The Borel singularities of the sectors $\Psi^{(\bm{n})}$ lie at
positions $t=\bm{\ell}\cdot\bm{A}$ in the Borel plane, with
$\bm{\ell}\in\mathbb{Z}^k$ with entries bounded from below by
\eq{multiParameterBridgeEq}.  Thus \eq{borelSingularities} becomes
\begin{equation}\label{eq:BorelSingularitiesMultiParTrans}
  \cB[\Psi^{(\bm{n})}](t+\bm{\ell}\cdot\bm{A}) 
  = \frac{a}{2\pi \img\, t} + \mathsf{S}_{\bm{n}\to \bm{n}+\bm{\ell}}\ \cB[\Psi^{(\bm{n}+\bm{\ell})}](t)\frac{\log(t)}{2\pi \img} + \text{holomorphic}.
\end{equation}
where the {\em Borel residues}
$\mathsf{S}_{\bm{n}\to \bm{n}+\bm{\ell}}$ can be computed in terms of
the Stokes parameters $S_{\bm{\ell}}^{(i)}$ using
\eq{multiParameterBridgeEq}.

In \sec{1Dalienchain}, we explained how, in the one-parameter case, the
bridge equation translates to a set of allowed motions along an
{\em alien chain} of non-perturbative sectors.  Furthermore, we gave 
 computational rules for the computation of Stokes
discontinuities and large order formulae for the asymptotic behaviour
of transseries.  The natural extension for multi-parameter transseries
is to think of the sectors $\Psi^{(\bm{n})}$ as living on a
$k$-dimensional {\em alien lattice}.  The computational rules outlined
in the previous subsections are then to a large extent unaltered, the main
exception being that we have a richer structure of allowed resurgence
motions on the (multi-dimensional) alien lattice.  

\begin{figure}
  \centering
  \includegraphics[width=0.6\linewidth]{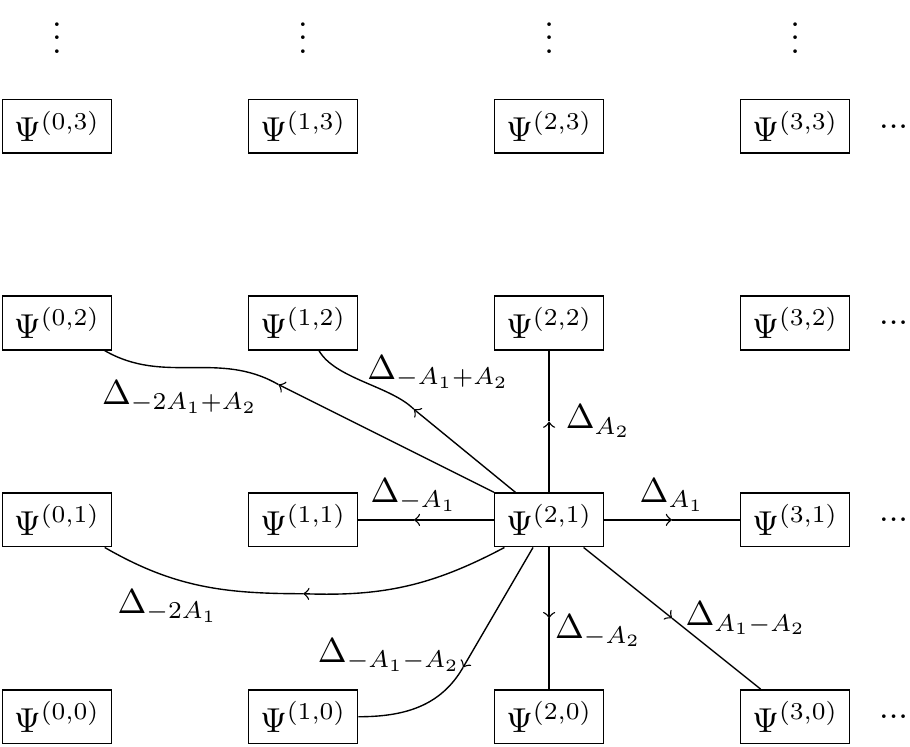}
  \caption{Two-dimensional alien lattice. We show all
    allowed motions of a single step starting from node
    $\Psi^{(2,1)}$. Compared to the allowed motions for the
    one-dimensional alien chain, we observe a much richer structure of
    allowed resurgence motions.}
  \label{fig:alienLatticeExample}
\end{figure}

As an example, we consider the two-dimensional case $k=2$ (see
\fig{alienLatticeExample}), and consider all the motions consisting of
a single step starting from the node $\Psi^{(2,1)}$.
Similar to what happens for the one-parameter transseries, the
constraints in the bridge equation \eq{multiParameterBridgeEq} lead to
only two types of forward motion in the $\bm\ell=(1,0)$ and
$\bm\ell=(0,1)$ directions: $\De_{A_1}$ and $\De_{A_2}$. In other
words, starting from node $\Psi^{(2,1)}$, one can only
reach\footnote{As will shortly be made clear, the concept of reaching
  a node means that the corresponding sector then occurs in the large
  order description of the coefficients of the original sector.} node
$\Psi^{(2,3)}$ by acting with $\De_{A_2}$ twice.  For pure backwards
motion we have, as before, more options. In the example in
\fig{alienLatticeExample}, we see the allowed purely backwards motions
consisting of a single step, obtained by acting with $\De_{-A_1}$,
$\De_{-A_2}$, $\De_{-2A_1}$ and $\De_{-A_1-A_2}$. This means that from
$\Psi^{(2,1)}$ one can reach node $\Psi^{(0,1)}$ using either
$\De_{-2A_1}$ or twice $\De_{-A_1}$.

The example shows that we now also have a type of mixed forward and
backward motion, obtained by acting with $\De_{-A_1+A_2}$,
$\De_{-2A_1+A_2}$ and $\De_{A_1-A_2}$. However, similarly to pure
forward motion, the constraint of the bridge equation limits the
forward part of the motion in particular to minimal step sizes. E.g.\
acting with $\De_{-A_1+2A_2}$ does not lead to an allowed motion.
Finally, we should emphasize that for paths of multiple steps it is
not allowed to mix single steps with different directions, similar to
the fact that one could not mix forward and backward motion in the
one-parameter case.  The reason for this is that for the computation
of the Stokes automorphism, \eq{StokesAutomorphism}, in a singular
direction
\begin{equation}\label{eq:singularDirection}
  \theta_{\bm\ell} = \arg(\bm{\ell}\cdot \bm{A})\,,
\end{equation}
one only requires alien derivatives in the $\bm\ell$-direction.  To
clarify this with an example, note that to reach node $\Psi^{(0,3)}$ starting from
$\Psi^{(2,1)}$ one can act with $\De_{-A_1+A_2}$ twice, but the path
where one first acts with $\De_{-2A_1+A_2}$ and subsequently with
$\De_{A_2}$ is not allowed since that combination does not occur in
any Stokes automorphism.

For the computational rules, the concepts of step, path, their
length, etc.\ stay unchanged in the multi-parameter setting. However,
as we now have a vector of Stokes coefficients, the weight of a step,
\eq{weightStep}, becomes an inner product
\begin{equation}
  w(\cS(\bm{n}\to\bm{m})) = \bm{m}\cdot \bm{S}_{\bm{m}-\bm{n}}.
\end{equation}
The expression for Stokes' automorphism acting on node $\Psi^{(\bm{n})}$
in the singular direction
$\theta_{\bm\ell} = \arg(\bm{\ell}\cdot \bm{A})$, is now given by the
sum over all paths linking the nodes $\Psi^{(\bm{n}+m\bm{\ell})}$, with
$m>0$. Therefore, we can write in general
\begin{equation}
  \underline{\mathfrak{S}}_{\theta_{\bm\ell}}\Psi^{(\bm{n})}
  = \Psi^{(\bm{n})}+\sum_{m>0}\text{SF}_{(\bm{n}\to \bm{n}+m\bm{\ell})}e^{-m\frac{\bm{\ell}\cdot\bm{A}}{\al}}\Psi^{(\bm{n}+m\bm{\ell})}.
\end{equation}
Likewise, the large order relation \eq{1DlargeOrderEquation} becomes
\begin{equation}\label{eq:MultiParameterLargeOrderFactor}
  f_k^{(\bm{n})}
  \sim \sum_{\bm{\ell}\neq\bm{n}} 
  \frac{\text{SF}_{(\bm{n}\to\bm{ \ell})}}{2\pi\img}
  \chi_{(\bm{n}\to\bm{\ell})}(k),
\end{equation}
with
\begin{equation}\label{eq:largeOrderFactorMultipParameter}
  \chi_{(\bm{n}\to\bm{\ell})}(k)
  =\sum_{h=0}^\infty  f_h^{(\bm{\ell})}
  \frac{\Ga(k+\beta_{\bm{n}}-h-\beta_{\bm{\ell}})}{((\bm{\ell}-\bm{n})\cdot\bm{A})^{k+\beta_{\bm{n}}-h-\beta_{\bm{\ell}}}}
\end{equation}
the generalization of the large order factor,
\eq{1DlargeOrderFactor}. These generalizations to multi-parameter
transseries will all play a role when we study the Adler function.

As a final note on multi-parameter transseries, we
want to mention that it might be the case that despite what we have said, one finds that the
action of an operator like $\De_{2A_i}$, for some value of $i$, is non-zero.  One possibility is that there is
an additional transseries parameter related to an exponential
transmonomial $e^{-\frac{2A_i}{\al}}$. As a result, a sector with
exponential $e^{-\frac{2A_i}{\al}}$ lies on top of a sector with
$\left(e^{-\frac{A_i}{\al}}\right)^2$. In \cite{Borinsky:2022knn},
which studies the all-order resurgence of factorially divergent series
associated to a renormalon in six-dimensional scalar $\phi^3$ theory,
such a transseries was actually found with three transseries
parameters and exponential transmonomials $e^{-\frac{A_i}{\al}}$,
$e^{-\frac{2A_i}{\al}}$ and $e^{-\frac{3A_i}{\al}}$.  A second way in
which the action of $\De_{2A_i}$ does not vanish is that it might be
the case that the problem at hand does not allow for a bridge equation
of the form \eq{multiParameterBridgeEq}. In other words, the bridge
equation does not have the constraint that $\De_{2A_i}$ vanishes.  We
will come back to this in our discussion of the Adler function in
\secs{adlerLO}{adlerNLO}. 

\label{page:resonance}
Next, we discuss transseries with logarithmic transmonomials $\log(\al)$. Note that in the literature, see e.g.~\cite{Garoufalidis:2010ya,Aniceto:2011nu,Aniceto:2018bis,Borinsky:2022knn}, both multi-parameter transseries and logarithmic factors often occur in the case of {\em resonant} transseries, which are transseries where multiple $A_i$ add up to 0 in such a way that logarithmic factors must occur to solve the problem at hand. Although for the Adler function we have two exponents $A_1=-A_2$ and logarithmic factors do occur, in its transseries the two effects are not related and our transseries is not resonant.

One can
of course add such logarithmic transmonomials to the general multi-parameter
transseries \eq{multiParameterTransseries}. However, despite the fact
that we need such a multi-parameter transseries for the Adler function, we
will see in \sec{adlerNLO} that the part of the transseries with
logarithms effectively looks like a one parameter
transseries.
Therefore, here we only discuss how to extend the one-parameter transseries of
\eq{transseries} by including logarithmic transmonomials.  Together with
the details given so far for multi-parameter transseries, it is then
straightforward to generalize this to the case of multi-parameter
transseries with logarithms.

The one-parameter transseries Ansatz with logarithms becomes:
\begin{equation}\label{eq:ansatzWithLogs}
  F(\al,\si)
  = \sum_{n=0}^\infty \si^n e^{-n\frac{A}{\al}} \Psi^{(n)}(\al),
  \quad\text{with}\quad
  \Psi^{(n)}(\al) 
= \sum_{p=0}^{p_n}\log^p(\al)\ 
\sum_{h=0}^\infty f_h^{(n)[p]}\al^{h+\beta_n^{[p]}},
\end{equation}
where we included an expansion in logarithmic powers of $\al$ up to some finite logarithmic power $p_n$.
Note that in doing this, we add a new transmonomial $\log(\alpha)$ to the
transseries, but {\em not} a new transseries parameter $\hat{\sigma}$
in addition to $\sigma$. The reason for this is that the addition of
logs generally does not change the location of singularities in the Borel plane,
so there are no new Stokes automorphisms that would act on such a
parameter. For similar reasons, we consider all of the $f_h^{(n)[p]}$
to belong to the same non-perturbative sector and we will always draw
them as a single box in alien chains and lattices.

The derivation of the bridge equation, \eq{resurgenceEqs}, is
unaltered (and thus the allowed motions along the alien chain are the
same), but when we apply Cauchy's residue theorem to obtain large
order relations, we get additional and more complicated integrals. To
be precise, when we expand \eq{Cauchy3} around $\al=0$, we
additionally need to perform integrals of the form
\begin{align}
  \int_0^\infty dy\ y^s e^{-m\frac{A}{y}} \log^p(y)\,,
\end{align}
for some $s$, $m$ and $p$. These integrals can be
evaluated exactly, and it is straightforward to show that this yields
the following large order relation for the perturbative coefficients
\begin{align}
  f_k^{(0)[0]} 
  &\sim \sum_{\ell=1}^\infty \frac{S_1^\ell}{2\pi\img}\sum_{h=0}^\infty
    f_h^{(\ell)[0]}\frac{\Ga(k-h-\beta_\ell^{[0]})}{(\ell A)^{k-h-\beta_\ell^{[0]}}}\nn\\
  &\hspace{2cm}
    +\sum_{\ell=1}^\infty \frac{S_1^\ell}{2\pi\img}\sum_{h=0}^\infty \sum_{p=1}^{p_\ell}
    f_h^{(\ell)[p]}\frac{\big[\log(\ell A)-\partial_g\big]^p\Ga(g)}{(\ell A)^{k-h-\beta_\ell^{[p]}}}\Bigg|_{g=k-h-\beta_\ell^{[p]}}\,,
    \label{eq:largeOrderWithLogs}
\end{align}
where the first term is analogous to \eqref{eq:largeOrderRelation}
and the second follows from the logarithmic transmonomials.  One
can derive similar equations for the large order behaviour of the
coefficients of the non-perturbative sectors.

Although most of our discussion in \sec{resurgence} was focused on the
case of a single pole and a $\log$-branch cut, we finish this section
with a short discussion on other types of singularities in the Borel
plane that might appear. Indeed  we will need this in the upcoming
sections for the Adler function.  In fact, the characteristic
exponents $\beta_{\bm{n}}$ we added in the transseries Ansatz already
allow for higher order poles in the Borel plane. Looking at
\eqs{1DlargeOrderFactor}{MultiParameterLargeOrderFactor}, we notice
that these characteristic exponents play a role in the large order
behaviour of the perturbative and non-perturbative coefficients.  To
see how this translates to the Borel plane, consider a formal power
series $F(\al)$ and its Borel transform
\begin{equation}\label{eq:valueBetaCoefficientHigherOrderPoles}
  F(\al) = \sum_{n=0}^\infty \Ga(n-\beta)\,\al^{n+1}
  \qquad\implies\qquad
  \cB[F](t) = \frac{\Ga(-\beta)}{(1-t)^{-\beta}},
\end{equation}
where we assumed that $\beta<0$. In particular, $\beta=-1,-2,-3,...$
correspond to a single, double, triple pole etc. in the Borel plane.

Furthermore, the sectors with logarithms in the transseries Ansatz
\eq{ansatzWithLogs} yield new types of singularities. As an example, consider the $p=1$ terms of \eq{largeOrderWithLogs} with $\partial_g\Ga(g) = \Ga(g)\psi(g)$ (with $\psi(z)=\td{}{z}\log\Ga(z)$ the digamma function), i.e.\ consider the formal series
\begin{equation}\label{eq:logTransSingularityExample1}
  G(\al) = \sum_{k=0}^\infty \Ga(k-\beta)\psi(k-\beta)\,\al^{k+1}
\end{equation}
with $\beta<0$ again.
Its Borel transform is given
by
\begin{equation}\label{eq:logTransSingularityExample}
  \cB[G](t)
  = -\Ga(-\beta)\frac{\log(1-t)-\psi(-\beta)}{(1-t)^{-\beta}},
\end{equation}
and we indeed observe that these terms lead to a type of singularity we have not considered so far. We
will discuss the role of these Borel singularities for the Adler
function when we encounter them in \secs{adlerLO}{adlerNLO}.

\section{Adler function with one bubble chain (\texorpdfstring{$\ord{1/N_f}$)}{ }}
\label{sec:adlerLO}
We briefly discussed, in \sec{uvirrenormalons}, 
how bubble chain diagrams cause perturbative series to
show factorial growth. Here, in \sec{flavour} we formalize this with a
discussion of the flavour expansion in QED and QCD. We then turn to
the Adler function in \sec{adler} and compute the
diagrams that contain one bubble chain and are leading in the flavour expansion. Subsequently we analyze the resulting
asymptotic series using the resurgence techniques developed in
\sec{resurgence}: in \sec{resurgenceLO} for the leading and subleading non-perturbative orders and in \sec{resurgenceLOall} for all other nonvanishing orders. In \sec{discussionLO} we briefly discuss the results we found and explain how they set us up for the $\ord{1/N_f^2}$ investigations in the rest of the paper.

\subsection{Flavour expansion}
\label{sec:flavour}
Let us consider the flavour expansion of QED/QCD with $N_f$ massless fermions. We can write an observable in perturbation theory as
\begin{equation}
F(\al) = \sum_{n=0}^\infty c_n\,\al^{n}, 
\end{equation}
where, in general, the coefficients $c_n$ can be written as an expansion in $N_f$,
\begin{equation}\label{eq:coefficientsExpandedInNf}
c_n = c_n^{(0)} + ... + c_n^{(n-1)}N_f^{n-1}.
\end{equation}
In this flavour expansion we take the large $N_f$ limit while keeping the 't Hooft coupling $N_f\al$ fixed. In this limit, the dominant contribution to $c_n$ is given by $c_n^{(n-1)}$ and we therefore reorder the perturbative expansion as an expansion in $1/N_f$:
\begin{equation}\label{eq:flavourExpansion}
F = \frac{1}{N_f}\sum_{n=0}^\infty c_{n}^{(n-1)}(N_f \al)^n
\left(
1+ \ORD{\frac{1}{N_f}}
\right).
\end{equation}
To see what diagrams go into each order of $1/N_f$, recall the
one-loop vacuum-polarization graph (fermion bubble)
\begin{equation}
\Pi_{\mu\nu}(k)
=\vcenter{\hbox{\includegraphics[width=.16\textwidth]{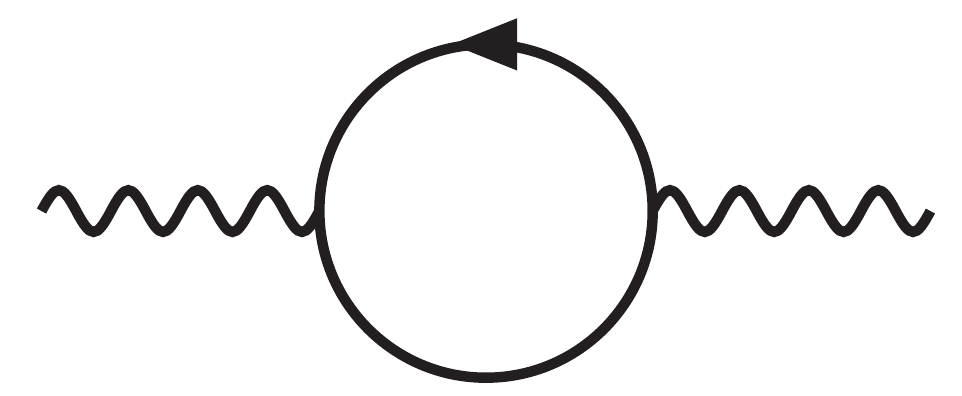}}}
= (k_\mu k_\nu-\eta_{\mu\nu}k^2)\ \pi(k^2),
\end{equation}
which we already calculated in \eqs{vacpolint}{vacpolint3}, where we found
\begin{equation}\label{eq:fermionBubble}
\pi(k^2) = -\al \beta_{0f} \bigg[\log\bigg(\frac{-k^2}{\mu^2}\bigg) + C\bigg].
\end{equation}
Note that we have canceled the UV divergence with a counterterm. 
Here $\beta_{0f}=T N_f/(3\pi)$ with $T=1$ in QED and $T=1/2$ in
QCD, while $C$ is a scheme dependent constant (e.g.\ $C=-\frac{5}{3}$ in the $\overline{\text{MS}}$-scheme, as in \sec{uvirrenormalons}). 
We notice that such a fermion bubble counts as $N_f\al =\ord{1}$ in the flavour expansion. 
Including the external lines, the effective photon propagator with $n$ such fermion bubbles reads
\begin{equation}\label{eq:bubbleChain}
\vcenter{\hbox{\includegraphics[width=.17\textwidth]{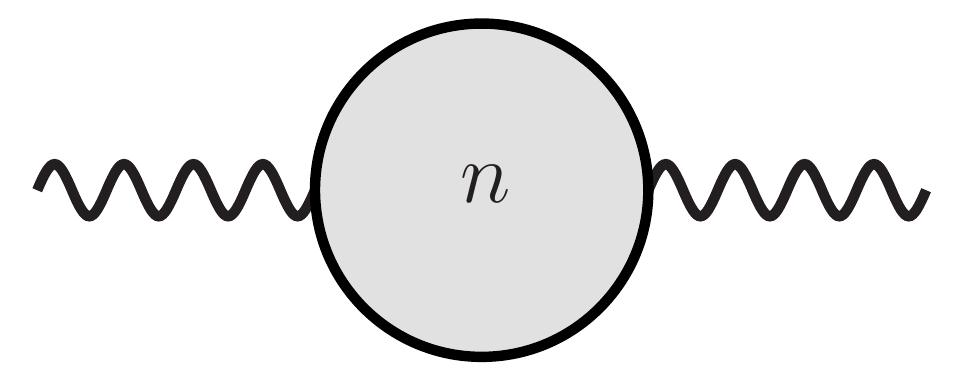}}}
\equiv\underbrace{\vcenter{\hbox{\includegraphics[width=.32\textwidth]{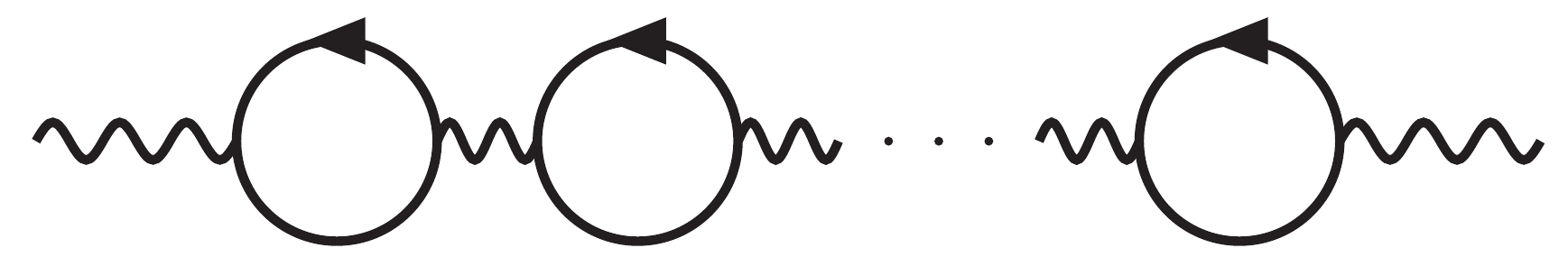}}}}_{n\ \text{fermion bubbles}}
=-\frac{\img}{k^2}\Big(\eta_{\mu\nu} - \frac{k_\mu k_\nu}{k^2}\Big)\Big[-\pi(k^2)\Big]^n,
\end{equation}
so that a bubble chain counts as $(N_f\al)^n =\ord{1}$ in the flavour expansion as well. 
In other words, the coefficients $c_n^{(n-1)}$ in \eq{flavourExpansion}, and also the coefficients at higher orders in the flavour expansion can be computed by replacing virtual photons/gluons with the bubble chain \eq{bubbleChain}. From \eq{fermionBubble} it then follows that this leads to the inclusion of logarithms in the Feynman integral.
As already explained in \sec{uvirrenormalons}, integrating over such logarithms in a Feynman integral leads to factorial growth. 

Diagrams including the sum over bubble chains can be computed using the effective (Dyson summed) propagator of such a bubble chain:
\begin{align}
D_{\mu\nu}(k) 
&= -\frac{\img}{k^2}\Big(\eta_{\mu\nu} - \frac{k_\mu k_\nu}{k^2}\Big) \frac{1}{1+\pi(k^2)}
-\img\xi \frac{k_\mu k_\nu}{k^4},
\end{align}
with $\xi$ the gauge fixing parameter.
However, instead of working with this Dyson-summed effective propagator, it will be more convenient to work with its Borel transform with respect to $\alpha$ (which appears in $\pi(k^2)$), which reads \cite{Beneke:1992ch}
\begin{align}
\cB\big[\al D_{\mu\nu}(k)\big] (u)
=-\frac{\img}{k^2}\bigg(\eta_{\mu\nu}-\frac{k_\mu k_\nu}{k^2}\bigg)  
\bigg(-\frac{\mu^2}{k^2}e^{-C}\bigg)^u -\img\xi\frac{k_\mu k_\nu}{k^4},
\label{eq:BorelChain}
\end{align}
where we conveniently rescaled the usual variable $t$ in the Borel plane to $u=-\beta_{0f}t$. 
Furthermore, before Borel transforming we inserted an overall factor
of $\al$ that will be convenient later, when we perform the actual
diagram calculations.
When we have $n_c>1$ bubble chains in a diagram, we can use the fact that the Borel transform of a product of series is obtained by taking the convolution of the individual Borel transforms:
\begin{equation}\label{eq:flavourExpansionConvoInt}
\cB\bigg[\prod_{j=1}^{n_c} \al D_{\mu_j\nu_j}(k_j)\bigg](u)
= \frac{1}{(-\beta_{0f})^{n_c-1}} \int_0^u \bigg[\prod_{j=1}^{n_c}du_j\bigg] \de\Big(u-\sum_{j=1}^{n_c}u_j\Big) \prod_{j=1}^{n_c} \cB\big[\al D_{\mu_j\nu_j}(k_j)\big] (u_j).
\end{equation}
This can be verified by taking the Laplace transform of both sides of
the equation with respect to $t$. In \sec{convoInt} we give more details on the effect of the convolution integral on the resurgence structure in the case of general $n_c$. In \sec{adlerNLO}, where we discuss the Adler function at order $1/N_f^2$, we will only need the case $n_c=2$.

For the case of QED we notice that positive $t$ corresponds to negative $u$,
since $\beta_{0f}$ is positive. For QCD one completes $\beta_{0f}$ to the
full first coefficient of the $\beta$-function $\beta_0 = -(11-2 N_f/3)/(4\pi)$ \cite{Beneke:1998ui,Parisi:1978bj,Mueller:1984vh}, so that now positive $t$
corresponds to positive $u$ (thus, infrared renormalons appear on the
positive $u$ axis for QCD, and on the negative axis for QED).
In practice this implies that one can perform a QED calculation to obtain the non-abelian counterpart in QCD by replacing $\beta_{0f}$ with $\beta_{0}$ and by adding the appropriate SU$(3)$ color factors. In the literature, this procedure is also known as the process of naive non-abelianization \cite{Broadhurst:1994se,Ball:1995ni,Beneke:1994qe}. 
We should mention here that for the large $N_f$ limit, $\beta_0$ changes sign (this happens for $N_f>16$) after which QCD is no longer asymptotically free. 
Hence, for QCD, instead of a large $N_f$ expansion one uses a large
$\beta_0$ expansion,  and we rewrite \eq{coefficientsExpandedInNf} as an expansion in $\beta_0$
\begin{equation}
c_n = \tilde{c}_n^{(0)} + ... + \tilde{c}_n^{(n-1)}\beta_0^{n-1}.
\end{equation}
Thus, in the large $\beta_0$ expansion for QCD, the expansion of a generic observable \eq{flavourExpansion} becomes
\begin{equation}\label{eq:flavourExpansionQCD}
F= \frac{1}{\beta_0}\sum_{L=0}^\infty \tilde{c}_{n}^{(n-1)}\left(\beta_0 \al\right)^n
\left(1+\ORD{\frac{1}{\beta_0}}\right).
\end{equation}
where we now define the 't Hooft coupling as $\beta_0 \al$ for the QCD case. In what follows, we refer to the large $N_f$ expansion for both QED and QCD, even though we use the large $\beta_0$ expansion in QCD. We perform our calculations in QED and in the end use the procedure of naive non-abelianization to convert our results to obtain the QCD result.

\subsection{Adler function}
\label{sec:adler}
\begin{figure}
\centering
\begin{subfigure}{4cm}
    \includegraphics[width=\textwidth]{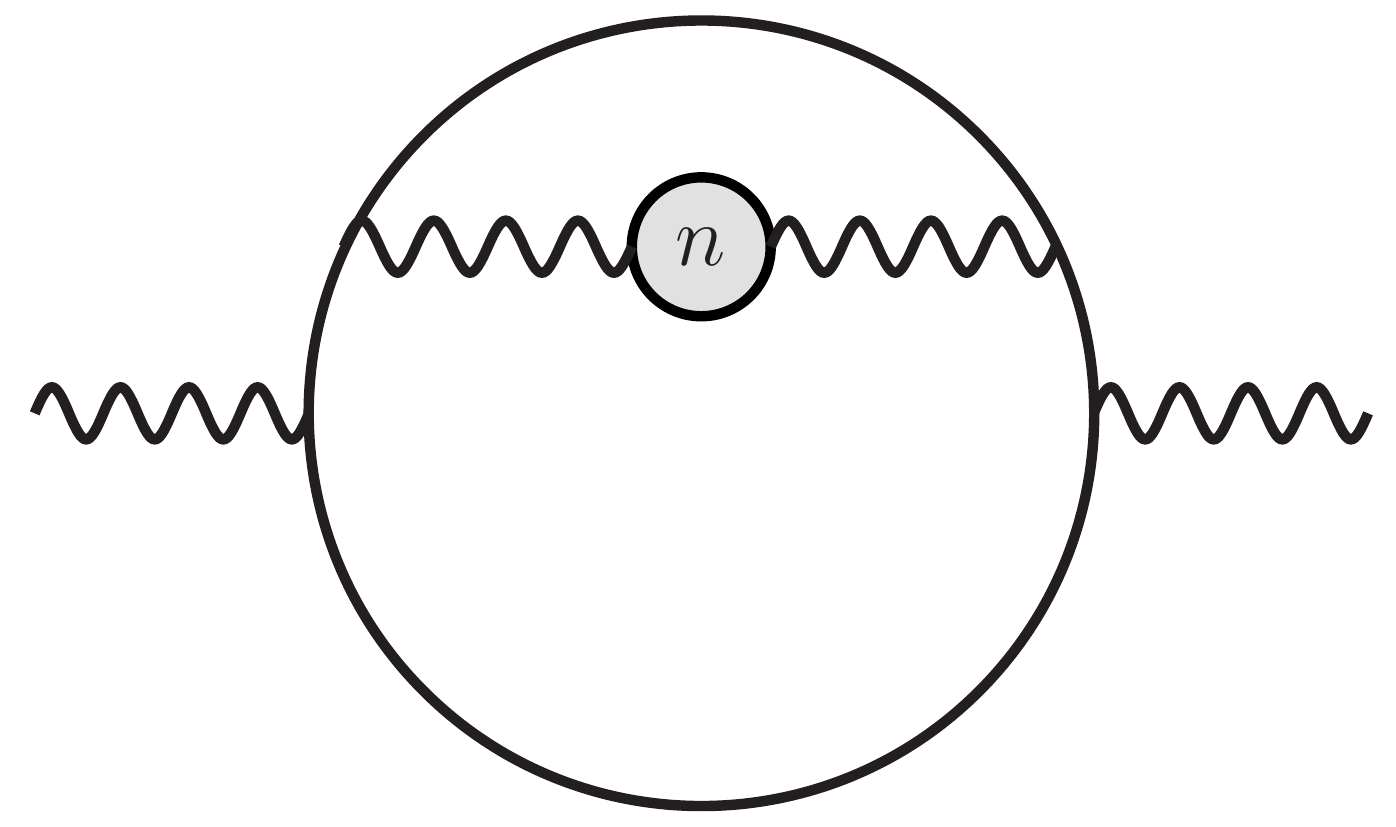}
    \caption{}
    \label{fig:adlerLO1}
\end{subfigure}
\begin{subfigure}{4cm}
    \includegraphics[width=\textwidth]{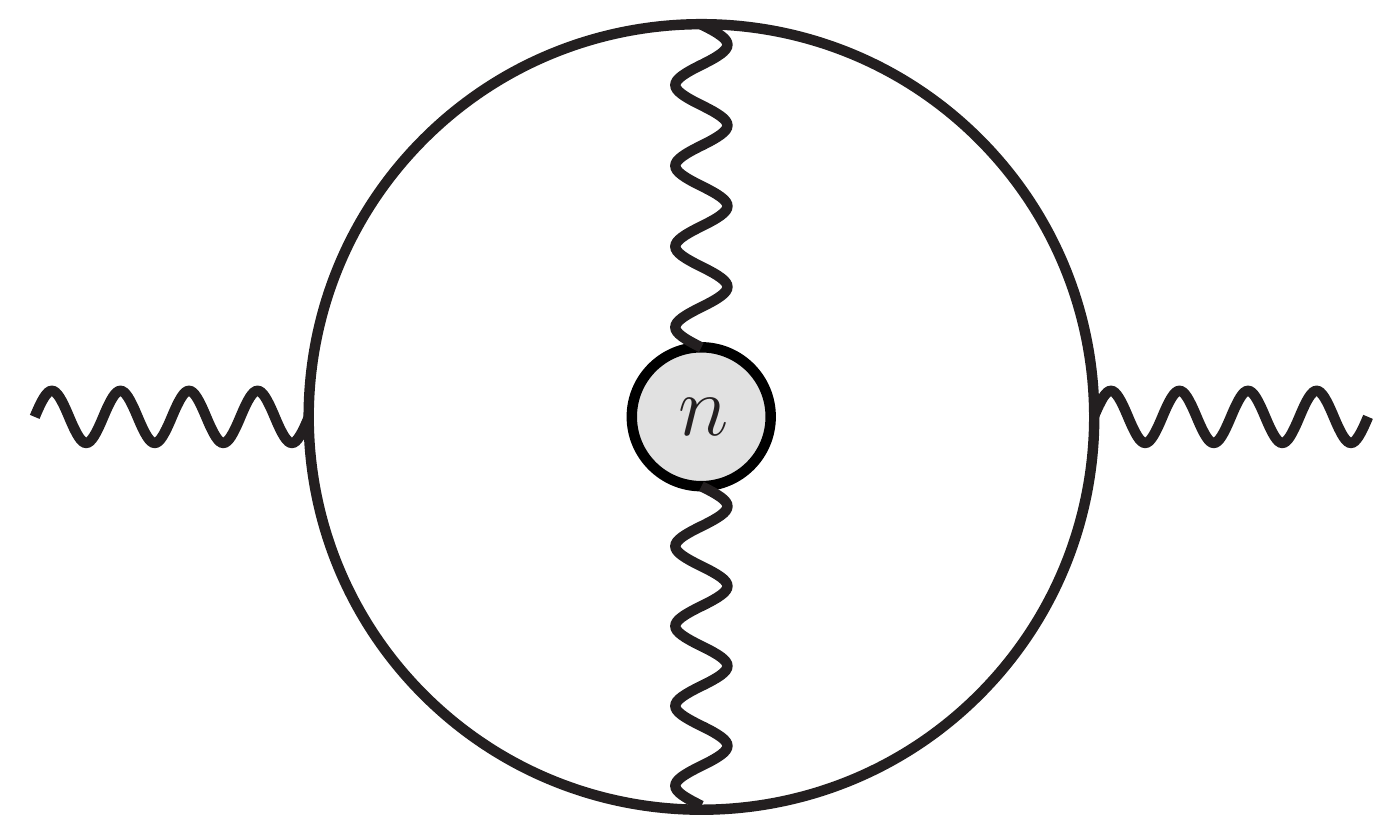}
    \caption{}
    \label{fig:adlerLO2}
\end{subfigure}
\caption{Diagrams at $\ord{1/N_f}$ in the flavour expansion. Diagram $(a)$ contributes with a factor of 2 to account for the similar diagram where the bubble chain connects below the vertices.}
\label{fig:AdlerLO}
\end{figure}

We consider the Fourier transform of the correlation function of two vector currents $j_\mu=\bar \psi\ga_\mu \psi$ of massless quarks, which can be written as 
\begin{equation}
(-i)\int d^4x\, e^{-iqx} \mae{0}{T\{j_\mu(x)j_\nu(0)\}}{0} = (q_\mu q_\nu - \eta_{\mu\nu}q^2)\Pi(Q^2),
\end{equation}
with $Q^2=-q^2$. The Adler function is defined as
\begin{equation}
\label{eq:pitoadler}
D(Q^2) = 4\pi^2Q^2\td{\Pi(Q^2)}{Q^2},
\end{equation}
which is once again an expansion in the coupling $\alpha$. Understanding its asymptotic behaviour in perturbation theory and how
this is related to non-perturbative expansions will be the main focus
for the rest of this work.  We use the flavour expansion to isolate
the relevant Feynman diagrams that lead to asymptotic series.  In this
section we focus on the diagrams in \fig{AdlerLO} that contribute
to the Adler function at leading order in the flavour expansion. A
brief discussion on the contribution of these diagrams to
the Adler function was already given in \sec{uvirrenormalons}.  In
\sec{adlerNLO}, we will then discuss (a subset of) the diagrams at
next-to-leading order in $1/N_f$ for the Adler function.

Instead of calculating diagrams directly, we shall calculate their
Borel transform, as the bubble chain then simplifies to an analytic, 
regularized photon propagator (recall \eq{BorelChain}). In what
follows, we will work in Landau gauge,
i.e.\ $\xi=0$, and in $d=4$ dimensions since the Adler function is
UV finite after counterterms for the fermion bubbles are included.
As we will explain in the next subsection, the
perturbative coefficients of the Adler function itself can then be
easily extracted by applying an inverse Borel transform.  

\begin{figure}
\centering
\includegraphics[width=.7\linewidth]{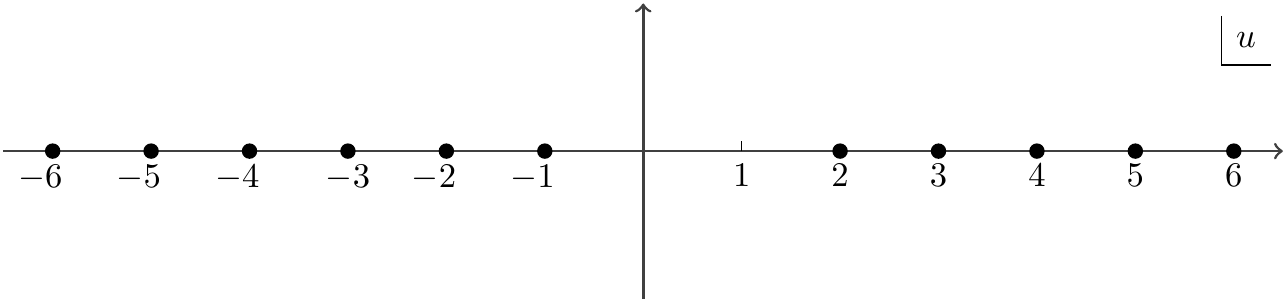}
\caption{Singularities of $\cB[D_\LO](u)$. The UV-renormalons lie at $u=-1,-2,-3,...$ and the IR-renormalons at $u=2,3,4,...$. Characteristic for the Adler function is that the singularity at $u=1$ is missing (see the discussion below \eq{AdlerNLOReadPoleStructure}).}
\label{fig:BorelSingularities}
\end{figure}

The calculation of the diagram in \fig{adlerLO1} is relatively
straightforward. Using results of \app{masters} and \ref{app:momentumIntegrals} one can show that
the two-loop integral factorizes into one-loop ones. With the one-loop master integral \eq{6} we find
\begin{equation}\label{eq:LOdiagram1}
\cB[\Pi_a(Q^2)](u)
= \frac{1}{2\pi^2}\bigg(\frac{Q^2}{\mu^2}e^{C}\bigg)^{-u}
\frac{1}{u(u+1)(u-1)^2(u-2)^2}.
\end{equation}
where for notational simplicity we ignored an overall factor of the colour Casimir factor $C_F$ that we will reinstate in our final expressions. In order to calculate the diagram of \fig{adlerLO2}, one needs the two-loop scalar master integral given in \eq{resultZakarov}.
By similar methods as for the diagram in \fig{adlerLO1} we derive
\begin{align}\label{eq:LOdiagram2}
\cB[\Pi_b(Q^2)](u)
&= -\frac{1}{6\pi^2}\bigg(\frac{Q^2}{\mu^2}e^{C}\bigg)^{-u}
\bigg[
\frac{6}{u(u+1)(u-1)^2(u-2)^2}\nn\\
&\hspace{2cm}+
\frac{\psi^{(1)}(\tfrac{4-u}{2})
-\psi^{(1)}(\tfrac{3-u}{2})
+\psi^{(1)}(\tfrac{u+1}{2})
-\psi^{(1)}(\tfrac{2+u}{2})}
{u(u-1)(u-2)}
\bigg],
\end{align}
with $\psi^{(1)}(z)=\td{^2}{z^2}\log\Ga(z)$ the trigamma function.
Taking the two diagrams together and using \eq{pitoadler} to translate
the result for $\Pi(Q^2)$ to that of the Adler function $D(Q^2)$, we
obtain the Borel transform of the Adler function at leading order (LO) in the flavour expansion (which has also recently been derived in \cite{Mikhailov:2023lqe}):
\begin{align}
\cB[D_{\LO}(Q^2)](u)
&=4\pi^2Q^2\td{}{Q^2}\Big[\cB[\Pi_b(Q^2)](u) + 2 \cB[\Pi_a(Q^2)](u)\Big]\nn\\
&= \frac{2}{3}\bigg(\frac{Q^2}{\mu^2}e^C\bigg)^{-u}
\frac{\psi^{(1)}\big(\tfrac{4-u}{2}\big) 
-\psi^{(1)}\big(\tfrac{3-u}{2}\big) 
+\psi^{(1)}\big(\tfrac{1+u}{2}\big) -\psi^{(1)}\big(\tfrac{2+u}{2}\big)}{(u-1)(u-2)}.\label{eq:LOAdler}
\end{align}
In the next subsection we use this result to perform a resurgence
analysis, as the expression readily allows for an expansion around $u=0$. Note that this result was already known in the equivalent form \cite{Broadhurst:1992si}
\begin{align}
\cB[D_{\LO}(Q^2)](u)
&= \frac{32}{3}\bigg(\frac{Q^2}{\mu^2}e^C\bigg)^{-u}
\frac{1}{2-u}
\sum_{n=2}^\infty \frac{(-1)^n n}{(n^2-(1-u)^2)^2}\label{eq:Broadhurst}\\
&= \frac{32}{3}\bigg(\frac{Q^2}{\mu^2}e^c\bigg)^{-u}
\frac{1}{2-u} \sum_{n=2}^\infty \frac{(-1)^n}{4(1-u)} \bigg[\frac{1}{(n-1+u)^2}-\frac{1}{(n+1-u)^2}\bigg],\label{eq:AdlerNLOReadPoleStructure}
\end{align}
where the second form of this equation allows one to easily read off
the pole structure. We see that there exists an infinite set of both
UV ($u<0$) and IR ($u>0$) singularities (for QCD) at integer values of $u$, all
of which are double poles, except for the singularity at $u=2$ which
is a single pole and the singularity at $u=1$ which is
absent\footnote{Note that although there appears to be a pole at $u=1$
  in \eq{AdlerNLOReadPoleStructure}, it  vanishes because the
  expansion around $u=1$ of the terms in square brackets starts at
  order $\ord{u-1}$.} (see also \fig{BorelSingularities}). This also agrees with the calculation that led to \eq{11}, which presented the leading IR and UV poles in the Borel plane. We should
mention here that the singularity at $u=1$ really is absent:
it is present in \eq{LOdiagram1} and \eq{LOdiagram2} separately, but cancels when we take the two diagrams together.  The
fact that the IR renormalon at $u=1$ is absent is characteristic for
the Adler function and is already expected on physical grounds, as there is no dimension-2 operator in the OPE for the Adler function with massless fermions \cite{Parisi:1978az,Mueller:1984vh}.

\subsection{Resurgence analysis: first two non-perturbative sectors}
\label{sec:resurgenceLO}
Our goal is to construct the transseries
for the Adler function. For the associated resurgence analysis we need the perturbative coefficients of the leading order
Adler function $D_{\LO}(Q^2)$ itself, i.e. we need to do an inverse
Borel transform on the results obtained in the previous subsection. In this section, and the sections hereafter, we will work mostly with the variable $u=-\beta_{0}t$ as the actual Borel parameter instead of $t$, so that singularities in the Borel plane are conveniently placed at integer positions. 
Therefore, we expand the Borel transform \eq{LOAdler} around $u=0$ and
write the result as
\begin{equation}
\cB[D_{\LO}(Q^2)](u) 
= \sum_{n=0}^\infty \frac{d_n}{\Ga(n+1)} u^n,
\end{equation}
after which the perturbative coefficients $d_n$ of $D_{\LO}$ can be read off:
\begin{align}
\hat{D}_{\LO}(\al) 
&= \sum_{n=0}^\infty d_n\hat\al^{n+1}\label{eq:AdlerLOPerturbativeSector1}\\
&= \hat\al 
+ \left(\frac{23}{6}-4\zeta_3\right)\hat\al^2
+ \left(18-12\zeta_3\right)\hat\al^3
+ \left(\frac{201}{2}-42\zeta_3-60\zeta_5\right)\hat\al^4
+...\nn
\end{align}
Here $\hat\al=-\beta_{0}\al$ is the variable conjugate to the Borel variable $u$. As the inverse Borel transform gives an additional factor of $\hat{\al}$, we also defined $\hat{D}(Q^2)=-\beta_0D(Q^2)$, to compensate for the additional factor of $-\beta_0$. Furthermore, to avoid logarithms of
$\frac{Q^2}{\mu^2}e^C$ that will make the analysis needlessly
complicated, we choose $\mu^2=Q^2e^C$. For notational convenience, we will drop the hats on $\hat{D}$ and $\hat{\al}$ in what follows. When we give the full transseries expression in
the end of this section, we will reinstate the factors of
$\beta_{0}$.

With the exact Borel transform \eq{LOAdler} one can easily compute the
first, say $n=1000$, perturbative coefficients. (However, for the resurgence
analysis we perform below, we found that $200$ coefficients was
enough.) With these coefficients at our disposal, we can start
thinking about what we may expect the transseries to look like. As
explained near \eq{BorelSingularitiesMultiParTrans}, for
a $k$-parameter transseries the Borel singularities of the
transseries sector lie at positions $u=\bm{\ell}\cdot\bm{A}$ with
$\ell\in\bN^k$ and $\bm{A}=(A_1,...,A_k)$ the non-perturbative
exponents.  As the poles of $\cB[D_\LO](u)$ lie at both positive and
negative integer values of $u$, a minimal Ansatz for the transseries
is a two-parameter transseries with $A_1 = 1$ and $A_2 = -1$ the
non-perturbative exponents.  Therefore, we write the perturbative
sector as
\begin{equation}
  \label{eq:7}
D_{\LO}^{(0,0)}(\al) 
\equiv \sum_{n=0}^\infty d_n^{(0,0)}\al^{n+1},
\end{equation}
with $d_n^{(0,0)}\equiv d_n$ the perturbative coefficients of \eq{AdlerLOPerturbativeSector1}. This will now be the $(0,0)$-sector of a two-parameter transseries
\begin{equation}
\label{eq:5}
D_{\LO}(\al) 
= \sum_{n=0}^\infty\sum_{m=0}^\infty \si_1^n\si_2^m e^{-n\frac{A_1}{\al}}e^{-m\frac{A_2}{\al}}D_{\LO}^{(n,m)}(\al).
\end{equation}
We should emphasize here that this is a {\em minimal} transseries Ansatz and, recalling our discussion in \sec{generalizations}, it might be the case that one needs additional transseries parameters on top of the parameters $\si_1$ and $\si_2$, as well as further (e.g.\ logarithmic) transmonomials.
We shall discuss the interpretation of the transseries parameters in \sec{discussionLO} and will indeed find logarithmic transmonomials when we extend our analysis to order $1/N_f^2$ in section \sec{adlerNLO}.

In order to test the Ansatz \eq{5}, and construct the non-perturbative sectors
\begin{equation}
D_{\LO}^{(n,m)}(\al) = \al^{\beta_{nm}}\sum_{h=0}^\infty d_h^{(n,m)}\al^h
\end{equation}
we will use resurgent large order relations. Note that we use the conventional notation $\beta_{nm}$ (and sometimes for readability $\beta_{n,m}$) for the starting orders in the non-perturbative sectors; of course these orders (that always have two indices) should not be confused with the $\beta$-function and its coefficients. Recall that we can
readily write large order relations using the allowed motions on the
alien lattice, which for our two-parameter Ansatz looks as shown in
\fig{alienLatticeExample}. Since our starting node is the
$(0,0)$-sector, we know that we can only have forward steps in the
$(1,0)$ and $(0,1)$ direction, so that only the $(\ell,0)$ and
$(0,\ell)$ sectors will play a role. In other words, for the large
order behaviour of the perturbative coefficients, we get
\begin{align}
  \label{eq:8}
d_k^{(0,0)}
\sim \sum_{\ell=1}^\infty \frac{S_{1,0}^\ell}{2\pi\img}\sum_{h=0}^\infty d_h^{(\ell,0)}\frac{\Ga(k-h-\beta_{\ell,0})}{(\ell A_1)^{k-h-\beta_{\ell,0}}}
+ \sum_{\ell=1}^\infty \frac{S_{0,1}^\ell}{2\pi\img}\sum_{h=0}^\infty d_h^{(0,\ell)}\frac{\Ga(k-h-\beta_{0,\ell})}{(\ell A_2)^{k-h-\beta_{0,\ell}}},
\end{align}
which is a relation between the perturbative coefficients $d_k^{(0,0)}$ and the non-perturbative coefficients $d_h^{(\ell,0)}$ and $d_h^{(0,\ell)}$. 
As we will not determine the Stokes constants $S_{1,0}$ and $S_{0,1}$ in this paper, and to avoid writing down ubiquitous factors of $2\pi\img$, we will absorb these factors in the non-perturbative coefficients. Therefore we write
\begin{equation}
\tilde{d}_h^{(\ell,0)} \equiv \frac{S_{1,0}^\ell}{2\pi\img} d_h^{(\ell,0)}
\qquad\text{and}\qquad
\tilde{d}_h^{(0,\ell)} \equiv \frac{S_{0,1}^\ell}{2\pi\img} d_h^{(0,\ell)}\,.
\end{equation}
We shall see in a moment that ordering the elements of \eq{8} by their
size gives us a way to extract the non-perturbative coefficients,
keeping in mind that this relation is valid in the large $k$ limit. 

\begin{figure}[t]
\centering
\begin{subfigure}{7cm}
    \includegraphics[width=\textwidth]{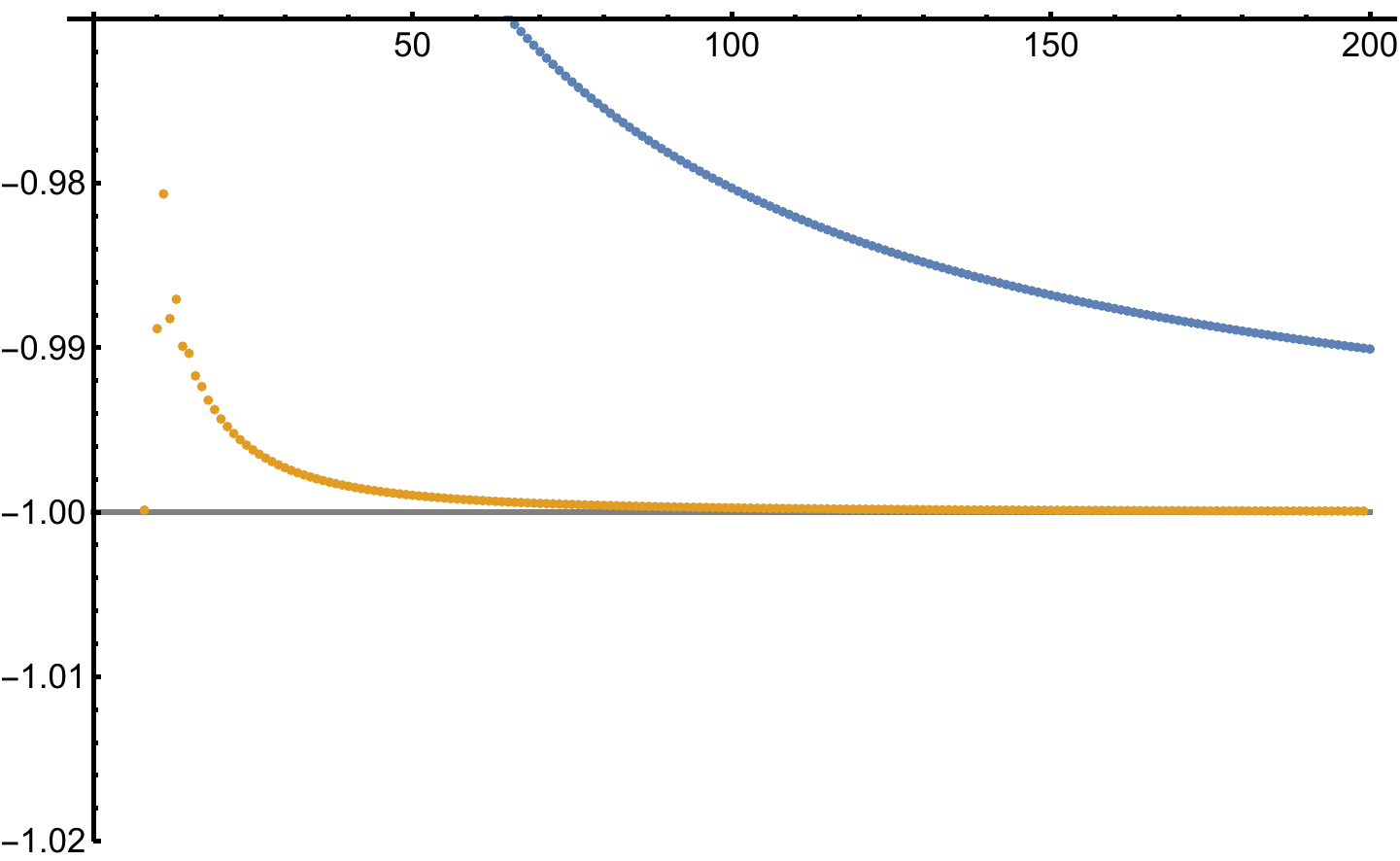}
    \caption{}
    \label{fig:ratioA}
\end{subfigure}
\hspace{.5cm}
\begin{subfigure}{7cm}
    \includegraphics[width=\textwidth]{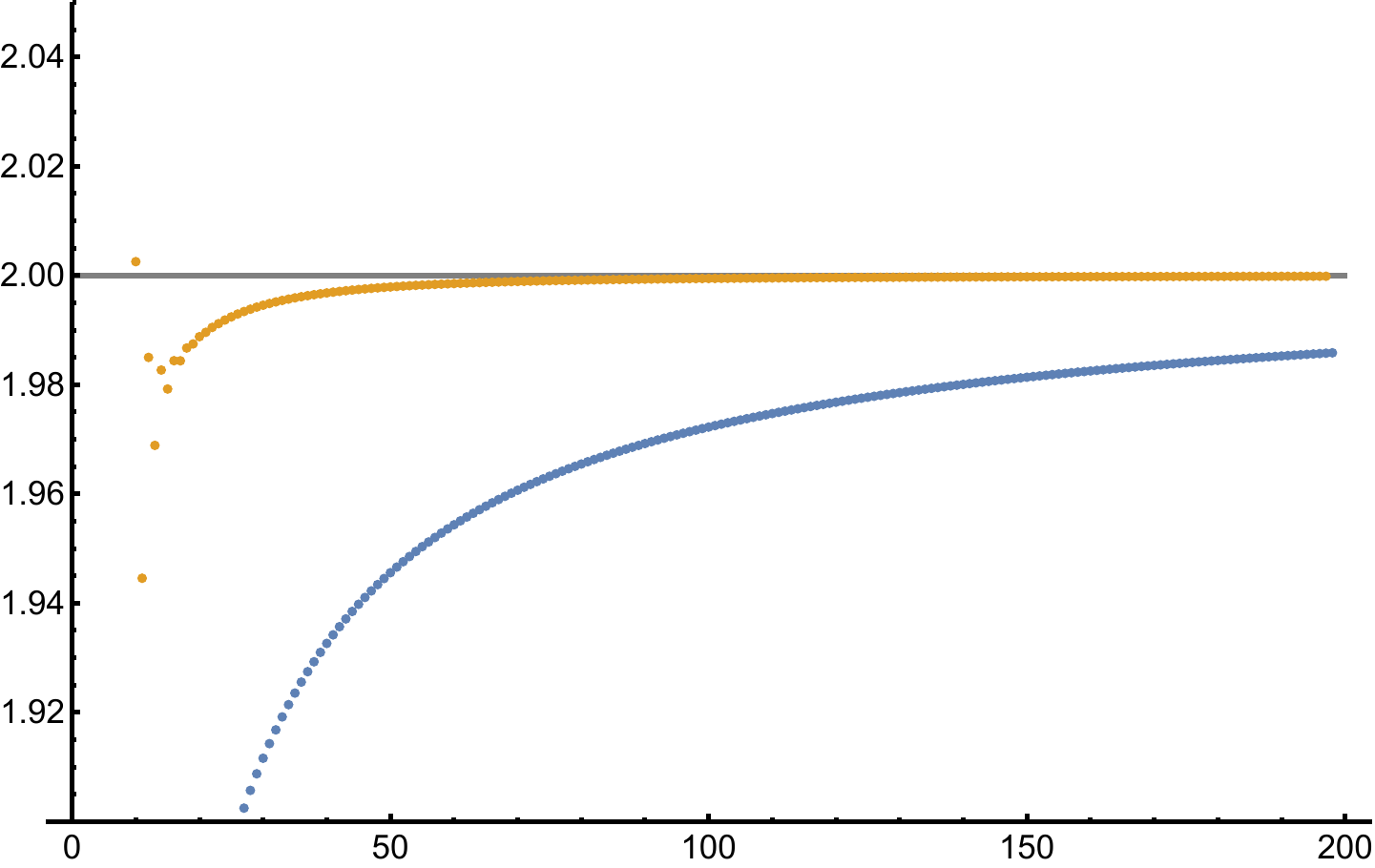}
    \caption{}
    \label{fig:ratioBeta}
\end{subfigure}
\caption{Fig.~(a) shows the sequence $\bA(k)$, \eq{ratioA} (top curve, blue),
  together with its first Richardson transform (bottom curve, orange) to accelerate
  the convergence. Similar, (b) shows the sequence $\bB(k)$ (bottom curve, blue),
  \eq{ratioBeta}, and its first Richardson transform (top curve, orange). We observe that
  already the first Richardson transform significantly accelerates the
  convergence. As explained in the text, higher order Richardson
  transforms converge even faster and become indistinguishable from
  the gray, horizontal lines denoting the expected values $A_2=-1$ and
  $\beta_{0,1}=-2$.}
\label{fig:ratioLO}
\end{figure}

First, we observe that nodes in the alien lattice that are further away from the
$(0,0)$-node, i.e.\ with larger $\ell$, have a more exponentially
suppressed contribution to the large order growth of $d_k^{(0,0)}$.
As $A_1=1$ and $A_2=-1$ are equal in size, the leading order growth
comes from the $(1,0)$ and $(0,1)$ sectors. However, we saw above that
the pole in the Borel plane at $u=1$ is missing, meaning that the
$(1,0)$-contribution is actually absent, so hence the leading
growth will be coming from the $(0,1)$ sector:
\begin{align}\label{eq:largeOrderSector1}
d_k^{(0,0)}
&\sim 
\sum_{h=0}^\infty \tilde{d}_h^{(0,1)}\frac{\Ga(k-h-\beta_{0,1})}{A_2^{k-h-\beta_{0,1}}}
+ \ord{2^{-k}} \\
&= 
\frac{\Ga(k-\beta_{0,1})}{A_2^{k-\beta_{0,1}}}
\bigg[ 
\tilde{d}_0^{(0,1)} 
+ \frac{A_2\, \tilde{d}_1^{(0,1)}}{k} 
+ \frac{A_2(\beta_{0,1}+1) \tilde{d}_1^{(0,1)}+A_2^2\, \tilde{d}_2^{(0,1)}}{k^2} 
+...
\bigg]
+ \ord{2^{-k}}\,, \nn
\end{align}
where $\ord{2^{-k}}$ refers to contributions from the $(\ell,0)$ and $(0,\ell)$ non-perturbative
  sectors with $\ell \geq 2$. 
In the second line we factored out the leading order growth in
$k$ and explicitly wrote the first two $1/k$ corrections.
To show that this expression indeed captures the large order
behaviour of the perturbative coefficients, and in order to find the various
as yet unknown coefficients that appear in this large order relation, we
can perform so called {\em ratio tests} (see e.g.\ \cite{Marino:2007te, Garoufalidis:2010ya, Aniceto:2011nu}) on the known
perturbative coefficients $d_k^{(0,0)}$. To see this, we start with $A_2$ and
$\beta_{0,1}$, and consider
\begin{equation}\label{eq:ratioA}
\bA(k) \equiv k\,\frac{d_k^{(0,0)}}{d_{k+1}^{(0,0)}} \sim A_2 +\ORd{\frac{1}{k}}
\end{equation}
and 
\begin{equation}\label{eq:ratioBeta}
\bB(k) \equiv k\log\bigg[\frac{f_{k+1}}{f_k}\bigg] 
\sim - \beta_{0,1}+\ORd{\frac{1}{k}}
\qquad\text{with}\qquad
f_k = \frac{d_k^{(0,0)} A_2^k}{\Ga(k)},
\end{equation}
which should converge to $A_2$ and $-\beta_{0,1}$ in the large 
$k$-limit, respectively.  In \figs{ratioA}{ratioBeta} we have plotted
these two sequences  together with their first {\em Richardson
transform} for the first 200 perturbative coefficients.
The Richardson transform method accelerates convergence of
sequences of the form
\begin{equation}
s_0=\lim_{k\to\infty}\bS(k)
\qquad\text{with}\qquad
\bS(k) = s_0+\frac{s_1}{k}+\frac{s_2}{k^2}+...
\end{equation}
where the coefficients $s_n$ are not known explicitly. We  denote
the $N$th Richardson transform of $\bS(k)$ by RT$[\bS](k,N)$. We refer
the interested reader to \app{richardson} for more details. 

Above \eq{7} we expressed the expectation that the non-perturbative exponent $A_2$ is equal
to $-1$. Recalling our discussion around
\eq{valueBetaCoefficientHigherOrderPoles}, also the value
$\beta_{0,1}=-2$ for the characteristic exponent was  expected since
the leading singularity in the Borel plane at $u=-1$ is a double pole. Using Richardson transforms, we confirm these expectations to good numerical precision. For example, RT$[\bA](190,10)$
agrees to 23 decimal places with $A_2=-1$, and we obtain similar
precision for $\beta_{(0,1)}= -2$.

\begin{figure}
\centering
\begin{subfigure}{7cm}
    \includegraphics[width=\textwidth]{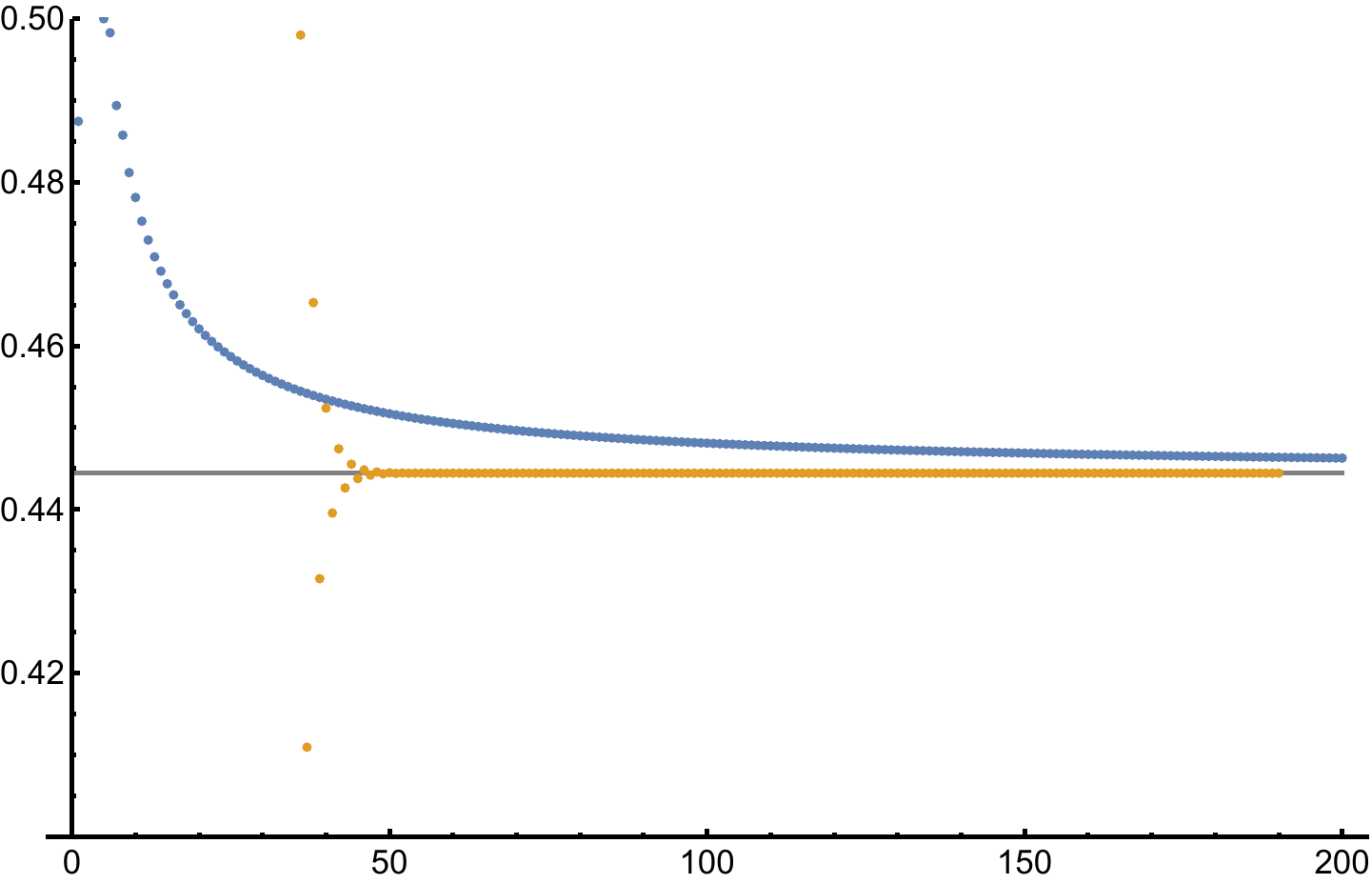}
    \caption{}
    \label{fig:LO_growth}
\end{subfigure}
\hspace{.5cm}
\begin{subfigure}{7cm}
    \includegraphics[width=\textwidth]{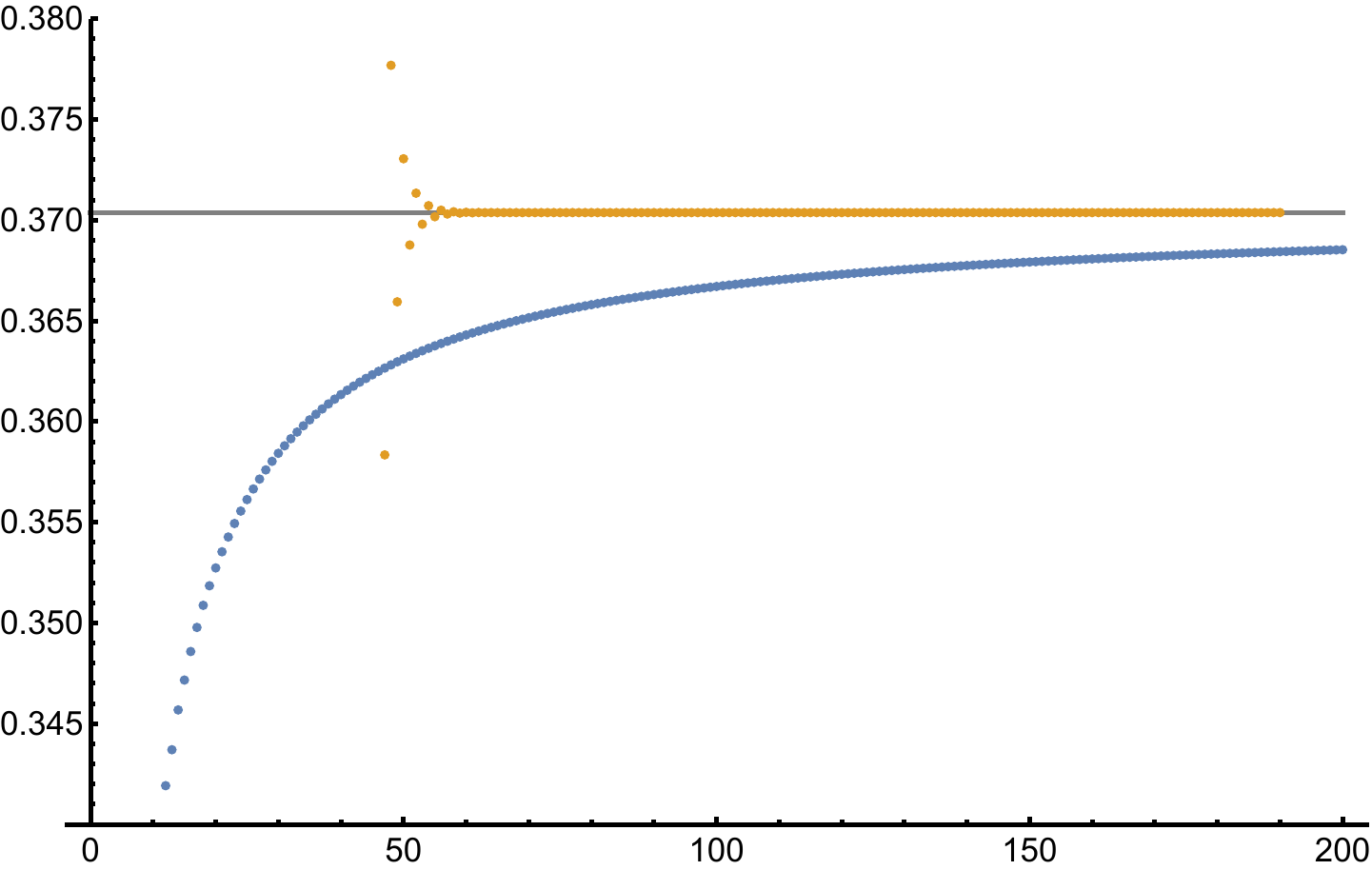}
    \caption{}
    \label{fig:NLO_growth}
\end{subfigure}
\caption{Fig. $(a)$ and $(b)$ show the sequence $\bD_0(k)$ (top curve left, blue),
  \eq{ratioCoef1}, and $\bD_1(k)$ (bottom curve right, blue), \eq{ratioCoef2}, respectively. To
  accelerate the rate of convergence, the other curves show the 10th Richardson
  transform in orange, which shows good convergence to the exact values
  $\frac49$ and $\frac{10}{27}$, respectively, denoted by the gray horizontal
  lines.}
\label{fig:sector01}
\end{figure}

We can now systematically extract the non-perturbative coefficients
$d_h^{(0,1)}$ in \eq{largeOrderSector1}. First, we rewrite that equation in the form
\begin{equation}\label{eq:ratioCoef1}
\bD_0(k)
\equiv\frac{d_k^{(0,0)} A_2^{k-\beta_{0,1}}}{\Ga(k-\beta_{0,1})} 
\sim 
\tilde{d}_0^{(0,1)}+\ORD{\frac{1}{k}}.
\end{equation}
In \fig{LO_growth} we have plotted the sequence together with its 10th Richardson transform, and we conclude that 
\begin{equation}
\tilde{d}_0^{(0,1)} = \frac{4}{9}
\end{equation}
to great precision, e.g.\ RT$[\bD_0](190,10)$ agrees with $4/9$ to 26 decimal
places.  We should notice here that we determined the value of
$\tilde{d}_0^{(0,1)}$, in which the unknown Stokes constant $S_{0,1}$
is absorbed. This is actually a generic phenomenon in the resurgence of
transseries with free parameters such as $\si_1$ and $\si_2$ in \eq{5}; one can see from that expression that rescaling these
parameters will rescale the expansion coefficients in the
non-perturbative sectors. Moreover, as $\si_i$ is rescaled, the overall Stokes constants which determine the jump in $\si_i$ are similarly rescaled. As a
result, only `scale invariant' combinations of the expansion
coefficients and the Stokes constants have a physical
meaning. Computing the Stokes constants (after fixing this scale invariance) is a different enterprise, which we will not
pursue in this paper.

Next, we can plot the sequence 
\begin{equation}\label{eq:ratioCoef2}
\bD_1(k)
\equiv k\left[\bD_0(k) -\frac{4}{9}\right]
\sim A_2\, \tilde{d}_1^{(0,1)}+\ORd{\frac{1}{k}}
\end{equation}
and from \fig{NLO_growth} we deduce that this sequence converges to 
\begin{equation}
A_2\, \tilde{d}_1^{(0,1)} = \frac{10}{27}.
\end{equation}
which can again be verified up to 26 decimal places by doing 10
Richardson transforms. We can repeat this process and in fact derive a
closed form for all of the $1/k$ contributions in the large order
relation:
\begin{align}
d^{(0,0)}_k
&\sim \frac{\Ga(k+2)}{(-1)^{k}}
\bigg[
\frac49
+ \frac{10}{27}\frac{1}{k}\sum_{n=0}^\infty \bigg(\frac{-1}{k}\bigg)^n
\bigg]
+ \ord{2^{-k}}\nn\\
&= \frac{\Ga(k+2)}{(-1)^{k}}
\bigg[
\frac49
+ \frac{10}{27}\frac{1}{k+1}
\bigg]
+ \ord{2^{-k}}.
\end{align}
Using \eq{largeOrderSector1}, we can also rewrite this in terms of the non-perturbative coefficients $d_h^{(0,1)}$, which are seen to be 
\begin{equation}\label{eq:LOAdler(0,1)coeffs}
\tilde{d}_0^{(0,1)}= \frac{4}{9},
\qquad
\tilde{d}_1^{(0,1)}= -\frac{10}{27},
\qquad
\tilde{d}_{h\geq2}^{(0,1)}= 0.
\end{equation}
Most of the coefficients vanish and as a result, the
$(1,0)$ sector is not an asymptotic series but a finite one. We
will revisit this point more extensively later.

So far, from the first 200 perturbative coefficients that we computed numerically and analyzed, we have extracted the complete first non-perturbative sector $D_\LO^{(0,1)}$.  For the leading order Adler function, we know the
exact Borel transform \eq{LOAdler} and can therefore 
check our large order relations. That is, we have that
\begin{align}
\cB[D_{\LO}^{(0,0)}](u)\Big|_{u=-1}
&= \frac{4/9}{(u+1)^2} + \frac{10/27}{(u+1)} + \text{regular terms},
\end{align}
which indeed agrees with \eq{LOAdler(0,1)coeffs}. (This is also the
reason that we wrote the above numerical estimates with an equal
sign.) Note that this leading UV behaviour was already indicated in \eq{11}, but here we determined the precise coefficient, as well as the subleading term. We  emphasize that this is a rather unique situation. In
many resurgence examples one does not have the luxury of knowing the exact
Borel transform. In fact, in \sec{adlerNLO} when we study the Adler
function at the next order in the flavour expansion, we cannot compute the Borel transform exactly
anymore. Clearly, however, we can study
the perturbative coefficients numerically (using similar ratio tests as
the ones above), and extract the non-perturbative sectors without
ambiguity.

\begin{figure}
\centering
\begin{subfigure}{7cm}
    \includegraphics[width=\textwidth]{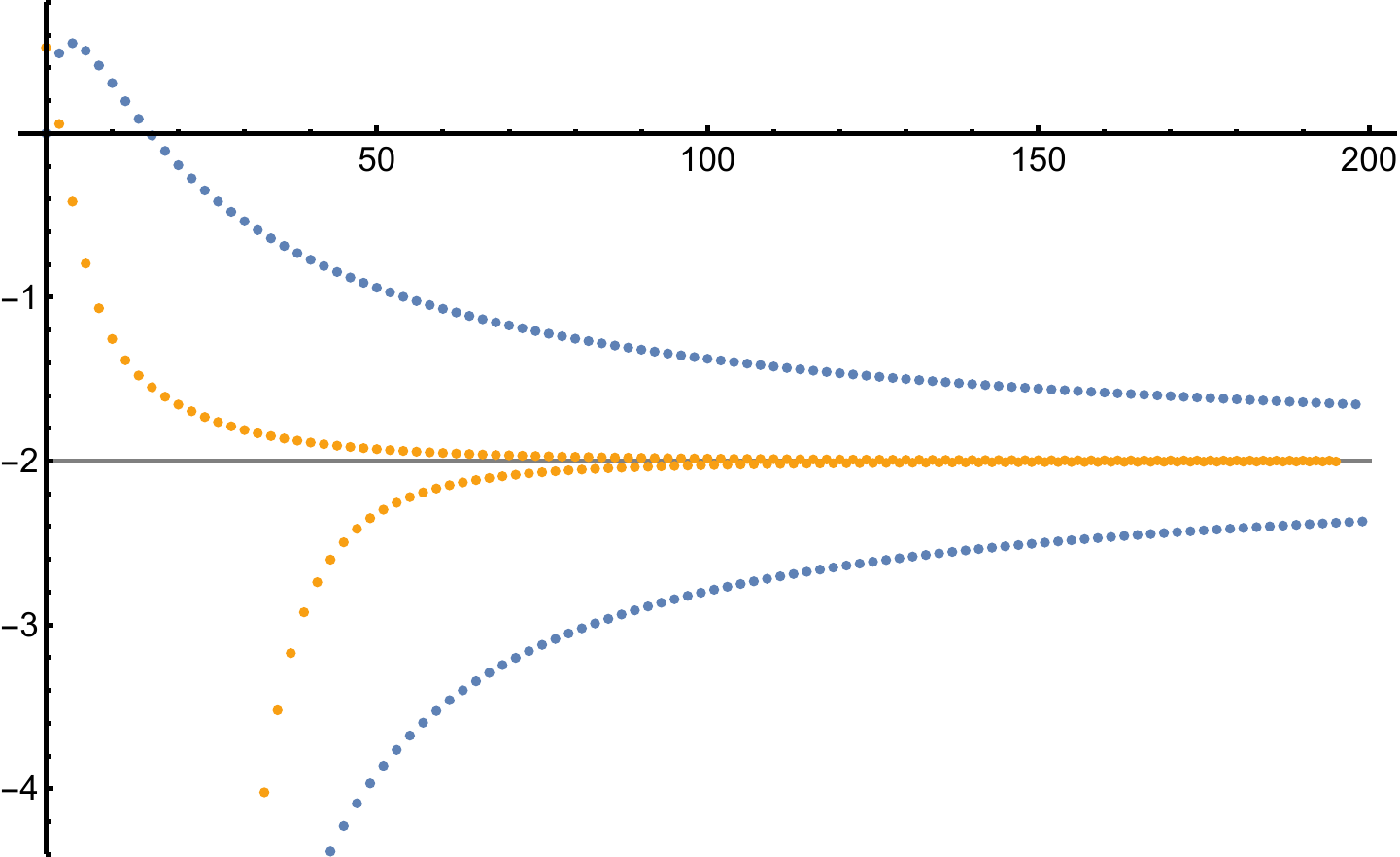}
    \caption{}
    \label{fig:sector2A}
\end{subfigure}
\hspace{.5cm}
\begin{subfigure}{7cm}
    \includegraphics[width=\textwidth]{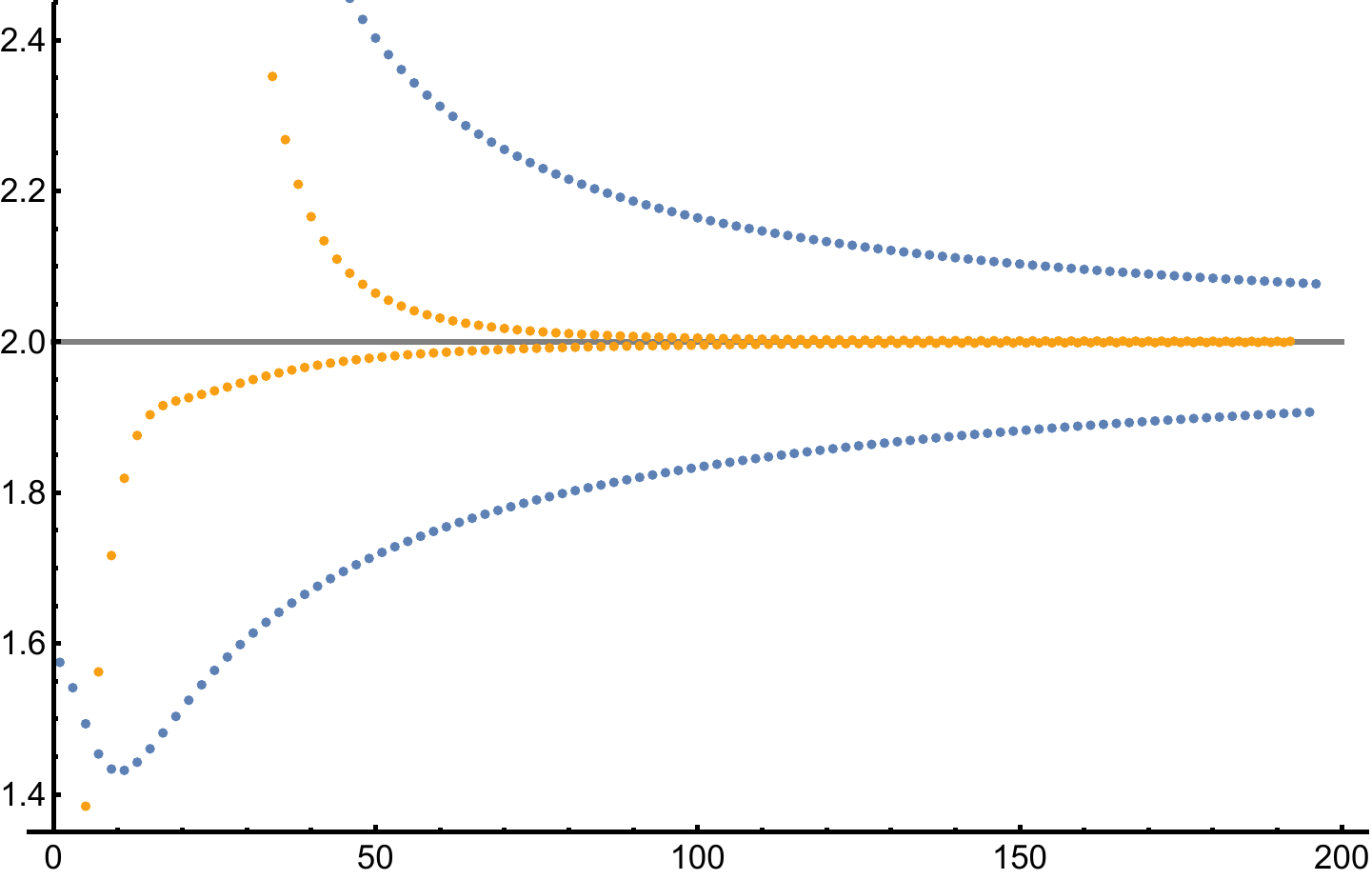}
    \caption{}
    \label{fig:sector2beta}
\end{subfigure}
\caption{In Fig. $(a)$ and $(b)$, we show the sequences $\bA(k)$ of \eq{ratioA} (outer two curves, in blue), and $\bB(k)$ of \eq{ratioBeta}, but now constructed from $\de_k^{(1)}$ given in \eq{LOsubtractSector1}. We also show their 2nd Richardson transforms. (Middle two curves, in orange.) Note that as the parity of $k$ is important, we took the Richardson transforms separately on even and odd $k$.}
\label{fig:sector2}
\end{figure}

Having obtained the complete first non-perturbative sector, we can now subtract the corresponding leading order growth from the large order expression for the perturbative sector coefficients. That is, we construct the sequence 
\begin{equation}\label{eq:LOsubtractSector1}
\delta_k^{(1)} 
\equiv d_k^{(0,0)}-\frac{\Ga(k+2)}{(-1)^{k}}
\bigg[
\frac49
+ \frac{10}{27}\frac{1}{k+1}
\bigg]
\end{equation}
where the asymptotic growth should now be dominated by the non-perturbative sectors $(2,0)$ and $(0,2)$, i.e.
\begin{align}\label{eq:largeOrderLOsector2}
\de_k^{(1)}
\sim 
\sum_{h=0}^\infty \tilde{d}_h^{(2,0)}\frac{\Ga(k-h-\beta_{2,0})}{(2 A_1)^{k-h-\beta_{2,0}}}
+
\sum_{h=0}^\infty \tilde{d}_h^{(0,2)}\frac{\Ga(k-h-\beta_{0,2})}{(2 A_2)^{k-h-\beta_{0,2}}} 
+ \ord{3^{-k}}.
\end{align}
Since the Borel transform has poles at both $u=-2$ and $u=+2$, we now
see two sectors appearing in this formula. Note that also
$A_1=-A_2=1$, i.e. they are equal but opposite in sign. Even though the rest of the
$(\ell,0)$ and $(0,\ell)$ sectors are quite different from
each other, as we shall see later, the parity of $k$ will
have an important effect on the right hand side of this expression. In
particular, in order to use Richardson transforms to speed up the
convergence of series, we have to separate even and odd $k$ and
perform Richardson transforms on them separately.  In
\figs{sector2A}{sector2beta}, we show the sequences of \eq{ratioA} and
\eq{ratioBeta}, but now constructed from $\de_k^{(1)}$.  Notice that
the ratio test \eq{ratioBeta} can be taken, if we assume that the starting orders
$\beta_{2,0}=\beta_{0,2}\equiv\beta_2$ are the same. 
Even though we find in \eq{LOsector2parameters} below that strictly speaking $\beta_{2,0}\neq \beta_{0,2}$, this is not an issue since we can compute with $\beta_2 = \min(\beta_{2,0},\beta_{0,2})$ and allow for a finite number of vanishing leading coefficients in one of the two series.  Here, we
conclude that $2A_2 = -2A_1=-2$ and $\beta_{2}=-2$.
Following \eqs{ratioCoef1}{ratioCoef2}, but now also taking the parity of $k$ into account by using ratio tests for even and odd $k$ separately, we obtain
\begin{equation}\label{eq:LOsector2parameters}
\tilde{d}_0^{(2,0)} = 0,
\qquad
\tilde{d}_1^{(2,0)} = 2
\qquad \mbox{and} \qquad
\tilde{d}_0^{(0,2)} = -\frac29,
\qquad
\tilde{d}_1^{(0,2)} = \frac{7}{54},
\end{equation}
with all other coefficients equal to zero. 
This can be compared to the expansions around $u=-2$ and $u=2$ respectively, and we find the expected agreement with the exact Borel transform:
\begin{align}
\cB[D_{\LO}^{(0,0)}](u)\Big|_{u=-2}
&= -\frac{2/9}{(2+u)^2} - \frac{7/54}{(2+u)} + \text{regular terms}\nn\\
\cB[D_{\LO}^{(0,0)}](u)\Big|_{u=2}
&= \frac{2}{(2-u)} + \text{regular terms}\,,
\end{align}
where this leading IR behaviour was indicated in \eq{11}, here given with the precise coefficient.
Notice that in the second line the leading singularity around $u=2$ is a single pole and
therefore one should expect $\beta_{2,0}=-1$ instead of $-2$, as mentioned above. Indeed,
we found that $\tilde{d}_0^{(2,0)}=0$ and one can conclude that the series effectively starts at order $-1$.

\subsection{Resurgence analysis: all non-perturbative sectors}
\label{sec:resurgenceLOall}
Having used resurgence to recover the expressions for the first and second
non-perturbative sectors of the Adler function transseries, we can
repeat the analysis of the previous subsection to also obtain other
sectors. Since the procedure is very similar, we will be more brief here about the techniques and will focus more on the results and on the general structure that emerges. 

To find higher sectors, we recursively probe the next sector by
subtracting from the large order expression for the perturbative coefficients all sectors we have found
so far. In this way we find the following pattern:
\begin{equation}\label{eq:LOallOrderLargeOrderFormula}
\de_k^{(\ell)}
\equiv \de_k^{(\ell-1)} - \frac{\Ga(k+2)}{(-\ell)^{k+2}}\bigg(s_k^{(\ell)}+\frac{t_k^{(\ell)}}{k+1}\bigg),
\end{equation}
where $\ell$ labels the sectors, and the coefficients $\de_k^{(\ell)}$ are the ones whose large order
behaviour can be used to probe sector $\ell+1$.  Each sector provides
two nonzero coefficients, $s_k^{(\ell)}$ and $t_k^{(\ell)}$, whose
values for $1\leq\ell\leq8$ are given in \tab{adlerHigherSectors}. As
before, we obtained these numbers numerically, using
ratio tests. However, for these specific diagrams, we fortunately have
an analytic Borel transform result, and as in the previous section our
numbers can be directly checked with the expansion of the Borel
transform around the poles. Thus, our numbers are not only
approximately equal to the given fractions, but
turn out to be exact.
\begin{table}
\centering
\renewcommand{\arraystretch}{1.7}
\begin{tabular}{|c|c|c|}
\hline
$\ell$ & $s_k^{(\ell)}$ & $t_k^{(\ell)}$ \\ 
\hline
1 & $\frac49$ & $\frac{10}{27}$\\
2 & $-\frac29$ & $-\frac{7}{27}+4(-1)^k$\\
3 & $\frac{2}{15}-\frac43(-1)^k$ & $\frac{9}{50}-6(-1)^k$ \\
4 & $-\frac{4}{45}+\frac49(-1)^k$ & $-\frac{88}{675}+\frac{40}{27}(-1)^k$\\
\hline
\end{tabular}
\hspace{.2cm}
\begin{tabular}{|c|c|c|}
\hline
$\ell$ & $s_k^{(\ell)}$ & $t_k^{(\ell)}$ \\ 
\hline
5 & $\frac{4}{63}-\frac{2}{9}(-1)^k$ & $\frac{130}{1323}-\frac{35}{54}(-1)^k$\\
6 & $-\frac{1}{21}+\frac{2}{15}(-1)^k$ & $-\frac{15}{196}+\frac{9}{25}(-1)^k$\\
7 & $\frac{1}{27}-\frac{4}{45}(-1)^k$ & $\frac{119}{1944}-\frac{154}{675}(-1)^k$\\
8 & $-\frac{4}{135}+\frac{4}{63}(-1)^k$ & $-\frac{304}{6075}+\frac{208}{1323}(-1)^k$\\
\hline
\end{tabular}
\caption{This table gives the values of $s_k^{(\ell)}$ and $t_k^{(\ell)}$ for $1\leq\ell\leq8$ contributing to the large order behaviour of the perturbative coefficients via \eq{LOallOrderLargeOrderFormula}. As explained in the text, these values can be extracted numerically from the perturbative coefficients, but as these values can also be extracted from the exact Borel transform, this table is exact.}
\label{tab:adlerHigherSectors}
\end{table}

A closer examination of these numbers reveals the following general pattern:
\begin{equation}
s_k^{(\ell)} 
= \frac83(-1)^\ell
\begin{cases}
\frac{-1}{(\ell+1)(\ell+2)} 
&\quad\text{for}\quad \ell<3\\
\frac{-1}{(\ell+1)(\ell+2)} 
+ (-1)^k\frac{1}{(\ell-2)(\ell-1)} 
&\quad\text{for}\quad \ell\geq3.
\end{cases}
\end{equation}
Similarly, we find a closed form for $t_k^{(\ell)}$:
\begin{equation}
t_k^{(\ell)} 
= \frac83(-1)^\ell
\begin{cases}
-\frac{10}{27} &\quad\text{for}\quad \ell=1\\
-\frac{7}{72}+\frac32(-1)^k &\quad\text{for}\quad \ell=2\\
\frac{-\ell(2\ell+3)}{(\ell+1)^2(\ell+2)^2} 
+ (-1)^k\frac{\ell(2\ell-3)}{(\ell-2)^2(\ell-1)^2} 
&\quad\text{for}\quad \ell\geq3.
\end{cases}
\end{equation}
Observe that in all these results, there are terms without $k$-dependent signs as well as terms with $(-1)^k$ factors, respectively probing the coefficients of the $(\ell,0)$ sectors (with non-perturbative exponent $\ell A_1 = +\ell$) and the $(0,\ell)$ sectors (with non-perturbative exponent $\ell A_2 = -\ell$).

Also notice that we have a closed form with a fixed pattern for all $\ell\geq3$, while 
$\ell=1$ and $\ell=2$ are different. This can be traced back to the
fact that the LO Adler function is the sum of the two diagrams of
\fig{AdlerLO}, where the first diagram only contributes
singularities at $u=-1$, $u=1$ and $u=2$ to the Adler function. Here
we recall \eq{LOdiagram1}, and note that the singularity at $u=0$ from that expression vanishes
when we take the derivative w.r.t $Q^2$ to obtain the Adler function.

Combining all results, we arrive at the first main result of this
paper: the complete asymptotic expansion of the perturbative
coefficients of the Adler function at LO in the flavour expansion,
\begin{equation}
  d_k^{(0,0)}
  = \frac{\Ga(k+2)}{(-1)^k}
  \sum_{\ell=1}^\infty\frac{1}{\ell^{k+2}}\bigg(
  s_k^{(\ell)}+\frac{t_k^{(\ell)}}{k+1}
  \bigg).
\end{equation}
The sum over $\ell$ converges, and for each term within the sum we only have
finitely many coefficients rather than asymptotic expansions, so this is now
not a large order formula but an exact form for the
coefficients (hence the equal sign). At the next order in $1/N_f$,  
in \sec{adlerNLO}, we will mostly lack exact expressions, but resurgence
will still provide us with
very precise large order relations, that involve non-perturbative sectors.

\begin{figure}
    \centering
    \includegraphics[width=.6\linewidth]{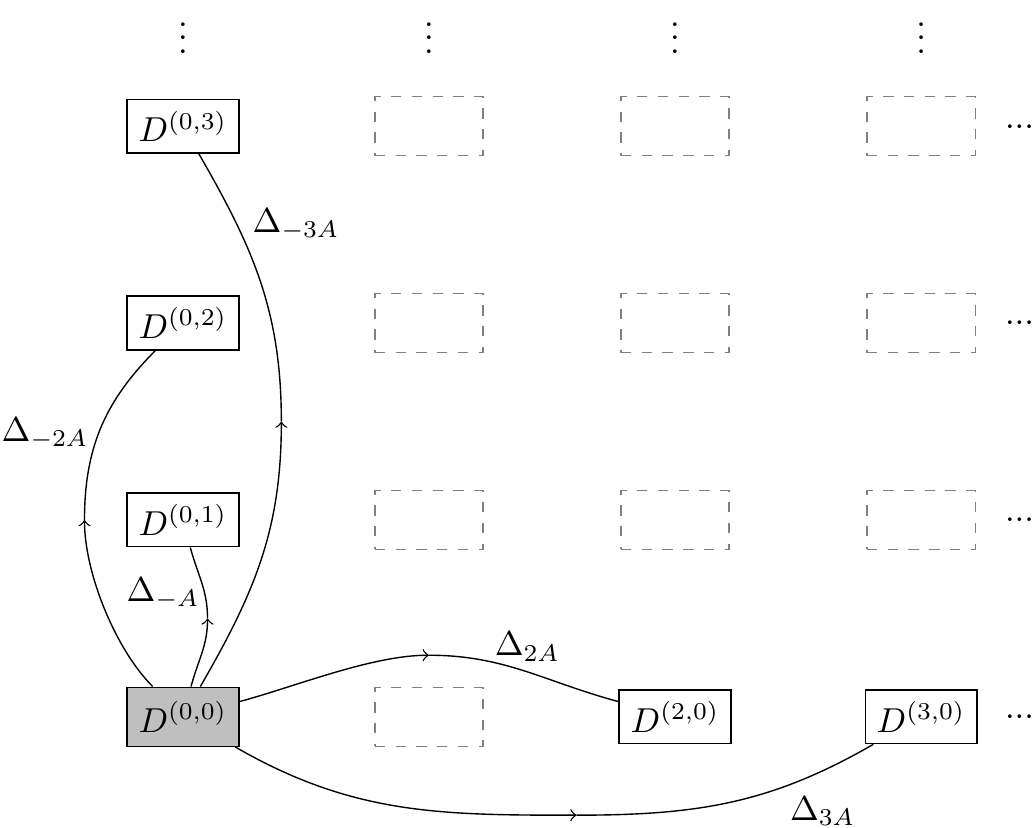}
    \caption{Alien lattice for the LO Adler function. The only asymptotic sector is the $(0,0)$-sector, which we denoted by a filled box to distinquish it from the $(\ell,0)$ and $(0,\ell)$ sectors which are not asymptotic. The dashed boxes are sectors that completely vanish.}
    \label{fig:alienLatticeLOAdler}
\end{figure}

We have seen now that for the Adler function at $\ord{1/N_f}$ the non-perturbative
sectors consist of only one or two coefficients, i.e. the $1/k$ expansions
are not asymptotic. This means that these sectors do not lead to further resurgence.  Thus we have now carried out this resurgence analysis
to its natural end: there is no resurgence of non-perturbative sectors at any other
sector than the $(0,0)$-sector. In other words, the picture of
the alien lattice in this case simplifies to that of \fig{alienLatticeLOAdler}.

This structure implies that the ${(\ell,0)}$-sector with $\ell>1$ is not reached by repeatedly applying $\De_A$, i.e.\ for $\ell>1$, $\De_{A}^\ell D_\LO^{(0,0)}=0$. Instead one needs non-vanishing alien derivatives for larger steps forward: $\De_{\ell A} D_\LO^{(0,0)}\sim D_\LO^{(\ell,0)}$. Similarly, we have for $\ell>1$ that $\De_{-A}^\ell D_\LO^{(0,0)}=0$, but $\De_{-\ell A} D_\LO^{(0,0)}\sim D_\LO^{(0,\ell)}$. We will comment more on the implications of this in the next subsection.

Finally, observe that at order $1/N_f$, the alien `lattice' in fact is hardly a lattice: all internal $(n,m)$ sectors with $n \neq 0$ and $m \neq 0$ vanish. In \sec{adlerNLO}, we will see that this is a result of the relative simplicity of the expressions at order $1/N_f$, and that at higher orders such internal sectors {\em will} appear.

As all non-perturbative sectors containing both non-zero powers of $\si_1$ and $\si_2$ vanish, we can write  the complete transseries \eq{5} in the following form:
\begin{equation}
D_{\LO}(\al,\si_1,\si_2) 
= D_{\LO}^{(0,0)}(\al)
+ D_\LO^{\text{IR}}(\al,\si_1)
+ D_\LO^{\text{UV}}(\al,\si_2),
\end{equation}
where we labeled the different pieces according to their QCD nature. That is,  $D_\LO^{\text{IR}}$ contains all IR renormalons in QCD, meaning that these are the UV renormalons in QED. Likewise, $D_\LO^{\text{UV}}$ contains all UV (IR) renormalons in QCD (QED).
For the IR sectors, we get 
\begin{align}
D_\LO^{\text{IR}}(\al,\si_1)
&= -\frac{4\pi\img\,C_F}{\beta_{0}}\left(\frac{\si_1}{S_{1,0}}\right)^2\,e^{\frac{2}{\beta_{0}\al}}\label{eq:DLOIR1}\\
&\hspace{.5cm}
- \frac{2\pi\img\,C_F}{\beta_{0}}\sum_{\ell=3}^\infty
\left(\frac{\si_1}{S_{1,0}}\right)^\ell\, e^{\frac{\ell}{\beta_{0}\al}} 
\bigg(
\frac{\frac83(-1)^\ell}{(\ell-1)(\ell-2)}\frac{-1}{\beta_{0}\al} 
+ \frac{\frac83(-1)^\ell(2\ell-3)}{(\ell-1)^2(\ell-2)^2}
\bigg)\,,\nn
\end{align}
and for the UV sectors
\begin{align}
D_\LO^{\text{UV}}(\al,\si_2)
&= -\frac{2\pi\img\,C_F}{\beta_{0}}\frac{\si_2}{S_{0,1}}\,e^{\frac{-1}{\beta_{0}\al}}
\bigg(\frac{4}{9}\frac{-1}{\beta_{0}\al} -\frac{10}{27}\bigg)\label{eq:DLOUV1}\\
&\hspace{.5cm}
-\frac{2\pi\img\,C_F}{\beta_{0}}\left(\frac{\si_2}{S_{0,1}}\right)^2\,e^{\frac{-2}{\beta_{0}\al}}
\bigg(-\frac{2}{9}\frac{-1}{\beta_{0}\al} +\frac{7}{54}\bigg)\nn\\
&\hspace{.5cm}
- \frac{2\pi\img\,C_F}{\beta_{0}}\sum_{\ell=3}^\infty
\left(\frac{\si_2}{S_{0,1}}\right)^\ell\, e^{\frac{-\ell}{\beta_{0}\al}} 
\bigg(
\frac{-\frac83(-1)^\ell}{(\ell+1)(\ell+2)}\frac{-1}{\beta_{0}\al} 
+ \frac{\frac83(-1)^\ell(2\ell+3)}{(\ell+1)^2(\ell+2)^2}
\bigg)\,.\nn
\end{align}
In these expressions, we switched back to the original real coupling constant $\al$ -- recall the discussion below \eq{AdlerLOPerturbativeSector1}, where we switched to the variable $\hat{\al} = - \beta_0\al$ -- and reinstated all factors of $\beta_0=-(11-2N_f/3)/(4\pi)$, again expressed in QCD variables. We observe the overall factor $1/\beta_0$, and that the coupling constant $\al$ always comes with a factor $\beta_0$. Recalling \eq{flavourExpansionQCD}, this was to be expected as the flavour expansion is an expansion in $1/\beta_0$. Furthermore,  in \eqs{DLOIR1}{DLOUV1} we reinstated factors of $2\pi\img$ and the Stokes constants $S_{1,0}$ and $S_{0,1}$.
We will discuss the role of the transseries parameters $\si_1$ and $\si_2$ in the next subsection.

The sums over the sector number $\ell$ converge and can be carried out explicitly. 
We define
\begin{equation}
x_1=\frac{\si_1}{S_{1,0}}\,e^{\frac{1}{\beta_{0}\al}}
\qquad\text{and}\qquad
x_2=\frac{\si_2}{S_{0,1}}\,e^{-\frac{1}{\beta_{0}\al}}\,,
\end{equation}
to highlight the fact that the transseries parameters, the exponential factors  and the Stokes constants $S_{1,0}$ and $S_{0,1}$ always group together. 
Performing the sums in \eq{DLOIR1} yields
\begin{equation}\label{eq:DLOIR2}
D_\LO^{\text{IR}}(\al,\si_1)
= \frac{16\pi\img\,C_F}{3\beta_{0}}\left[
x_1^2\left(\frac{1}{\beta_{0}\al}-\frac{7}{4}\right)
-x_1(1+x_1)\left(\frac{\log(1+x_1)}{\beta_{0}\al}
+\text{Li}_2(-x_1)\right)
\right]\,,
\end{equation}
where Li$_2$ is the dilogarithm. For \eq{DLOUV1} we get
\begin{align}
D_\LO^{\text{UV}}(\al,\si_2)
&= \frac{2\pi\img\,C_F}{\beta_{0}}x_2
\bigg(\frac{4}{9}\frac{1}{\beta_{0}\al} +\frac{10}{27}\bigg)
+\frac{2\pi\img\,C_F}{\beta_{0}}x_2^2
\bigg(\frac{4}{9}\frac{1}{\beta_{0}\al} -\frac{8}{27}\bigg)\label{eq:DLOUV2}\\
&\hspace{1cm}
+ \frac{16\pi\img\,C_F}{3\beta_{0}}
\frac{1}{x_2}\left(1+\frac{1}{x_2}\right)
\left(\frac{\log^{(>3)}(1+x_2)}{\beta_{0}\al}
+\text{Li}_2^{(>3)}(-x_2)
\right)
,\nn
\end{align}
where we defined
\begin{equation}
\log^{(>n)}(1+x)
\equiv\log(1+x)+\sum_{j=1}^n\frac{(-x)^n}{n}
\quad\text{and}\quad
\text{Li}_2^{(>n)}(-x)
\equiv\text{Li}_2(-x)-\sum_{j=1}^n\frac{(-x)^n}{n^2}
\end{equation}
as the logarithm and polylogarithm with the first $n$ terms of their Taylor expansion removed. Although \eq{DLOUV2} appears to generate factors such as
$1/x_2$, these in fact cancel as the first three coefficients of the Taylor
expansion of the $\log$ and Li$_2$ are subtracted.

\subsection{Discussion}
\label{sec:discussionLO}
At this point, we have studied the full single bubble-chain part of the Adler function. Before moving on to the main goal of our investigation, the multi-bubble chain contributions, let us discuss some finer details of the structure that has appeared.

First of all, one may wonder how many free parameters the transseries structure of the Adler function really has. This is not a question that can be answered from the Borel plane structure alone; once one finds a singularity at a location $A$ it is natural to include a free parameter $\sigma$ in front of the $e^{-A/\alpha}$ transmonomial, but it is then not clear whether the $e^{-nA/\alpha}$ sectors should come with powers $\sigma^n$ or with new parameters. (See e.g.~\cite{Borinsky:2022knn} where, as we discussed below \eq{largeOrderFactorMultipParameter} the latter is the case.) 

Let us consider only the `forward' Stokes automorphism, coming from the IR singularities on the positive real $u$-axis. Since these singularities are evenly spaced, one may expect a single parameter to be sufficient to describe these sectors. However, the fact that e.g. $\De_A^2D_\LO^{(0,0)}=0$, but $\De_{2A}D_\LO^{(0,0)}\sim D_\LO^{(2,0)}$ shows that the transseries we find has a structure that differs from the simplest one-parameter transseries structure known from many toy models. There, the forward Stokes automorphism is simply $\sigma \to \sigma + S$ which implies in particular that $\De_{2A}D_\LO^{(0,0)} = 0$. This means that our forward Stokes automorphism at least has a bridge equation beyond that of such simple toy models. Unfortunately, since e.g.\ a differential equation for the Adler function as needed for the procedure in \sec{bridgeEquation} is not known, we cannot derive such a bridge equation which would allow us to read off the number of parameters.

Thus, our transseries may be a many-parameter one, but on physical
grounds we still expect the number of `true' parameters to be two. The
reason is that there are only two Stokes automorphisms that play a
role -- and as we shall see, this continues to be the case when we
include further bubble chains -- an IR one along the positive real
$u$-axis and a UV one along the negative real $u$-axis. A single parameter
could in principle suffice to describe the jump across every single Stokes line, and there is no additional physical effect that would lead to singularities on other rays in the $u$-plane, so we conjecture that two parameters is in fact enough. Interestingly, this is similar to the recent work of \cite{Gu:2022fss} where in a different setting a transseries was found with many degrees of freedom (parameterized by variables $\tau_k$ there) but with only a single transseries parameter for each Stokes line.

When it comes to the {\em values} of the transseries parameters, these are difficult to determine and require further physics input. This is beyond the scope of this paper, where we are mainly interested in the underlying structure. In principle, one would need to numerically sum non-perturbative sectors of the Adler function transseries and compare these to experimental data to read off values for $\sigma_1$ and $\sigma_2$. See~\cite{Maiezza:2021mry} for an effort in this direction for the IR Stokes automorphism.

Recall that in \eq{DLOUV2}, a variable $x_2$ appears which comes from the UV singularities and grows exponentially large when $\alpha \to 0$. This is a common feature for transseries that have instanton actions of opposite sign, $\pm A$. An expansion in such a large variable $x_2$ may not seem very physically meaningful. One possible solution to this is of course that in $x_2 = \sigma_2 e^{-\frac{1}{\beta_{0f}\alpha}}$ the parameter $\sigma_2$ vanishes, or is small compared to the transmonomial it multiplies so that $x_2$ is (very) small. Even if this were not the case, $x_2$ still has a definite value at any given energy scale, and so an expression like \eq{DLOUV2} still makes sense, even though its small $\alpha$ expansion may not. In fact, often a sum over different powers of $x_2$ is itself not asymptotic but {\em converges}, as in our examples above (see also \cite{Marino:2023epd} for recent progress on this topic) and therefore one does not always need such an expansion to make sense. Also note that part of the large order growth of the original perturbative series is still determined by $\si_2$-dependent sectors, so even when they have coefficient zero, these sectors do play a role.

As a final remark, let us address the fact that many of the non-perturbative sectors we have found only have a finite number of terms -- and therefore, these sectors in particular have no asymptotic growth of their own. This may seem to limit the use of resurgence techniques to only the perturbative sector, but we expect this to be an artifact of the simplicity of the single bubble chain approximation. In fact, in the next sections we shall see that many non-perturbative sectors do become true asymptotic expansions when further bubble chains are included. This is reminiscent of the `Cheshire cat resurgence' of \cite{Kozcaz:2016wvy} where asymptotic growth of sectors can disappear in particular parameter limits\footnote{We quote the authors of \cite{Kozcaz:2016wvy}: ``All of the characteristics of resurgence remains even when its role seems to vanish, much like the lingering grin of the Cheshire Cat''.}. This may also happen in $1/N_f$ expansions -- see for example the interesting approach of \cite{DiPietro:2021yxb} where it was found that in the Gross-Neveu model, only at finite $N_f$ the full resurgent structure and asymptotic growth of sectors becomes visible.

\section{Using Borel convolution integrals for resurgence}
\label{sec:convoInt}
We saw in the previous section that the flavour
expansion is a useful framework to isolate perturbative series
that show factorial growth due to individual diagrams, i.e.\
renormalons.  At higher orders in $1/N_f$, an important
ingredient of the flavour expansion is the convolution integral
\eq{flavourExpansionConvoInt}, which we repeat for convenience  
\begin{equation}\label{eq:convoIntrepeated}
\cB\bigg[\prod_{j=1}^{n_c} \al D_{\mu_j\nu_j}(k_j)\bigg](u)
= \frac{1}{(-\beta_{0f})^{n_c-1}} \int_0^u \bigg[\prod_{j=1}^{n_c}du_j\bigg] \de\Big(u-\sum_{j=1}^{n_c}u_j\Big) \prod_{j=1}^{n_c} \cB\big[\al D_{\mu_j\nu_j}(k_j)\big] (u_j).
\end{equation}
In section \ref{sec:adlerNLO}, we shall
apply this integral to go beyond the leading order in
the flavour expansion, but before doing so, we discuss in this
section what the effect is of the convolution integral on the
resurgence structure.  
In particular, we will see momentarily that one does not need to know the particular exact Borel transforms, but that the structure of the alien derivatives and their calculus can be used instead. Although the techniques are a straightforward application of the framework outlined in \sec{resurgence}, in physics applications this is, as far as we are aware, a novel way to study renormalon effects.

The results we present in this section have twofold use. First, they
set up the calculation of particular $\ord{1/N_f^{2}}$ diagrams that we
shall encounter in the next section. The results, however, are more general
and also apply to more complicated convolution integrals which one
would encounter at higher orders in $1/N_f$. We explain, by means of examples that are relevant for the next section, how the convolution integral
 `builds' resurgent functions. The results will also show
glimpses of the full resurgent structure that would emerge upon 
including all orders in $1/N_f$.

\subsection{Resurgence of the convolution integral}
Given two asymptotic formal power series $F^{(0)}(\al)$ and $G^{(0)}(\al)$, we can define a new asymptotic power series $\Psi^{(0)}(\al)$ by taking the product:
\begin{equation}
\Psi^{(0)}(\al)
\equiv F^{(0)}(\al) \, G^{(0)}(\al).
\end{equation}
As explained in more detail in \sec{resurgence}, and shown extensively
in \sec{adlerLO}, the resurgence properties of $\Psi^{(0)}$ can
be recovered from the singularities of the Borel transform.  Here, the
convolution integral appears as the Borel transform
of $\Psi^{(0)}$ is computed as
\begin{equation}\label{eq:convo}
\cB[\Psi^{(0)}](u) 
= \int_0^u du_1\, \cB[F^{(0)}](u)\cB[G^{(0)}](u-u_1)\,,
\end{equation}
and the resurgence structure can be found using the relation
\eq{borelSingularities} between the different non-perturbative sectors $\cB[\Psi^{(n)}](u)$.  However, except for simple examples,
performing the convolution integral might be a difficult task. 
We can study instead the resurgence properties of $\Psi^{(0)}$ by using the
alien derivatives.  Acting with an arbitrary alien
derivative $\De_\w$ yields
\begin{align}\label{eq:alienDerivativeFG}
\De_\w\Psi^{(0)}
= \De_\w\left(F^{(0)}(\al)\right)G^{(0)}(\al)
+ F^{(0)}(\al)\De_\w\left( G^{(0)}(\al)\right),
\end{align}
since the alien derivative satisfies the Leibniz rule.  Thus
the calculation of $\De_\w\Psi^{(0)}$ has shifted to the
calculation of $\De_\w F^{(0)}$ and $\De_\w G^{(0)}$.  Consequently, if
 the resurgence structure of $F^{(0)}$ and $G^{(0)}$ is
known, one can  compute the resurgence structure of
$\Psi^{(0)}$.  As alluded to above, we will see in \sec{adlerNLO} that, with one
exception, the convolution integral \eq{convo} cannot be
computed exactly. It will be easier to obtain the
resurgence properties of $F^{(0)}$ and $G^{(0)}$ instead of that of
$\Psi^{(0)}$. Therefore, to prepare our discussion in
\sec{adlerNLO}, and to show how \eq{alienDerivativeFG} can be employed
in practice, we discuss in this section a few relevant examples.
In these examples the convolution integral can be
computed exactly and therefore these computations act as a check on the method
of alien derivatives.

\subsection{Convolution of pure factorial growth}
Let us begin with the simplest possible model. Consider an asymptotic formal power series where the perturbative coefficients show pure factorial growth:
\begin{equation}
F^{(0)}(\al) = \sum_{n=0}^\infty\Ga(n+1)\al^{n+1}
\qquad\implies\qquad
\cB[F^{(0)}](u)=\frac{1}{1-u}.
\end{equation}
As the complete asymptotics of $F^{(0)}$ is determined by the pole of
$\cB[F^{(0)}](u)$ in the Borel plane, we know that in the full
transseries there is just one possible non-perturbative sector,
$F^{(1)}$, which consists of only a single term
$S_1f_0^{(1)}(\al) = 2\pi\img$. As we will encounter many different Stokes
constants for different resurgent functions, we adopt the notation
in which we put the function name as a superscript on the Stokes
constants, e.g.\ $S_1^F$.  

Although the resurgence of $F^{(0)}$ is
relatively simple, i.e. 
\begin{equation}\label{eq:delta1F}
\De_1 F^{(0)} = 2\pi\img\,,
\end{equation}
with all other alien derivatives vanishing, we show in this section that it can be used as a
building block to build more complicated resurgent functions.  By
taking powers of $F^{(0)}$, we will show schematically how such a
resurgence structure builds up.

Consider the formal power series
\begin{align}
\Psi^{(0)}(\al)
= \Big(F^{(0)}(\al)\Big)^2,
\end{align}
where its Borel transform is computed as
\begin{equation}\label{eq:convoIntFsq}
\cB\Big[\Psi^{(0)}\Big](u) 
= \int_0^udu_1\ \frac{1}{1-u_1}\frac{1}{1-u+u_1} 
= -2\frac{\log(1-u)}{2-u}.
\end{equation}
Having the exact Borel transform, we can read off the resurgence
structure using \eq{borelSingularities}. Since we know that
\begin{equation}
\cB\Big[\Psi^{(0)}\Big](u) 
= -S_1^\Psi\cB\Big[\Psi^{(1)}\Big](u-1)\frac{\log(1-u)}{2\pi\img}\,,
\end{equation}
we can read off the result
\begin{equation}
S_1^\Psi\cB\Big[\Psi^{(1)}\Big](u-1)
=\frac{4\pi\img}{1-(u-1)}.
\end{equation}
Transforming back to the $\al$-plane, this corresponds to the formal power series 
\begin{equation}\label{eq:convoExample1Psi1}
S_1^\Psi\Psi^{(1)}(\al)=4\pi\img\sum_{n=0}^\infty\Ga(n+1)\al^{n+1}\,.
\end{equation}
We see that in this `squared model', the
coefficients of the leading non-perturbative sector $\Psi^{(1)}$ are
no longer a single term but now show pure factorial growth themselves.

Though we were able to perform the convolution integral exactly
and therefore immediately read off the resurgence structure, it is
instructive to obtain the same result using alien derivatives.  On the
one hand, the bridge equation, \eq{resurgenceEqs}, tells us that
$\De_1\Psi^{(0)}=S_1^\Psi\Psi^{(1)}$. On the other hand, we have by
direct calculation
\begin{equation}
\De_1\Psi^{(0)}
= 2F^{(0)}\De_1F^{(0)} 
= 2\sum_{n=0}^\infty 2\pi\img\,\Ga(n+1)\al^{n+1},
\end{equation}
where we used \eq{delta1F}. We observe the same result as obtained from the exact Borel transform in \eq{convoExample1Psi1}. Likewise, for the second non-perturbative sector $\Psi^{(2)}$ we get 
\begin{equation}\label{eq:delta1sq}
2\Big(S_1^\Psi\Big)^2\Psi^{(2)}
= \De_1^2\Psi^{(0)}
= 2\De_1F^{(0)}\De_1F^{(0)} 
= 2 (2\pi\img)^2\,,
\end{equation}
which is a sector with a single coefficient.  Notice that this can also be related to the expansion of \eq{convoIntFsq} around $u=2$ which reads
\begin{equation}
\cB[\Psi^{(0)}(u)\Big|_{u=2} = \pm 2\frac{\pi\img}{2-u}+...\,,
\end{equation}
where the ellipsis denotes regular terms. Up to a sign ambiguity this agrees with \eq{delta1sq}. This ambiguity originates in the ambiguous expansion of the logarithm $\log(1-u)$ in \eq{convoIntFsq} around $u=2$. We extensively come back to this point in \sec{adlerNLOsecondsectors}, where we shall see how to resolve such ambiguities.

\begin{figure}
\centering
\includegraphics[width=.5\linewidth]{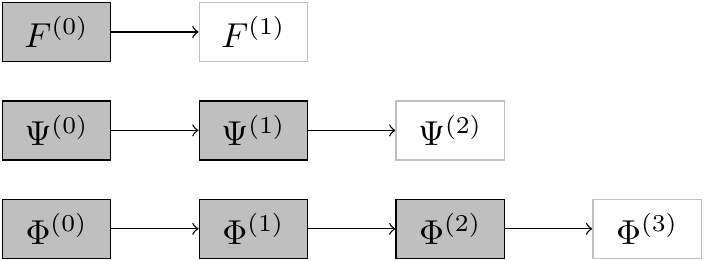}
\caption{Alien chains for the convolution of an asymptotic power series where the perturbative coefficients show pure factorial growth. The sectors with a filled box are true asymptotic sectors, whereas the white boxes are not and consist of a finite number of terms.}
\label{fig:AlienLatticeConvoIntExample1}
\end{figure}

This simple example easily generalizes to higher powers of $F^{(0)}(\al)$. 
For example, consider 
\begin{equation}
\Phi^{(0)}(\al)
\equiv \Big(F^{(0)}(\al)\Big)^3.
\end{equation}
The convolution integral to get the Borel transform
$\cB[F^{(0)}(\al)^3]$ can still be done exactly, but we stay in the
$\al$-plane and follow the second method we just used for
$\big(F^{(0)}(\al)\big)^2$. Acting once with $\De_1$ yields
\begin{equation}
S_1^\Phi \Phi^{(1)}
= \De_1\Phi^{(0)}
= 3\Big(F^{(0)}\Big)^2\De_1F^{(0)} 
= 6\pi\img \sum_{n=0}^\infty \al^{n+2}\sum_{h=0}^n\Ga(n-h+1)\Ga(h+1).
\end{equation}
Acting twice with $\De_1$ yields
\begin{equation}
2\left(S_1^\Phi\right)^2\Phi^{(2)}
= \De_1^2\Phi^{(0)}
= 6F^{(0)}\Big(\De_1F^{(0)}\Big)^2
= 6(2\pi\img)^2 \sum_{n=0}^\infty \Ga(n+1)\al^{n+1}.
\end{equation}
Finally we have
\begin{equation}
6\left(S_1^\Phi\right)^3\Phi^{(3)}
= \De_1^3\Phi^{(0)}
= 6\Big(\De_1F^{(0)}\Big)^3
= 6(2\pi\img)^3.
\end{equation}
In \fig{AlienLatticeConvoIntExample1}, we show the alien chain built
up with more sectors,  more of which are now asymptotic.

\subsection{Convolution of power series with a double pole}
\label{sec:convoDoublePole}
In the next section we shall consider diagrams at order $1/N_f^2$ in
the flavour expansion We will encounter two generalizations of the
above convolution integral of power series with pure factorial growth.
We introduce and discuss these already in this section. The first generalization is the
case that the large order relation of the formal power series that we
convolute with itself is determined by a double pole, treated in this
subsection.  In the next subsection we address the convolution of a power
series with (infinitely many) evenly separated poles in the Borel
plane.

Consider a power series $F^{(0)}(\al)$ where the large order growth of the coefficients is determined by a double pole in the Borel plane at $u=1$, e.g.\
\begin{equation}
F^{(0)}(\al) 
= \sum_{n=0}^\infty\Ga(n+2)\left(a+\frac{b}{n+1}\right)\al^{n+1}
\quad\implies\quad
\cB[F^{(0)}](u)=\frac{a}{(1-u)^2} + \frac{b}{1-u}.
\end{equation}
Notice that we keep the
option open for subleading growth coming from a $(1-u)^{-1}$ term. We use $a$ and $b$ as a compact notation for
$f^{(1)}_0$ and $f^{(1)}_1$ respectively.  We are again interested in
the resurgence properties of
\begin{equation}
\Psi^{(0)}(\al)\equiv \left(F^{(0)}(\al)\right)^2.
\end{equation}
The Borel transform of this power series is given by the convolution integral
\begin{align}
\cB[\big(F^{(0)}\big)^2](u) 
&= \int_0^udu_1
\left(
\frac{a}{(1-u_1)^2}+\frac{b}{1-u_1}
\right)
\left(
\frac{a}{(1-u+u_1)^2}+\frac{b}{1-u+u_1}
\right)\nn\\
&= 
2\frac{a^2+ab}{1-u}
-S_1^\Psi\cB\Big[\Psi^{(1)}\Big](u-1)\frac{\log(1-u)}{2\pi\img}\nn\\
&\hspace{1cm}
-2\left(\frac{2a^2}{(2-u)^2} + \frac{a^2+2ab}{2-u}\right) \,,\label{eq:convoIntf2LOandNLO}
\end{align}
where in the second equality we emphasized the resurgence of the
non-perturbative sector $\Psi^{(1)}$, which has the explicit form
\begin{align}
S_1^\Psi\cB\Big[\Psi^{(1)}\Big](u-1)
=4\pi\img\left(\frac{2a^2}{(2-u)^3}+\frac{2ab}{(2-u)^2}+\frac{b^2}{2-u}\right).\label{eq:cubicpole}
\end{align}
Notice that, studying the singularities of these expressions around $u=2$, it might look like the second non-perturbative sector $\Psi^{(2)}$ gets contributions from both the $\Psi^{(1)}$ sector and the last line in \eq{convoIntf2LOandNLO}. However, as
\begin{equation}
\log(1-u)\Big|_{u=2} 
= \pm \pi\img -(2-u)-\frac12(2-u)^2 + ...\,,
\end{equation}
the last line cancels against the (real part of the) expansion of the logarithm and $\cB[\Psi^{(1)}]$ around $u=2$. 

The inverse Borel transforms of \eq{cubicpole}, together with the residue of the
simple pole at $u=1$ in \eq{convoIntf2LOandNLO}, yield the
coefficients of the first non-perturbative sector
\begin{equation}
S_1^\Psi\Psi^{(1)}(\al)
= 4\pi\img(a^2+ab)+4\pi\img\sum_{n=0}^\infty 2\Ga(n+3)\left(
a^2 + \frac{2ab}{n+2} + \frac{b^2}{(n+2)(n+1)}
\right)\al^{n+1}.\label{eq:convoIntf2LOandNLO1instanton}
\end{equation}
where, in the sum, we factored out the leading order growth $\Ga(n+3)$ coming from the fact that \eq{cubicpole} has a cubic pole in the Borel plane (recall \eq{valueBetaCoefficientHigherOrderPoles}). 

Instead of performing the convolution integral as in
\eq{convoIntf2LOandNLO}, we can again use alien derivatives. Since 
\begin{equation}
\De_1F^{(0)}=2\pi\img\left(\frac{a}{\al}+b\right)\,,
\end{equation}
we find
\begin{align}
S_1^\Psi\Psi^{(1)}
&= 2F^{(0)}\De_1F^{(0)}
= 4\pi\img\left(\frac{a}{\al}+b\right)
\sum_{n=0}^\infty\Ga(n+2)\left(a+\frac{b}{n+1}\right)\al^{n+1}\nn\\
&= 4\pi\img(a^2+ab)+4\pi\img\sum_{n=0}^\infty 2\Ga(n+3)\left(
a^2 + \frac{2ab}{n+2} + \frac{b^2}{(n+2)(n+1)}
\right)\al^{n+1}\label{eq:readOff1instantonLOandNLO},
\end{align}
which is indeed the same as in \eq{convoIntf2LOandNLO1instanton}.

For this specific example there is another way to think about \eq{readOff1instantonLOandNLO}. 
Instead of identifying the coefficients $f_n^{(0)}$, we rewrite \eq{readOff1instantonLOandNLO} as
\begin{align}\label{eq:doublePoleConvoTrick}
S_1^\Psi\Psi^{(1)}
&= 4\pi\img\left(\frac{a}{\al}+b\right)
\sum_{n=0}^\infty F^{(0)}_n\al^{n+1}\nn\\
&= 4\pi\img\,a\,F^{(0)}_0 
+ 4\pi\img\,a\sum_{n=0}^\infty f^{(0)}_{n+1}\al^{n+1}
+ 4\pi\img\,b\sum_{n=0}^\infty f^{(0)}_n\al^{n+1}
\end{align}
Next, consider a generic series and its Borel transform
\begin{equation}
G(\al) = \sum_{n=0}^\infty g_n\al^{n+1}
\qquad\implies\qquad
\cB[G](u)=\sum_{n=0}^\infty\frac{g_n}{\Ga(n+1)}u^n.
\end{equation}
The derivative of the Borel transform yields
\begin{equation}
\cB[G]'(u)
=\sum_{n=0}^\infty\frac{g_{n+1}}{\Ga(n+1)}u^{n}
\qquad\implies\qquad
H(\al) = \sum_{n=0}^\infty g_{n+1}\al^{n+1},
\end{equation}
where in the last step we applied an inverse Borel transform. In other
words, the newly constructed  series $H(\al)$, with
coefficients those of  $G(\al)$ but shifted: $g_n\to g_{n+1}$,
corresponds to taking the derivative in the Borel plane.  Applied to
\eq{doublePoleConvoTrick}, this implies
\begin{equation}\label{eq:doublePoleConvoTrick2}
S_1^\Psi\cB[\Psi^{(1)}](u)
=4\pi\img\,a\,\cB[F^{(0)}]'(u)+4\pi\img\,b\,\cB[F^{(0)}](u).
\end{equation}
We shall employ this observation in \sec{adlerNLO}.

\subsection{Convolution with equidistant singularities}
\label{sec:convoEquidistant}
A final case we need is where we take the
convolution of perturbative series $F^{(0)}$ and $G^{(0)}$, where the
large order behaviour of their coefficients, $f_n^{(0)}$ and
$g_n^{(0)}$, come from singularities at $u=1,2,3,...$ in the Borel
plane.  Furthermore, we assume that we know both the resurgence
properties of $F^{(0)}$ and $G^{(0)}$, i.e. we know the action of the
alien derivative $\De_1$: 
\begin{equation}
\De_1F^{(0)}=S_1^FF^{(1)}\,,
\qquad
\De_1^2F^{(0)}=2(S_1^F)^2F^{(2)}\,,
\qquad 
\De_1^3F^{(0)}=6(S_1^F)^3F^{(3)}\,,
\end{equation}
etc., and similar for $G^{(0)}$, but now with the Stokes constant $S_1^F$ replaced by $S_1^G$. This is the standard structure of alien derivatives if one assumes that $F^{(0)}$ and $G^{(0)}$ are one parameter transseries for which one can derive a bridge equation as we did in \sec{bridgeEquation}. 
However, from the results obtained for the Adler function at order $1/N_f$ in the flavour expansion (recall our discussion in \sec{discussionLO}), we know that the Adler function does not necessarily has such a bridge equation. Therefore, we also allow for non-vanishing actions of $\De_2$, $\De_3$, $\De_4$ etc. 
In fact, when we apply the machinery of this section to the Adler function at order $1/N_f^2$, we indeed see that such `multiple steps forward' alien derivatives indeed occur. 

Using the Leibniz rule
repeatedly, we may now also obtain the resurgence structure of the product
\begin{equation}
\Psi^{(0)}
\equiv F^{(0)}G^{(0)}\,.
\end{equation}
In particular, to obtain the first non-perturbative sector $\Psi^{(1)}$, we can act with $\De_1$. This yields
\begin{align}
S_1^\Psi\Psi^{(1)}
&= \left(\De_1F^{(0)}\right)G^{(0)}
+ F^{(0)}\left(\De_1 G^{(0)}\right)
\,.
\end{align}
In order to obtain the sector $\Psi^{(2)}$, we can act with $\De_1^2$, for which $\Psi^{(2)}$ gets contributions from
\begin{align}
\frac12\De_1^2 \left(F^{(0)}G^{(0)}\right)
&= \frac12\left(\De_1^2F^{(0)}\right)G^{(0)}
+ \left(\De_1F^{(0)}\right)\left(\De_1G^{(0)}\right)
+ \frac12F^{(0)}\left(\De_1^2 G^{(0)}\right)\,,
\end{align}
and potentially also with a nonzero $\De_2$, for which the contributions come from
\begin{align}
\De_2\left(F^{(0)}G^{(0)}\right)
= \left(\De_2F^{(0)}\right)G^{(0)}+ F^{(0)}\left(\De_2 G^{(0)}\right)\,.
\end{align}
Likewise, an arbitrary sector $\Psi^{(n)}$ could get contributions from products of alien derivatives of the form $\De_1^n$, $\De_1^{n-2}\De_2$, $\De_1^{n-3}\De_3$,~...,~$\De_n$.

\subsection{Prefactor singularities}
\label{sec:prefactor}
A final subtlety that we need to address is one that we will encounter at several points in our computations. The Borel transforms in the convolution integral, \eq{convoIntrepeated}, may contain singular factors -- in practice: poles -- that only depend on the overall Borel plane variable $u$, not on the integration variables $u_i$. Such singular factors can therefore be taken outside the integral; see \eq{convolution} and the expressions below it for examples that we will encounter.

In the case of a transseries with non-perturbative exponentials $A_1 = -A_2$, as we have here, these overall singularities can be somewhat difficult to interpret. Clearly, singular prefactors play a role in the singularity structure of the final Borel transform, and therefore they will determine some of the structure of the full transseries solution that we are after. However, since the singularities are not obtained by acting with a specific alien derivative on one of the factors in the integrand, it is not always immediately clear to which transseries sectors the expansions around them belong. 

For example, an expansion in the Borel plane around $u=A_1$ could describe a $(1,0)$ sector, but also other $(1+n,n)$ sectors. When such an expansion comes from actions of alien derivatives on factors in the convolution integrand, one can simply read off which alien derivatives play a role and therefore which sectors appear. When an expansion comes from a prefactor singularity, this is not the case. As a result, in these cases one needs other arguments (e.g.\ comparing different expansions that involve the same higher nonperturbative sectors) to determine the precise transseries structure.

Some of our results on the transseries structure will therefore be conjectural, depending on such additional arguments, though in many cases we can also fully pin down the structure. We will even encounter situations where the extra singularities are a virtue rather than a nuissance -- occasionally, they help us to read off transseries coefficients that would otherwise have remained hidden in a regular expansion, but that now become part of the singular structure.

\bigskip
\noindent
To summarize what we learned in this section, it is clear that the resurgence structure of the convolution integral, \eq{convoIntrepeated}, can be obtained in two ways: either directly via the singularity structure of the Borel transform, or by applying alien derivatives acting on the constituents of the convolution integral. The examples discussed in this section were relatively simple and we were able to compute the convolution integral exactly. However, in general, harder problems it is difficult (if not impossible), to compute the convolution integral exactly. In fact, as we see in the next section, this is indeed the case for the Adler function at order $1/N_f^2$ in the flavour expansion, so that there we have to turn to the second method using alien derivatives.
Let us emphasize that this method is quite general, and certainly not only applicable to the resurgence properties of the Adler function. It may further open the door to study renormalons in physics in general.

\section{Adler function with two bubble chains (\texorpdfstring{$\ord{1/N_f^2}$)}{ }}
\label{sec:adlerNLO}
In \sec{adlerLO} we discussed how the minimal two-parameter
transseries Ansatz for the Adler function could be obtained using the
resurgence relations described in \sec{resurgence}. From a resurgence
point of view, the structure of this transseries at order $1/N_f$ was surprisingly
simple as the only asymptotic sector is the perturbative
$(0,0)$-sector. In particular, we saw that most of the
non-perturbative sectors are in fact vanishing (recall
\fig{alienLatticeLOAdler}), and those sectors that were not vanishing
only consist of one or two terms. As a result, we were even able to
sum the whole non-perturbative part of the transseries, leading to a
closed analytic form, \eqs{DLOIR2}{DLOUV2}.

From our discussion in \sec{convoInt}, we do not expect that this
relatively simple transseries structure is still present at higher orders in
the flavour expansion, i.e. after adding more bubble chains. (See e.g. \cite{Beneke:1995qq,Vainshtein:1994ff,Peris:1997dq,Dondi:2020qfj} for earlier work including two or more bubble chains.) Recalling \fig{AlienLatticeConvoIntExample1},
we observed that the convolution integral leads to asymptotic
non-perturbative sectors, and found that the alien chain or lattice gradually builds up
by taking more and more convolution integrals. We therefore expect
that already at the next order in the flavour expansion, i.e.\ at order
$1/N_f^2$, the transseries for the Adler function will contain
asymptotic non-perturbative sectors.  

\begin{figure}
\centering
\begin{subfigure}{3.5cm}
    \includegraphics[width=\textwidth]{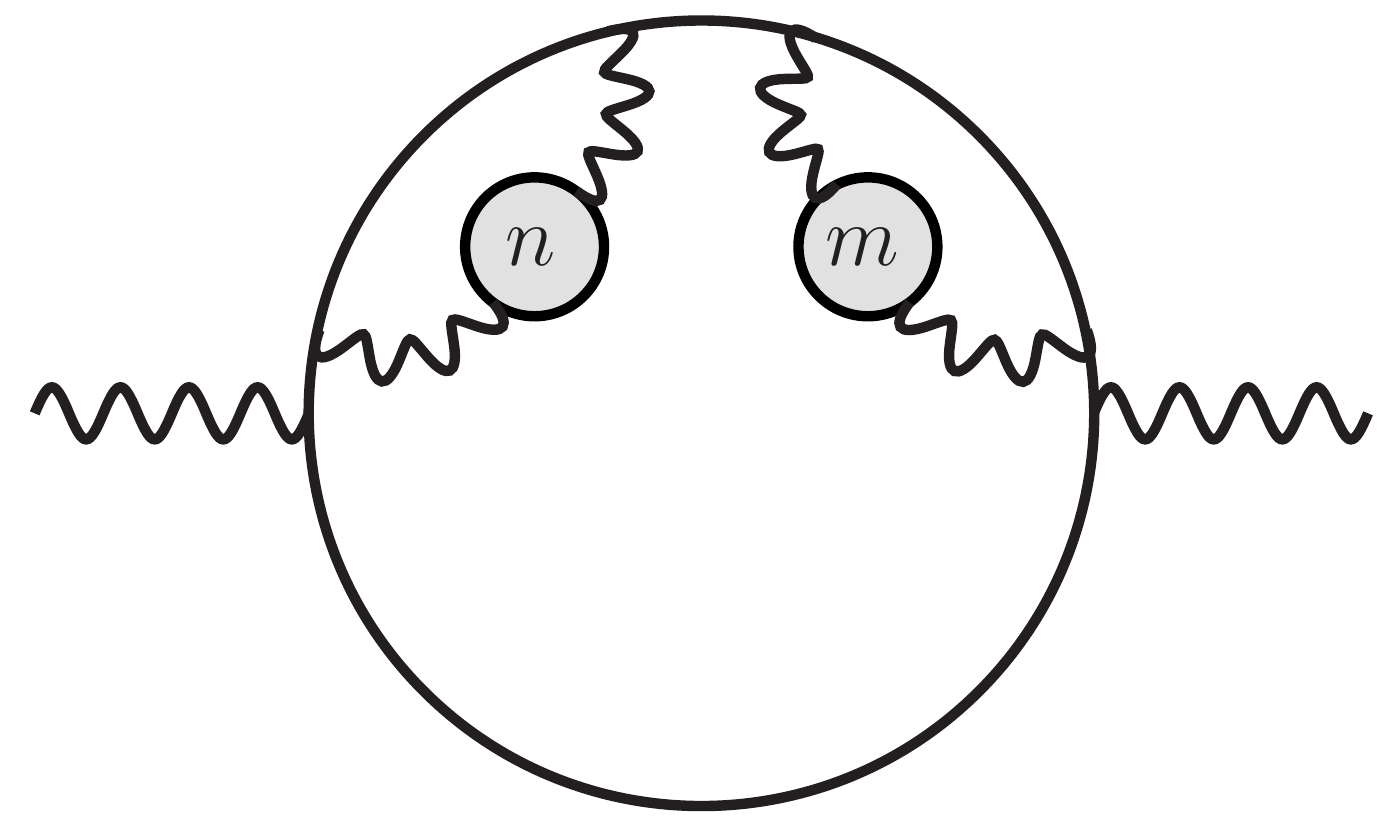}
    \caption{}
    \label{fig:adlerNLO1}
\end{subfigure}
\begin{subfigure}{3.5cm}
    \includegraphics[width=\textwidth]{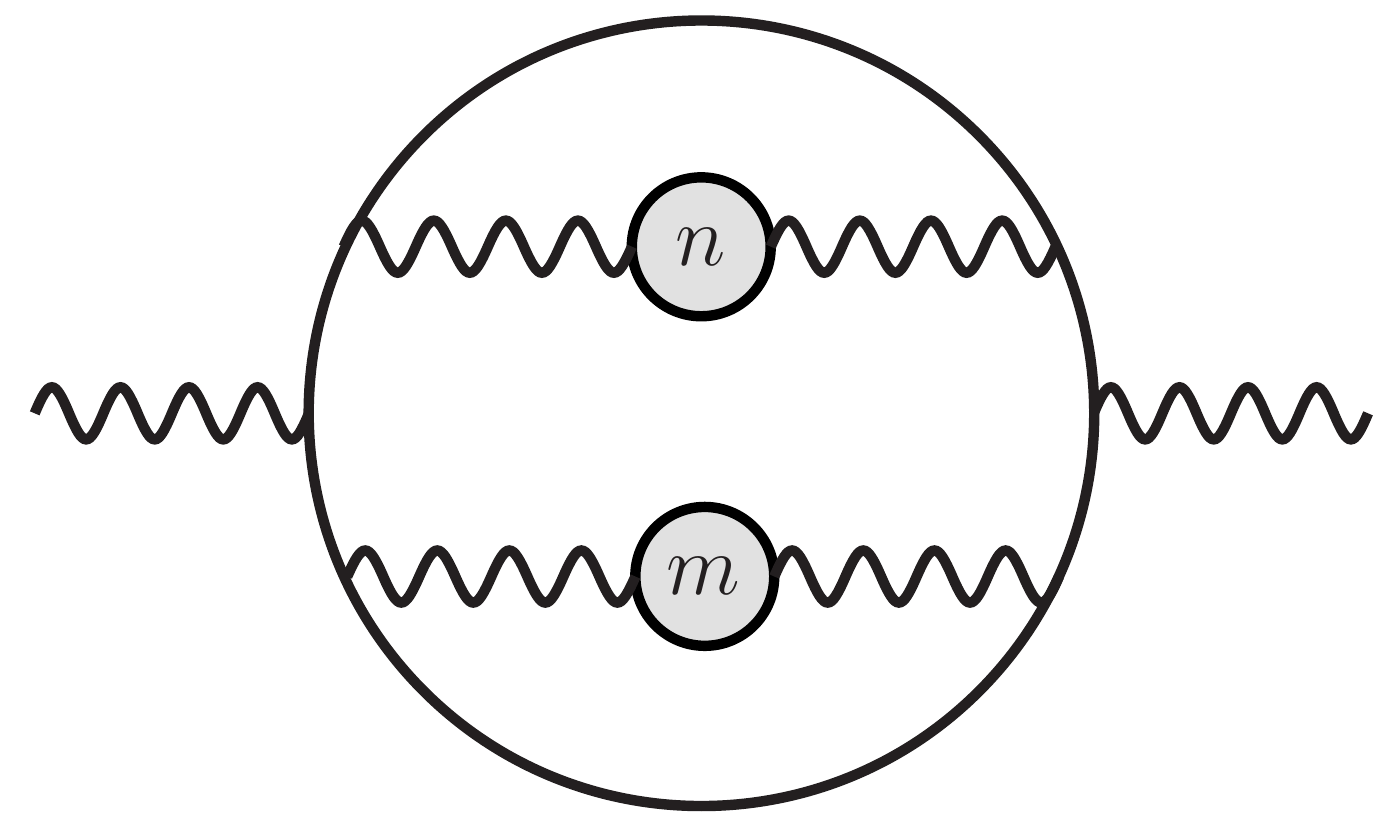}
    \caption{}
    \label{fig:adlerNLO2}
\end{subfigure}
\begin{subfigure}{3.5cm}
    \includegraphics[width=\textwidth]{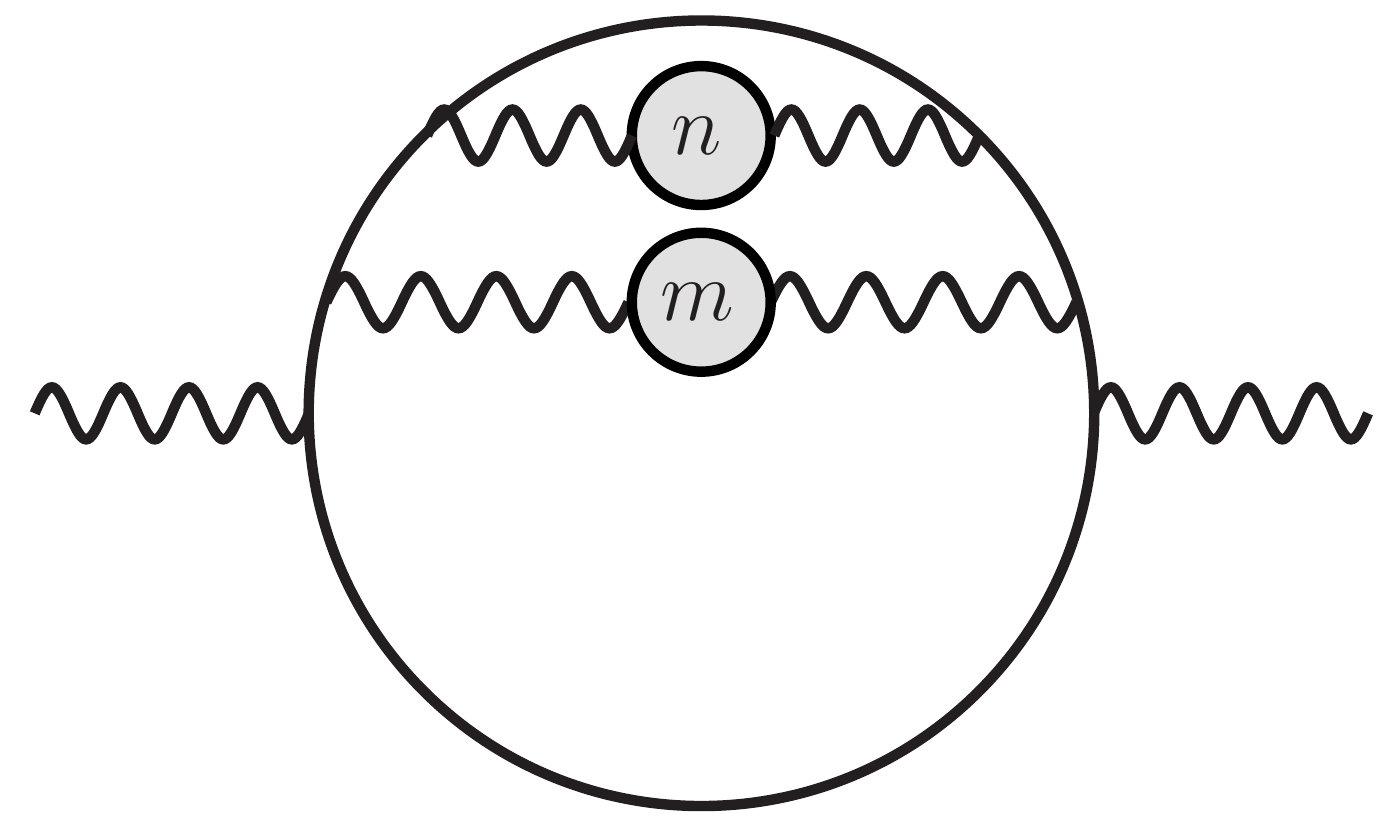}
    \caption{}
    \label{fig:adlerNLO3}
\end{subfigure}
\begin{subfigure}{3.5cm}
    \includegraphics[width=\textwidth]{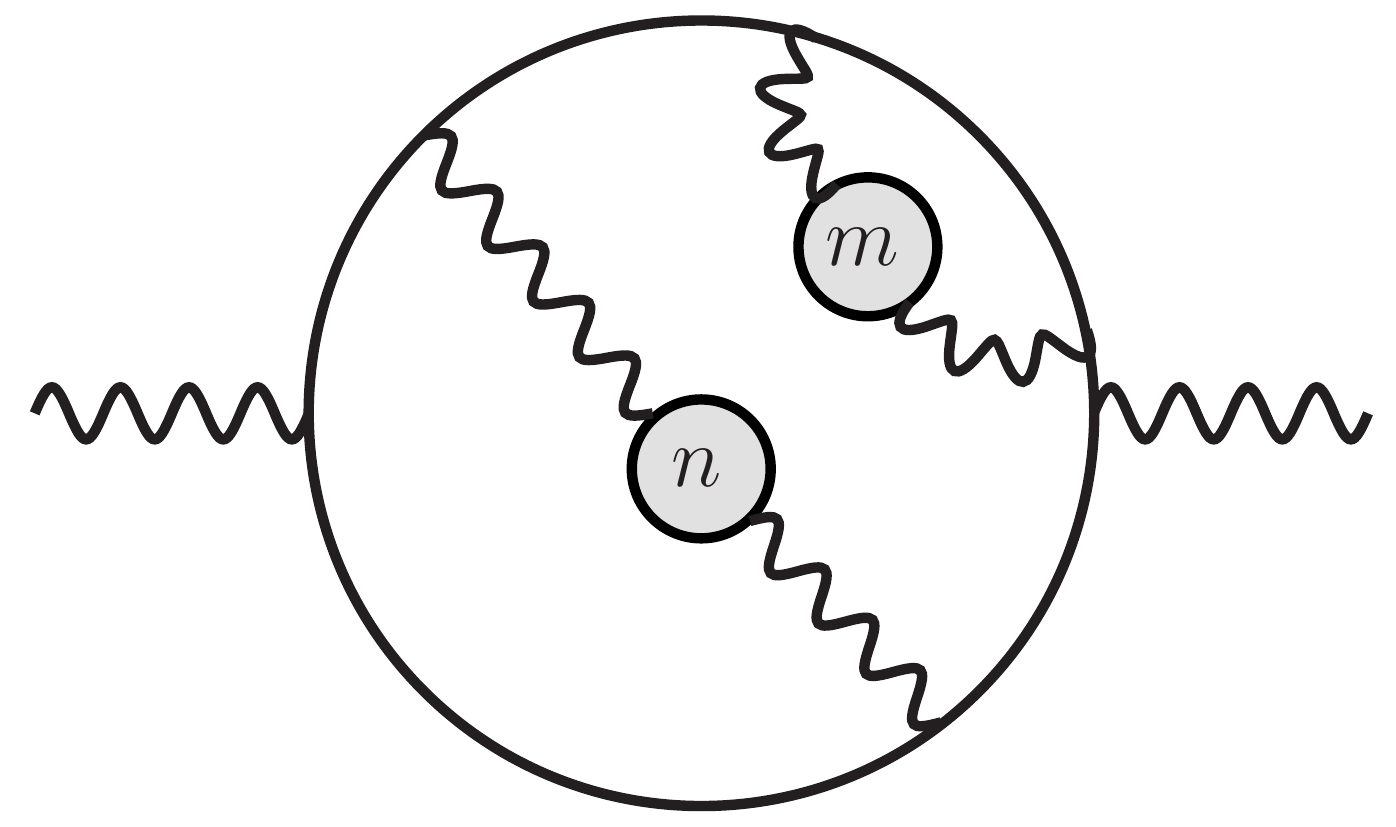}
    \caption{}
    \label{fig:adlerNLO4}
\end{subfigure}
\caption{Subset of diagrams at $\ord{1/N_f^2}$ that are computed in this section.}
\label{fig:AdlerNLO}
\end{figure}

In order to test these
expectations, and to get a first view of the resurgence structure
at higher orders in the flavour expansion, we discuss in this section
the Adler function at order $1/N_f^2$ (NLO). Unfortunately, a
complete calculation of the Adler function at this order is not
possible yet, because for some diagrams the master integrals for the Feynman
integrals appearing at this order are not yet known.  Therefore, we
focus in this section on  the set of planar diagrams shown in
\fig{AdlerNLO}, which we are able to compute. (These diagrams are leading in the SU$(3)$ color structure.) A brief review of the complete set of diagrams at
order $1/N_f^2$ can be found in \app{allDiagrams}.  

Note that the subset of diagrams of
\fig{AdlerNLO} is not gauge invariant, and our resurgence analysis in this
section is only on a diagram by diagram basis. Unless very specific cancellations occur between diagrams, however, one may expect most of the resurgence features that appear in individual diagrams to also appear in the sum of the full set of diagrams -- as indeed occurred at order $1/N_f$ -- and it is those resurgence features that we are after. In particular, we discuss the resurgence of the complete first two non-perturbative sectors, i.e.\ the $(1,0)$, $(0,1)$, $(2,0)$ and $(0,2)$ sectors. We show how these results can be extracted numerically from the perturbative coefficients only. To supplement these results, we use the techniques of the previous section, \sec{convoInt}, to get the key result of this section: the exact Borel transforms of these sectors. Using the same techniques, we even find that there are further $(n,m)$ sectors with both $n$ and $m$ nonzero, contary to what was the case at leading order in $1/N_f$. At the end of the section, we briefly discuss the remaining transseries sectors and summarize the structures we have found in \tab{NLOsummary}.

%

\subsection{Four diagrams at \texorpdfstring{$\ord{1/N_f^2}$}{ }}
Using \eq{BorelChain}, the Borel transform of the diagrams shown in
\fig{AdlerNLO} can be computed using an analytic regularized
propagator for each of the bubble chains. In terms of the convolution
integral, the Borel transform for each diagram is given by
\begin{align}
\cB[\Pi(Q^2)](u)
=\frac{-1}{\beta_{0f}} \bigg(\frac{Q^2}{\mu^2}e^{C}\bigg)^{-u}
\int_0^u du_1\,du_2\,
\de\big(u-u_1-u_2\big) \Pi(u_1,u_2)\,,
\label{eq:convolution}
\end{align}
where $\Pi(u_1,u_2)$ is the $Q^2$-independent part of these diagrams
in terms of the two Borel parameters $u_1$ and $u_2$ of the bubble
chains. Here and in the expressions that follow, we again ignore an overall colour factor $C_F^2$. For diagrams $(a)$-$(c)$ in \fig{AdlerNLO}, we managed to compute this $Q^2$-independent part exactly to all orders in $u_1$ and $u_2$, see App. \ref{app:masters} and \ref{app:momentumIntegrals} for more details, and the result
reads:
\begin{align}
\Pi_a(u_1,u_2)
&= -\frac{3}{2\pi^2}
\frac{1}{u(1\!-\!u)(2\!-\!u)}
\frac{1}{(u_1\!-\!2)(u_1\!-\!1)(u_1+1)}
\frac{1}{(u_2\!-\!2)(u_2\!-\!1)(u_2+1)}\,,\\
\Pi_b(u_1,u_2)
&= -\frac{3}{2\pi^2}
\frac{\Ga(u)}{(1-u)\Ga(3-u)}
\frac{\Ga(1-u_1)}{(u_1-2)\Ga(2+u_1)}
\frac{\Ga(1-u_2)}{(u_2-2)\Ga(2+u_2)}\,,\label{eq:resultDiagramb}\\
\Pi_c(u_1,u_2)
&= \frac{3}{2\pi^2}
\frac{1}{u(1-u)(2-u)}
\frac{\Ga(u)}{\Ga(3-u)}
\frac{u_1 \Ga(1-u_1)}{\Ga(2+u_1)}
\frac{\Ga(1-u_2)}{(u_2-2)\Ga(2+u_2)}\,,
\end{align}
where, after taking the $\delta$-function in the convolution integral
(\ref{eq:convolution}) into account, we have  $u=u_1+u_2$. For diagram $(d)$ we
do not have a such a closed form, but instead computed the expansion
in $u_1$ and $u_2$, for which the first few terms read
\begin{align}
\Pi_d(u_1,u_2)
&= \frac{201}{4}
-36\zeta_3
+\Big(\frac{315}{4}-54\zeta_3\Big)u_1
+\Big(\frac{3123}{16}-63\zeta_3-90\zeta_5\Big)u_1^2\nn\\
&+\Big(\frac{873}{8}-54\zeta_3\Big)u_2
+\Big(\frac{747}{4}-93\zeta_3-60\zeta_5\Big)u_2^2
+\frac{1}{u}\Big(18+\frac{9u_1}{4}+36u_1^2\Big)\nn\\
&+\Big(\frac{4539}{16}-144\zeta_3-90\zeta_5\Big)u_1u_2
+...
\end{align}
In \app{masters} we give additional details for the computation of the perturbative coefficients for this diagram.

In order to get the (Borel transform of the) Adler function, we can
perform the convolution integral (\ref{eq:convolution}) and take the
derivative with respect to $Q^2$ (recall \eq{pitoadler}).  The convolution
integral (\ref{eq:convolution}) for diagram $(a)$ can be computed
exactly, and we obtain
\begin{equation}
\cB[D(Q^2)](u) 
= \frac{1}{\beta_{0f}}\bigg(\frac{Q^2}{\mu^2}e^{C}\bigg)^{-u}\cB[D](u)\,,
\end{equation}
with the closed form expression
\begin{align}
\cB[D_a](u)
&=\frac{6}{(1-u)(2-u)}
\bigg[
\frac{\log(1+u)}{3(1-u)u(2+u)}
+\frac{\log(1-u)}{(3-u)(2-u)u}
\nn\\
&\hspace{7cm}
+\frac{2\log(1-\frac{u}{2})}{3(4-u)(3-u)(1-u)}\bigg].\label{eq:BorelTransNLOb}
\end{align}
However, for the other three diagrams we do not have the benefit  of
 a closed form. Instead, one can expand the $\Pi(u_1,u_2)$ for
these diagrams in $u_1$ and $u_2$ and perform the convolution integral
order by order. In this way we find
\begin{align}
\cB[D_a](u)
&= -\frac{3u}{4}
-\frac{3u^2}{2}
-\frac{81u^3}{32}
-\frac{215u^4}{64}
-\frac{2707u^5}{640}
+\ORd{u^6}\,,\label{eq:diagramaPerturbativeCoeffs}\\
\cB[D_b](u)
&= -\frac{3u}{4}
-\frac{3u^2}{2}
-\frac{85u^3}{32}
-\bigg(\frac{239}{64}-\frac{\zeta_3}{4}\bigg)u^4
-\bigg(\frac{3211}{640}-\frac{\zeta_3}{2}\bigg)u^5
+\ORd{u^6}\,,\\
\cB[D_c](u)
&= -\frac{3u}{8}
-\frac{13u^2}{16}
-\frac{95u^3}{64}
-\bigg(\frac{277}{128}-\frac{\zeta_3}{8}\bigg)u^4
-\bigg(\frac{759}{256}-\frac{11\zeta_3}{40}\bigg)u^5
+\ORd{u^6}\,,\\
\cB[D_d](u)
&= 3u
+\bigg(\frac{137}{16}-6\zeta_3\bigg)u^2
+\bigg(\frac{565}{32}-9\zeta_3\bigg)u^3
+\bigg(\frac{11219}{384}-\frac{38\zeta_3}{3}-\frac{65\zeta_5}{6}\bigg)u^4\nn\\
&\hspace{3cm}
+\bigg(\frac{27787}{640}-\frac{703\zeta_3}{48}+\zeta_3^2-\frac{65\zeta_5}{4}\bigg)u^5
+\ORd{u^6}\,,
\end{align}
where, even though we have the closed form in
\eq{BorelTransNLOb}, we added the first few coefficients for
$\cB[D_a](u)$. For diagrams $(a)$-$(c)$, we computed the coefficients
up to order $u^{150}$. Diagram $(d)$ is computationally more involved
and we managed to compute up to order $u^{18}$ \footnote{Our calculation exhausts the current datamine \cite{Blumlein:2009cf}, which we used to compute these coefficients.}.

\begin{figure}
\centering
\begin{subfigure}{7.5cm}
    \includegraphics[width=\textwidth]{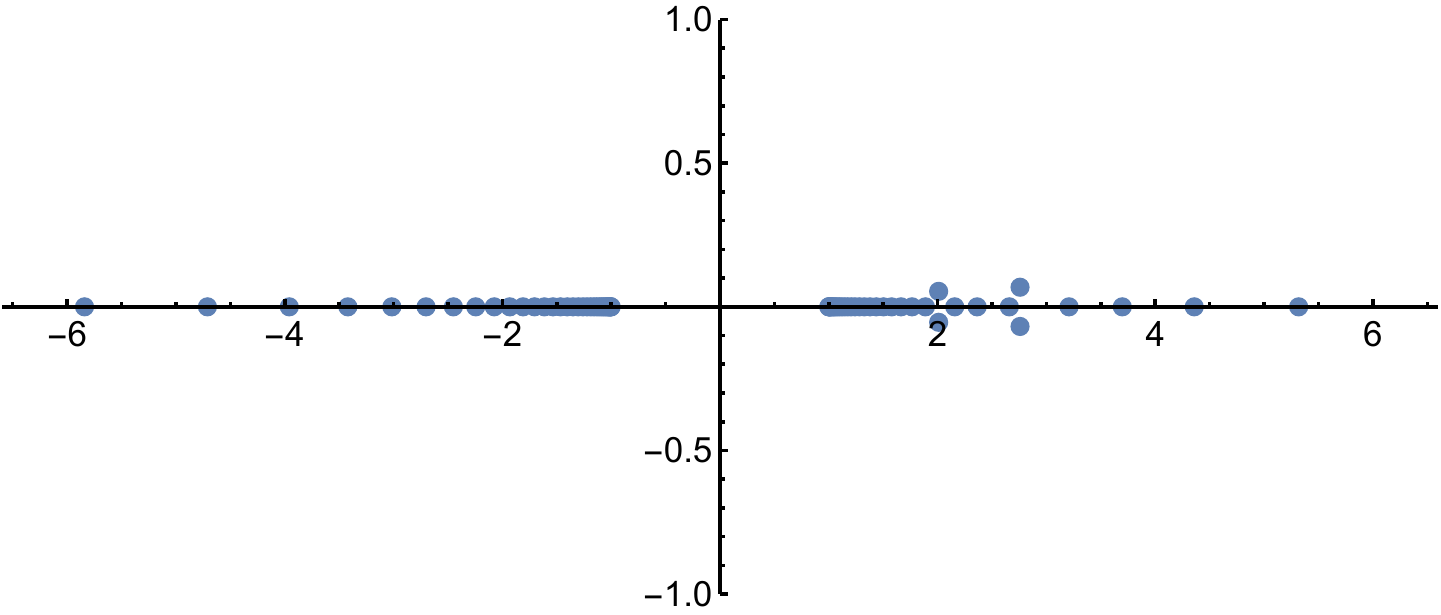}
    \caption{\vspace{1cm}}
    \label{fig:Pade1}
\end{subfigure}
\begin{subfigure}{7.5cm}
    \includegraphics[width=\textwidth]{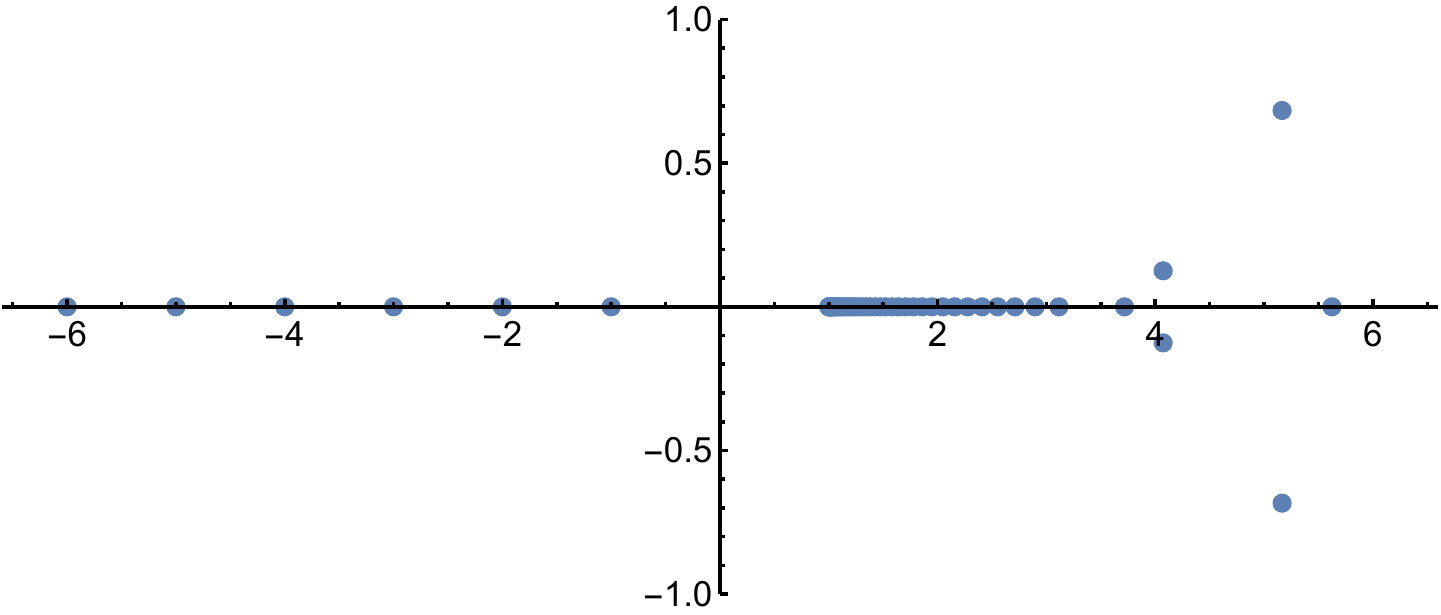}
    \caption{\vspace{1cm}}
    \label{fig:Pade2}
\end{subfigure}
\begin{subfigure}{7.5cm}
    \includegraphics[width=\textwidth]{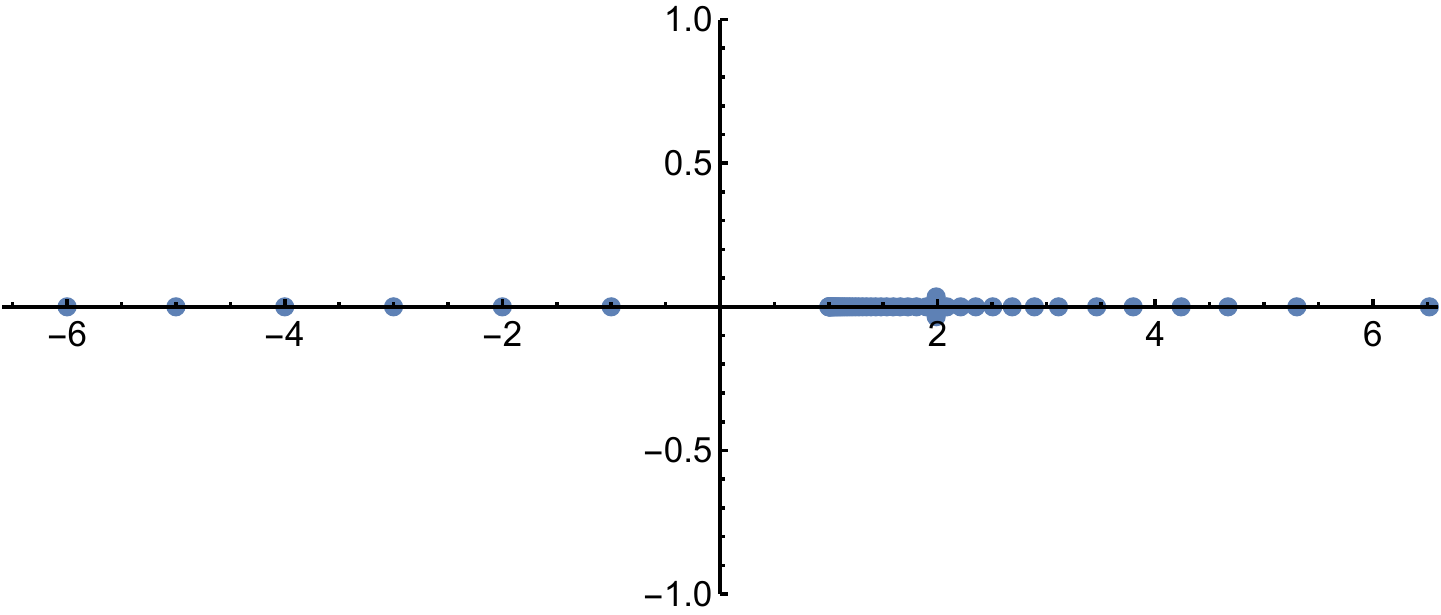}
    \caption{}
    \label{fig:Pade3}
\end{subfigure}
\begin{subfigure}{7.5cm}
    \includegraphics[width=\textwidth]{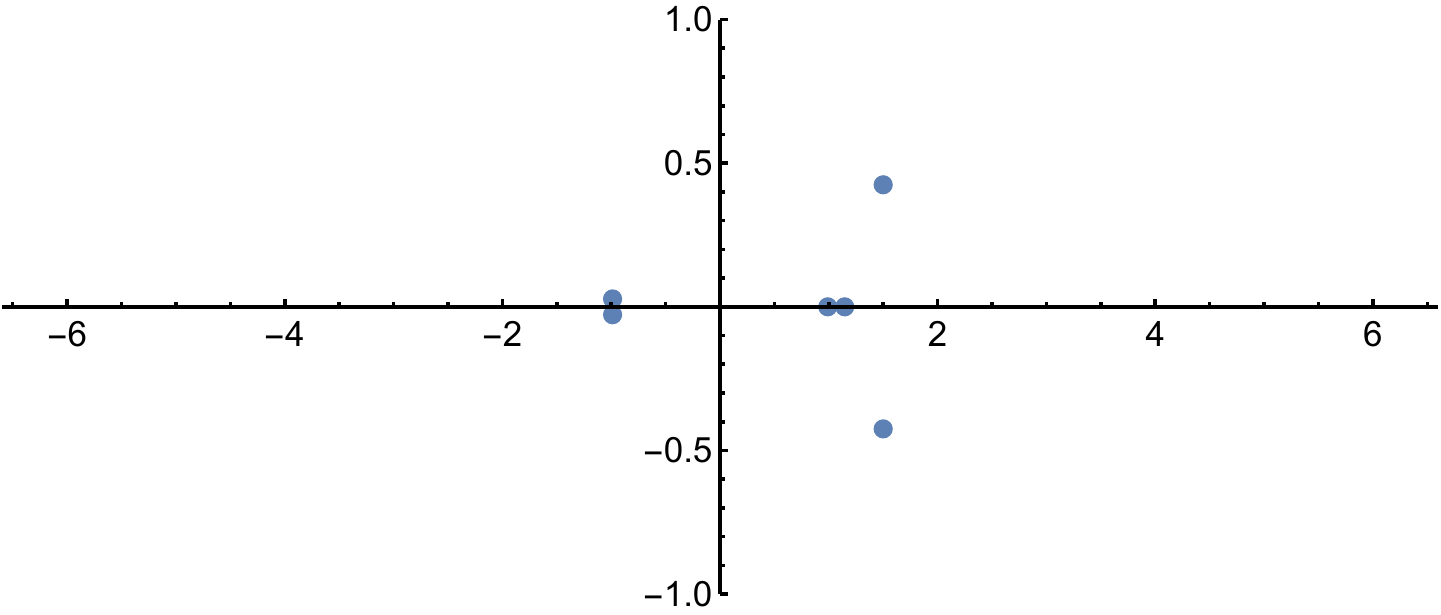}
    \caption{}
    \label{fig:Pade4}
\end{subfigure}
\caption{Plot (a)-(c) show the singularities of the diagonal order 75
  Pad\'e approximants of the diagrams (a)-(c) respectively. Branch
  cuts are mimicked by an accumulation of poles. As we only have 18
  coefficients for diagram (d), we show the Padé   poles of the diagonal order 9 Pad\'e approximant in plot (d).}
\label{fig:PadeNLO}
\end{figure}

Before we enter the detailed resurgence analysis, we can already
have a closer look at the singularities of each of those expressions
in the Borel plane. The singularities of diagram $(a)$ can
be read off immediately from its closed form, \eq{BorelTransNLOb}: 
branch cuts starting at $u=-1$, $u=1$ and $u=2$ and poles at $u=-2$,
$u=1$, $u=2$, $u=3$, and $u=4$.  For the other diagrams, for which we
only have a finite number of perturbative coefficients, we cannot
 read off the singularities in this way. Instead, we can use Padé
approximants to this end (see \app{borelPade} for a brief description). The poles
of these Padé approximants give a good indication of where the `true'
singularities in the Borel plane are located.  In order to see the
type of singularities that we can expect in the Borel plane of the
different diagrams, we plot the poles of the diagonal Padé
approximants in~\fig{PadeNLO}.

For diagrams $(b)$ and $(c)$, we see essentially the same pattern
arising as for diagram $(a)$: the emergence of poles in the UV
direction and a branch cut starting at $u=1$ and at $u=2$. Note that
of course a Padé approximant, being a rational function, cannot
produce an actual branch cut -- instead, the branch cut is mimicked by
the Padé approximants as an accumulation of poles. This well-known
effect occurs in many other examples that involve Padé approximants, see e.g.~\cite{STAHL1997139} for an early example.  We
also expect the emergence of branch cuts beyond $u=2$ for diagrams
$(b)$ and $(c)$, but more on this in a moment.  As we do not have many
terms for diagram $(d)$ it is hard to tell if the Padé poles near
$u=-1$ and $u=1$ are the start of a branch cut, or are Borel plane poles
instead.  
We also observe some `spurious' poles that have an imaginary component. A closer look reveals that these poles always come in pairs: they are the complex conjugate of each other. This is a common phenomenon for Padé approximants, and often reveals the fact that these mimic one pole without an imaginary component, see also \app{borelPade} for more details.

It is interesting to note that in the Borel transforms of the
individual diagrams, the singularity at $u=1$ does not disappear. This
was also the case for the LO ($\ord{1/N_f}$) Adler function:  when the diagrams
were taken separately they did have a singularity at $u=1$. However,
these singularities at $u=1$ cancelled when we added the diagrams.
 On physical grounds, we expect that this will still happen when we take all
diagrams together at every given order in the flavour expansion -- recall the discussion on the OPE at the end of \sec{adler}. 
Nevertheless we do {\em not} expect singularities at other values of $u$ to
cancel. This also did not happen in the case of the LO Adler function
where the singularities on a diagram by diagram basis indeed gave a realistic
picture of the singularities for the sum of the diagrams.

\subsection{Resurgence analysis: non-perturbative sectors \texorpdfstring{$(1,0)$}{ } and \texorpdfstring{$(0,1)$}{ }}
Our discussion of resurgence in \sec{resurgence} was mainly focused on
the case where the singularities of the Borel transform are simple
poles or logarithmic branch cuts. In many examples studied so far in the literature these are indeed
the only singularities that occur, but in the case of the Adler
function several other types of singularity are present. We
already encountered double poles in \sec{adlerLO}, and if we look at
the exact Borel transform for diagram $(a)$, we observe another 
type of singularity:
\begin{equation}\label{eq:logoverpole}
\cB[D_a](u)\Big|_{u=1} = -3\frac{\log(1-u)}{u-1}+...
\end{equation}
where the ellipsis denotes other
singularities (poles and logarithmic branch cuts) as well as regular
terms.  At the end of \sec{generalizations} we saw that a singularity
of the form in \eq{logoverpole} follows from perturbative
coefficients that grow as $\Ga(k+1)\psi(k+1)$, where $\psi(z)$ is the digamma function. Dividing by $\Ga(k+1)$, this means that the leading growth of the coefficients of the Borel transform $\cB[D_a](u)$ is given by $-3\log(k)$, where we read off the $-3$ from \eq{logoverpole} and used $\psi(k) = \log(k) + \ord{\frac{1}{k}}$. 

\begin{figure}
\centering
\includegraphics[width=.8\linewidth]{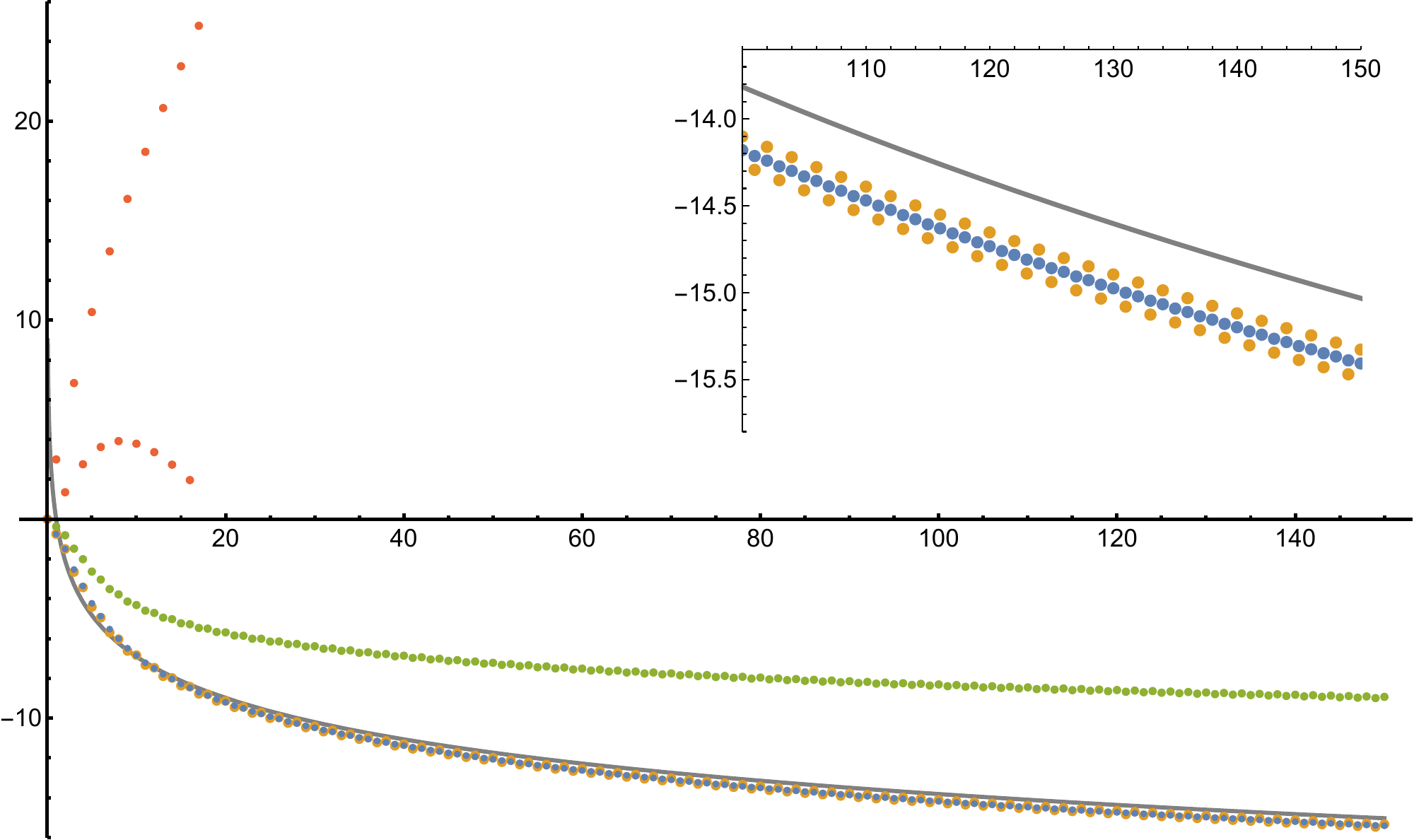}
\caption{Plot of the first 150 perturbative coefficient of $\cB[D_a]$
  (blue), $\cB[D_b]$ (orange), $\cB[D_c]$ (green). As the solid gray line we also
  show the function $-3\log(k)$ to highlight the fact that the
  coefficients for (a), (b) and (c) show logarithmic
  growth. Since the points denoting the
  coefficients for (a) and (b) practically overlap, we included a subfigure zooming in at part of those sequences. In red, above the horizontal axis, we show the 18
  coefficients for $\cB[D_d]$ we have calculated. As the even and odd
  coefficients seem to follow different curves it is unclear if
  these coefficients also show logarithmic growth. }
\label{fig:perturbativeCoefficientsNLO2}
\end{figure}

In \fig{perturbativeCoefficientsNLO2}, we show the perturbative
coefficients of $\cB[D_a]$ together with the
function $-3\log(k)$, and indeed this function matches the leading
growth of these coefficients quite well.  We observe similar
logarithmic growth for the coefficients $\cB[D_b](u)$ and
$\cB[D_c](u)$. Notice that the plotted points for the
coefficients of $\cB[D_b](u)$ nearly overlap with those for
$\cB[D_a](u)$, implying that the leading
growths of $D_b$ and $D_a$ are equal.  For the
coefficients of $\cB[D_d](u)$ it is more difficult to tell if the coefficients show
logarithmic growth, as we only have 18 coefficients where the even and
odd coefficients seem to follow notably different curves, so that we
only have 9 coefficients to determine the trend for each curve. This
makes determining the large order behaviour unrewarding. Nevertheless
it is a pleasant surprise to see from \fig{Pade4} how well the Pad\'e approximation already estimates
the location of the Padé poles in $u$ near $\pm 1$.

Comparing the logarithmic growth $\sim\Ga(k+1)\psi(k+1)$ with
\eq{largeOrderWithLogs}, which describes the large order relation for a
transseries Ansatz of the type \eq{ansatzWithLogs} with logarithms, we see that
at the level of the transseries this implies that logarithmic terms
are included in the expansion.  
As discussed in \sec{adlerLO} for the Adler function at LO in the flavour expansion, the form of the complete transseries
is further determined by the fact that the
Borel transforms of the diagrams have singularities at both negative
and positive integer values of $u$. This leads to a transseries with at
least two parameters.  Taking
these considerations into account, we make the following minimal
transseries Ansatz for diagrams $(a)$, $(b)$ and~$(c)$:
\begin{equation}
D_{\NLO}(\al)
=\sum_{n=0}^\infty\sum_{m=0}^\infty \si_1^n\si_2^m e^{-n\frac{A_1}{\al}}e^{-m\frac{A_2}{\al}}
\Big[
D_{\NLO}^{(n,m)[0]}(\al)
+\log(\al)D_{\NLO}^{(n,m)[1]}(\al)
\Big]\,,
\end{equation}
where as in the LO case\footnote{One often encounters the phenomenon of resonance -- see also the discussion on page \pageref{page:resonance} -- in
  transseries with logarithms and when multiple $A_i$ add up
  to $0$.  Although this is the case for the diagrams considered in
  this section, their transseries do not show resonance. See \cite{Borinsky:2022knn} for another example of this behaviour.} $A_1=-A_2=1$.  Note that for
diagram $(a)$, we do not have an infinite number of singularities in
the Borel plane, hence in that case the sums over $n$ and $m$ truncate.
Furthermore, note that as the sectors with logarithms lead to
singularities of the form  \eq{logoverpole}, we expect that the
sectors $D_\NLO^{(0,m)[1]}$ vanish as diagram $(a)$ does not contain
such a singularity at $u=-1$ and because the Pad\'e plots for diagram
$(b)$ and $(c)$ do not show branch cuts starting from negative
integers.  We will have more to say about diagram $(d)$ at the end of
this subsection.

Following what we did in \sec{adlerLO} for the LO Adler function, we
can translate the transseries Ansatz into large order relations for
the perturbative coefficients and study these numerically. This will
then lead to the extraction of non-perturbative sectors.  The singularity at $u=1$ does not disappear on a diagram by diagram basis,
which implies that the leading large order behaviour of the perturbative coefficients
follows from forward steps in the alien lattice in both the $(1,0)$ and
$(0,1)$ directions. Taking into account  that some
coefficients grow logarithmically, this yields the following natural
form for the large order relation for the perturbative coefficients:
\begin{align}
d_k^{(0,0)}
&\sim \frac{S_{1,0}}{2\pi\img}\sum_{h=0}^\infty \frac{\Ga(k-h-\beta)}{A_1^{k-h-\beta}}
\Big(d_h^{(1,0)[0]}+d_h^{(1,0)[1]}(\log(A_1)-\psi(k-h-\beta)\Big)
\nn\\
&\hspace{3cm}
+\frac{S_{0,1}}{2\pi\img}\sum_{h=0}^\infty
\frac{\Ga(k-h-\beta)}{A_2^{k-h-\beta}}
d_h^{(0,1)[0]}
+ \ord{2^{-k}}\,,
\end{align}
where, since we can allow a finite number
of the leading coefficients to vanish, we assumed
$\beta_{1,0}^{[0]}=\beta_{0,1}^{[0]}=\beta_{1,0}^{[1]}=\beta_{0,1}^{[1]}\equiv\beta$;
this entails no loss of generality (see also the discussion below \eq{largeOrderLOsector2}).
As we did in \sec{adlerLO}, we absorb the Stokes
constant and factors of $2\pi\img$ into the non-perturbative
coefficients:
\begin{equation}
\tilde{d}_h^{(\ell,0)[p]} \equiv \frac{S_{1,0}^\ell}{2\pi\img} d_h^{(\ell,0)[p]},
\qquad\qquad 
\tilde{d}_h^{(0,\ell)[p]} \equiv \frac{S_{0,1}^\ell}{2\pi\img} d_h^{(0,\ell)[p]}.
\end{equation}
For explicit results, let us first focus on diagram $(a)$ as the closed form of the Borel
transform, \eq{BorelTransNLOb}, allows us to read off the resurgence
structure, using \eqs{borelSingularities}{logTransSingularityExample}.
We extract the dominant growth
\begin{equation}
\tilde{d}_0^{(1,0)[1]}\frac{\Ga(k-\beta)\psi(k-\beta)}{A_1^{k-\beta}}
\sim \tilde{d}_0^{(1,0)[1]}\frac{\Ga(k-\beta)\log(k-\beta)}{A_1^{k-\beta}}
+\ORd{\frac1k}\,,
\end{equation}
from the expansion around \eq{logoverpole}, which yields the exact values
\begin{equation}\label{eq:dominantGrowthDiagrama}
\beta=-1\,,
\qquad
\tilde{d}_0^{(1,0)[1]} = -3\,,
\qquad\text{and}\qquad
\tilde{d}_{h>0}^{(1,0)[1]} = 0\,.
\end{equation}
Subtracting the leading growth from the large order formula yields
\begin{align}\label{eq:largeOrderWithoutPolyGamma}
\delta_k^{(0)}
&\equiv \frac{A_1^{k-\beta}}{\Ga(k-\beta)}
\Big(\tilde{d}_k^{(0,0)}-\tilde{d}_0^{(1,0)[1]}\frac{\Ga(k-\beta)\psi(k-\beta)}{A_1^{k-\beta}}\Big)\nn\\
&\sim \sum_{h=0}^\infty\frac{\Ga(k-\beta-h)}{\Ga(k-\beta)}A_1^h
\Big[\tilde{d}_h^{(1,0)[0]}+(-1)^{h+\beta-k}\tilde{d}_h^{(0,1)[0]}\Big]+...\,,
\end{align}
a type of growth that we already encountered in \sec{adlerLO}.
Taking similar ratio tests as we did there, and taking the parity
of $k$ into account, we find full asymptotic expansions:
\begin{align}
\tilde{d}_h^{(1,0)[0]}
&= 
\begin{cases}
\frac13-3\ga_E+\frac{13}{9}\log(2)
&\qquad h=0\\
\Ga(h)\Big(\left(3h+\frac{3}{2}\right)+(-1)^h\frac12+\frac{1}{2^{h}}\Big)
&\qquad h>0
\end{cases}\label{eq:sector(1,0)Diagrama}
\end{align}
for the $(1,0)$-sector, and 
\begin{align}
\tilde{d}_h^{(0,1)[0]}
&=
\begin{cases}
0
&\qquad h=0\\
\Ga(h)\Big(
\frac{1}{2}-(-1)^h\frac1{36}
-\frac{1}{2^{h+1}}\big(\frac23h+\frac49\big)
-\frac{1}{3^{h}}\frac14
\Big)
&\qquad h>0
\end{cases}\label{eq:sector(0,1)Diagrama}
\end{align}
for the $(0,1)$-sector. We have checked \eqs{sector(1,0)Diagrama}{sector(0,1)Diagrama}  numerically up to 20
decimal places for the first 15 coefficients by using Richardson
transforms. 
Note the gamma-functions $\Gamma(h)$ in these expressions, implying
that the first non-perturbative sectors are now indeed asymptotic
series, just like the perturbative sector.
As an additional check, we can read off the resurgence structure from the exact Borel transform \eq{BorelTransNLOb} using \eq{BorelSingularitiesMultiParTrans}. For example, near the singularity positioned at $u=1$ we get
\begin{align}\label{eq:borelDiagramaAround1}
\cB[D_a^{(0,0)}](u)\Big|_{u=1} 
&= -\frac{S_{1,0}}{2\pi\img}\bigg(
-3\frac{\log(1-u)+\psi(1)}{1-u} + \frac{\frac13+3\psi(1)+\frac{13}{9}\log(2)}{1-u}\nn\\
&\hspace{3cm}+ \cB[D^{(1,0)[0]}](u-1) \log(1-u)\bigg)+...\,,
\end{align}
where the ellipsis denote regular terms, and with 
\begin{equation}\label{eq:borelDiagramaAround1LogTerm}
\frac{S_{1,0}}{2\pi\img}\cB[D^{(1,0)[0]}](u-1)
= \frac{3}{(2-u)^2}+\frac{\frac32}{2-u} - \frac{1}{2u} + \frac{1}{3-u}\,.
\end{equation}
As already discussed above, the first term in \eq{borelDiagramaAround1} can be compared with \eqs{logTransSingularityExample1}{logTransSingularityExample} and yields the coefficients of \eq{dominantGrowthDiagrama}. This is the reason why we included the $\psi(1)$ term there, and then subtracted it again in the next term. Using $\psi(1)=-\ga_E$ in the second term in \eq{borelDiagramaAround1} yields the coefficient in the first line in \eq{sector(1,0)Diagrama}. Finally, an inverse Borel transform of \eq{borelDiagramaAround1LogTerm} gives the second line in \eq{sector(1,0)Diagrama}.  
In a similar way, one can read off the resurgence of the $D^{(0,1)[0]}$ sector, i.e. \eq{sector(0,1)Diagrama}, by expanding $\cB[D_a^{(0,0)}](u)$ around $u=-1$.

We discussed a diagram for which it was possible to find a
closed form Borel transform, but the true power of
resurgence analysis emerges when we study the diagrams for which we do
{\em not} have such an exact expression. To be precise, for diagrams $(b)$ and $(c)$, we can study
the large order behaviour of the perturbative coefficients for these
diagrams, compute coefficients in the transseries numerically and
can then usually infer their exact values. Furthermore, we can use
the discussion of \sec{convoInt} on convolution integrals and their
resurgence as a cross check. The interested reader can apply the methods of \sec{convoInt} to obtain the same results for diagram $(a)$.

Turning to diagram $(b)$, we find, by studying the large order behaviour of the perturbative coefficients, that $\beta_{1,0}=\beta_{0,1} = -1$ and that the non-zero coefficients are
\begin{align}
\tilde{d}_0^{(1,0)[1]}
&= -3\label{eq:101sectorDiagramb}\\
\tilde{d}_h^{(1,0)[0]}
&=
\begin{cases}
\frac13-3\ga_E+\frac{13}{9}\log(2)
&\qquad h=0\\
\Ga(h)\Big(
6-\frac{3}{2^{h}}
\Big)
&\qquad h>0
\end{cases}\label{eq:100sectorDiagramb}\\
\tilde{d}_0^{(0,1)[0]}
&= \frac15\Big(\log(2)-\log(3)\Big)\,.\label{eq:010sectorDiagramb}
\end{align}
We have checked these numbers numerically up to 12 decimal places for the
first 15 coefficients for $\tilde{d}_h^{(1,0)[0]}$, after which we inferred the exact expression -- an expression that we shall argue to be correct in a different way in a moment.  
From the coefficients, we deduce the following information. The leading order growth of this diagram is given by a $\log$-sector with a single coefficient $\tilde{d}_0^{(1,0)[1]}$. Furthermore, the non-perturbative $(1,0)$ sector is now an asymptotic sector, while the $(0,1)$ sector only contains just a single non-zero coefficient corresponding to a simple pole in the Borel plane (we already observed this in \fig{Pade2}).

Using the results of \sec{convoInt}, we can check these numerical results
using the method of alien derivatives acting on the convolution
integral. We therefore rewrite the convolution integral for $D_b^{(0,0)}$ (recall \eqs{convolution}{resultDiagramb}) as
\begin{align}\label{eq:convoMethodDiagramb}
\cB[D_b^{(0,0)}](u)
=\frac{-6\Ga(1+u)}{(1-u)\Ga(3-u)}\cB[\Psi^{(0)}](u)\,,
\end{align}
where we defined 
\begin{equation}\label{eq:convoMethodDiagramb2}
\cB[\Psi^{(0)}](u)
=\int_0^u du_1\, \cB[F](u_1)\cB[F](u-u_1),
\quad\text{with}\quad
\cB[F](u) 
= \frac{\Ga(1-u)}{(2-u)\Ga(2+u)}\,.
\end{equation}
The singularities of the Borel transform $\cB[F](u)$ are simple poles
at positive integers, except at $u=2$ which is a double
pole. Therefore, the resurgence structure of $F$ can be easily read off
from the expansions around these poles and we find
\begin{equation}
\De_1F = \frac12(2\pi\img).
\end{equation}
Using the procedure as outlined in \sec{convoInt}, we obtain
\begin{equation}
S_1^\Psi\Psi^{(1)}
=\De_1\Psi^{(0)}
= 2F\De_1F
= (2\pi\img) F(\al).
\end{equation}
In other words, in a neighborhood around $u=1$, the Borel transform of $\Psi^{(0)}$ looks like
\begin{align}
\cB[\Psi^{(0)}](u)\Big|_{u=1}
&= -S_1^\Psi\cB[\Psi^{(1)}](u-1)\frac{\log(1-u)}{2\pi\img}+...\nn\\
&= -\cB[F](u-1)\log(1-u)+...\nn\\
&= -\frac{\Ga(2-u)}{(3-u)\Ga(1+u)}\log(1-u)+...\,,
\end{align}
where the ellipsis denotes regular terms around $u=1$.
By adding back in the prefactor for $\cB[D_b^{(0,0)}](u)$, close to $u=1$ \eq{convoMethodDiagramb} becomes
\begin{align}
\cB[D_b^{(0,0)}](u)\Big|_{u=1}
&=\frac{-6\Ga(1+u)}{(1-u)\Ga(3-u)}\left[\cB[\Psi^{(0)}](u)\Big|_{u=1}+...\right]\nn\\
&= \frac{6\log(1-u)}{(1-u)(2-u)(3-u)}+...\nn\\
&\equiv - \tilde{d}_0^{(1,0)[1]}\frac{\log(1-u)}{1-u}
-S_{1,0}\cB[D_b^{(1,0)[0]}](u-1)\frac{\log(1-u)}{2\pi\img}+...\,,\label{eq:expansionAroundLog(1-u)b}
\end{align}
with $\tilde{d}_0^{(1,0)[1]}=-3$ already given in \eq{101sectorDiagramb} and where
\begin{equation}
\frac{S_{1,0}}{2\pi\img}\cB[D_b^{(1,0)[0]}](u) = \frac{6}{1-u} - \frac{3}{2-u}\,.
\end{equation}
Indeed, by performing an inverse Borel transform, we can now directly read off the coefficients
 $\tilde{d}_h^{(1,0)[0]}$ for $h>0$ already given in \eq{100sectorDiagramb}.
 
Note however that the prefactor in the first line of \eq{expansionAroundLog(1-u)b} itself has a pole at $u=1$ -- see also our discussion in \sec{prefactor}. As a result, a constant term from the regular part between square brackets (indicated by the dots) also contributes to the singular terms in the last line. Therefore, the coefficient $\tilde{d}_0^{(1,0)[0]}$, which should correspond to a simple pole at $u=1$ in the Borel plane, is not determined yet by studying the convolution integral. Furthermore, $\cB[F](u)$ in \eq{convoMethodDiagramb2} does not have singularities at negative integers, meaning that $\De_{-1}F = 0$. Therefore it does not seem possible to compute $\cB[D_b^{(0,0)}](u)$ locally near $u=-1$, and thus we can also not extract $\tilde{d}_0^{(0,1)[0]}$ using the method of convolution. We come back to this point in the next subsection, where we will see that by including further non-perturbative sectors in the analysis, some of these undetermined coefficients can still be found.

This concludes our discussion of diagram $(b)$. In a similar manner, for diagram $(c)$, a large order analysis of the perturbative coefficients yields $\beta_{1,0}=\beta_{0,1} = -1$ as well as the following non-zero expansion coefficients for the non-perturbative sectors:
\begin{align}
\tilde{d}_0^{(1,0)[1]}
&= -\frac{3}{2}\label{eq:101diagramc}\\
\tilde{d}_h^{(1,0)[0]}
&=
\begin{cases}
-\frac12-\frac{3}{2}\ga_E-\frac{1}{6}\log(2)
&\qquad h=0\\
\Ga(h)\Big(\left(3h+\frac{3}{2}\right)-(-1)^h\frac12+\frac{1}{2^{h}}\frac12\Big)
&\qquad h>0
\end{cases}\label{eq:100diagramc}\\
\tilde{d}_0^{(0,1)[0]}
&= \frac16+\frac12\Big(\log(2)-\log(3)\Big)\label{eq:010diagramc},
\end{align}
which we have checked numerically up to at least 13 decimal places for the
first 15 coefficients for $\tilde{d}_h^{(1,0)[0]}$.  We observe a similar pattern as for
diagram $(b)$: a $\log$-sector with a single coefficient $\tilde{d}_0^{(1,0)[1]}$, an asymptotic $(1,0)[0]$ sector, and a $(0,1)$ sector which contains only a single non-perturbative coefficient. 

Once again, we can also determine these coefficients by studying the
convolution integral more closely. Therefore, we write the convolution integral of diagram $(c)$ as
\begin{align}\label{eq:convoMethodDiagramc1}
\cB[D_c^{(0,0)}](u)
=\frac{-6\Ga(u)}{(1-u)(2-u)\Ga(3-u)}\cB[\Phi^{(0)}](u)\,,
\end{align}
where we defined 
\begin{equation}\label{eq:convoMethodDiagramc2}
\cB[\Phi^{(0)}](u)
=\int_0^u du_1\, \cB[F](u_1)\cB[G](u-u_1)\,,
\qquad\text{with}\qquad
\cB[G](u) 
= \frac{u\Ga(1-u)}{\Ga(2+u)}\,,
\end{equation}
and with $\cB[F]$ the same as in \eq{convoMethodDiagramb2}. Using that both $\De_1F=\De_1G=\frac12(2\pi\img)$ we obtain
\begin{equation}
S_1^{\Phi}\Phi^{(1)}
=\De_1\Phi^{(0)}
= F\De_1G+G\De_1F
= \frac12(2\pi\img) (F(\al)+G(\al)).
\end{equation}
A brief calculation now yields 
\begin{align}
\cB[D_c^{(0,0)}](u)\Big|_{u=1}
&=\frac{-6\Ga(u)}{(1-u)(2-u)\Ga(3-u)}\cB[\Phi^{(0)}](u)\Big|_{u=1}+...\nn\\
&= -\frac{\log(1-u)}{1-u}
\frac{6-3u(4-u)}{u(2-u)^2(3-u)}
+...\nn\\
&\equiv - \tilde{d}_0^{(1,0)[1]}\frac{\log(1-u)}{1-u}
-S_{1,0}\cB[D_c^{(1,0)[0]}](u-1)\frac{\log(1-u)}{2\pi\img}+...\,,\label{eq:expansionAroundLog(1-u)c}
\end{align}
with $d_0^{(1,0)[1]}=-\frac{3}{2}$ as given above in \eq{101diagramc}, and where
\begin{equation}
\frac{S_{1,0}}{2\pi\img}\cB[D_c^{(1,0)[0]}](u) = \frac{\frac12}{1+u}+\frac{3}{(1-u)^2}+\frac{\frac32}{1-u} + \frac{\frac{1}{2}}{2-u}\,.
\end{equation}
It is straightforward to show that the inverse Borel transform of this indeed yields the
coefficients $d_h^{(1,0)[0]}$ given in \eq{100diagramc}
for $h>0$. In the next subsection we see how the convolution integral method also allows to obtain information about the
coefficients $d_0^{(1,0)[0]}$ and $d_0^{(0,1)[0]}$.  

To summarize the results so far (see also \tab{NLOsummary}): although the convolution integral for diagram $(a)$ was the only one
we could compute exactly, our resurgence large order analysis, together with the power of the convolution analysis,
allowed us to make a transseries
Ansatz and extract the whole first non-perturbative $(1,0)$ and $(0,1)$ sectors for the diagrams $(a)$, $(b)$ and $(c)$. 
For all three diagrams, the leading order growth of the perturbative coefficients is governed by the non-perturbative coefficient $d_0^{(1,0)[1]}$. In particular, we found for diagram $(c)$ that $d_0^{(1,0)[1]}=-\frac{3}{2}$, while for both diagrams $(a)$ and $(b)$ the growth is twice as strong, i.e. $d_0^{(1,0)[1]}=-3$. We already observed this qualitatively in \fig{perturbativeCoefficientsNLO2}, where we saw that the growth of perturbative coefficients for diagram $(c)$ is indeed less than for the diagrams $(a)$ and $(b)$, which were more or less overlapping. However, at closer inspection, one sees that the points do not overlap exactly -- an artifact of the subleading growth dictated by the coefficients $d_h^{(1,0)[0]}$ and $d_h^{(0,1)[0]}$ that we have now also computed. For all three diagrams, the $(1,0)$ sector is asymptotic while only for diagram $(a)$ the $(0,1)$ sector is asymptotic. Furthermore, we notice that for diagram $(a)$ the coefficient $\tilde{d}_0^{(1,0)[0]}$ is the same as that of diagram $(b)$.

Finally, we want to offer some observations on diagram
$(d)$. A numerical resurgence analysis on this diagram is challenging
as we only have 18 coefficients, and we already observed that the even
and odd coefficients behave differently. In
\fig{perturbativeCoefficientsNLO2}, we see that the growth of the
upper curve is much stronger than that of the other three diagrams
$(a)$, $(b)$ and $(c)$. Although it looks like the curve bends slightly,
there are not enough coefficients to tell if this is the
beginning of logarithmic growth. It might as well be the case that 
the large order growth of the coefficients is different. Furthermore, the coefficients that we displayed have opposite sign to the coefficients of the other diagrams, though
the lower curve clearly bends downwards towards negative coefficients.

\subsection{Resurgence analysis: non-perturbative sectors \texorpdfstring{$(2,0)$}{ } and \texorpdfstring{$(0,2)$}{ }}
\label{sec:adlerNLOsecondsectors}
In the previous subsection we determined the non-perturbative $(1,0)$
and $(0,1)$ sectors for diagrams $(a)$-$(c)$, using both numerical
results and the method of alien derivatives acting on the convolution
integral.  In order to probe the second
non-perturbative sectors $(2,0)$ and $(0,2)$, we must
subtract the first sectors from the large order expressions. This is
not as straightforward as in the LO case (see
\eq{LOsubtractSector1}), since now the coefficients
$\tilde{d}_k^{(1,0)[0]}$ and $\tilde{d}_k^{(0,1)[0]}$ grow
factorially themselves, implying that we need to subtract an entire (divergent)
asymptotic series. To make this possible, we will apply Borel summation
on the first non-perturbative sector before subtracting it.  

To prepare for this we rewrite \eq{largeOrderWithoutPolyGamma} as
\begin{align}
\delta_k^{(0)}
&\sim \frac{\Ga(k-\beta)}{A^{k-\beta}}
\sum_{\ell=1}^\infty
\frac{1}{\ell^{k-\beta}}
e^{-\frac{\log(\ell)}{1/k}}
\sum_{h=0}^\infty \frac{\Ga(k-\beta-h)}{\Ga(k-\beta)}
(\ell A)^{h}\tilde{d}_h^{(\ell,0)[0]}
+(0,\ell)\text{-sectors}\nn\\
&= \frac{\Ga(k-\beta)}{A^{k-\beta}}
\sum_{\ell=1}^\infty
\frac{1}{\ell^{k-\beta}}
e^{-\frac{\log(\ell)}{1/k}}
\sum_{h=0}^\infty p_h^{(\ell,0)}\bigg(\frac{1}{k}\bigg)^h
+(0,\ell)\text{-sectors}\label{eq:largeOrderWithoutPolyGammaRewritten}
\end{align}
where $p_h^{(\ell,0)}$ and $p_h^{(0,\ell)}$ are the coefficients in
the $1/k$ expansion of the first line, obtained from expanding
the ratio of gamma functions. In particular, $p_h^{(\ell,0)}$ is a
linear combination of the $\tilde{d}_g^{(\ell,0)[0]}$ with $g \leq h$,
and similarly for $p_h^{(0,\ell)}$. The important observation is that
the above expression is itself a transseries in $1/k$ with
non-perturbative exponents proportional to $\log(\ell)$.  It is therefore natural to
perform Borel summation on the series
\begin{equation}\label{eq:largeOrderTransseriesSectors}
P^{(1,0)}(x) \equiv \sum_{h=0}^\infty p_h^{(1,0)}x^h
\qquad\text{and}\qquad
P^{(0,1)}(x) \equiv \sum_{h=0}^\infty p_h^{(0,1)}x^h\,,
\end{equation}
with $x=\frac1k$. Then, in the large order expressions we 
replace these series with the Borel-summed version\footnote{Recall that the constant term is
  not included in our definition of $\cB[p^{(1,0)}](t)$, which is why
  we have to add it separately.}
\begin{equation}
\cS[P^{(1,0)}](x) = p_0^{(1,0)} + \int_0^\infty dt\,\cB[P^{(1,0)}](t)e^{-t/x}\,,
\end{equation}
and similarly for $\cS[P^{(0,1)}](x)$. Once we have summed the leading asymptotic series in this way, we will be able to read off further subleading coefficients from the large $k$ behaviour of $\delta_k^{(1)}$.

\begin{figure}
\centering
\begin{subfigure}{7.5cm}
    \includegraphics[width=\textwidth]{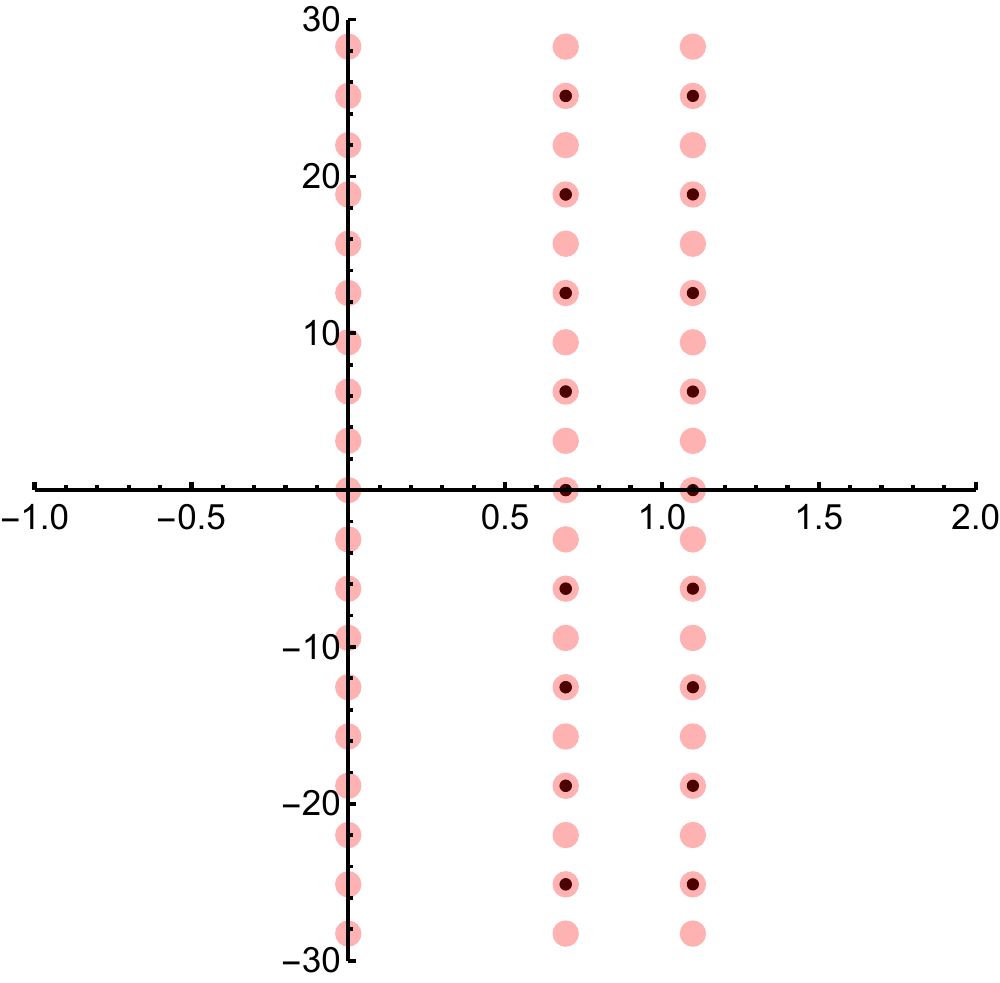}
    \caption{}
    \label{fig:BorelPade10}
\end{subfigure}
\begin{subfigure}{7.5cm}
    \includegraphics[width=\textwidth]{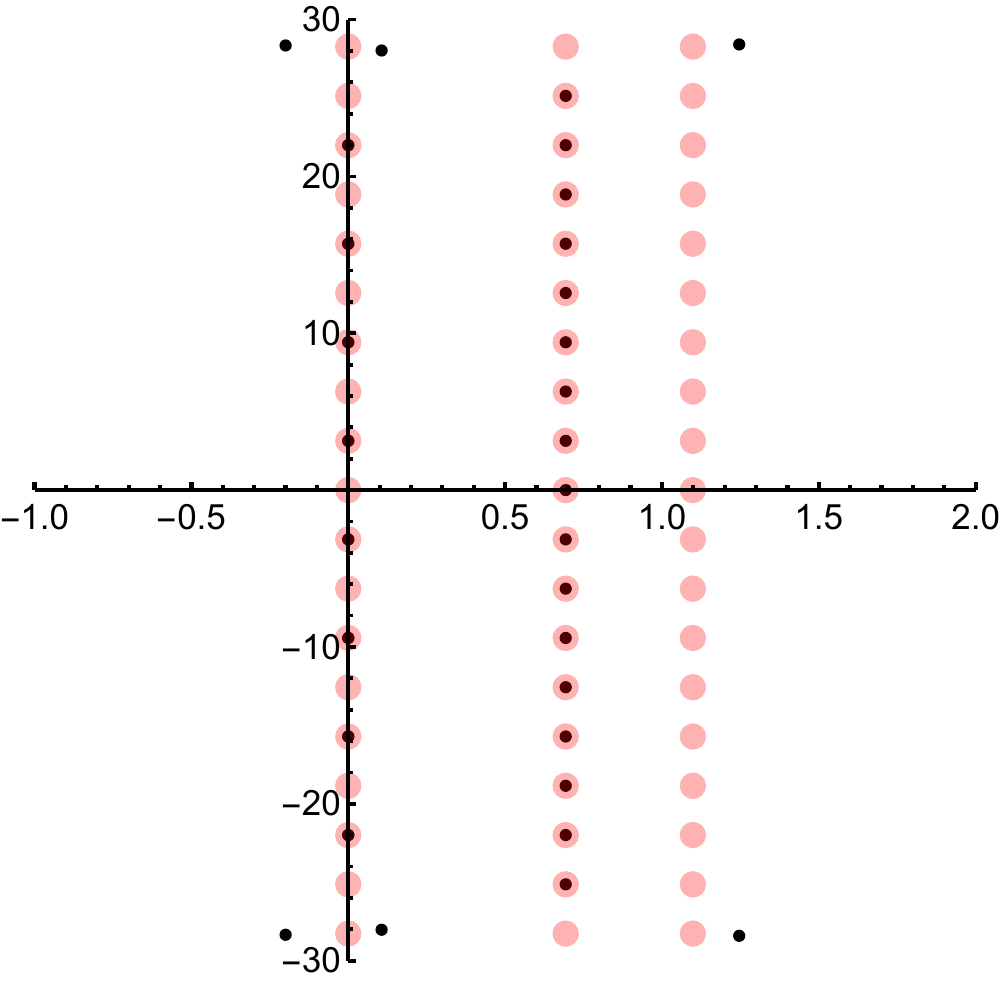}
    \caption{}
    \label{fig:BorelPade01}
\end{subfigure}
\caption{For diagram $(a)$, Figs. (a) and (b) show the Pad\'e poles (small black dots) of the 75th diagonal Padé approximant of $\cB[P^{(1,0)}]$ and $\cB[P^{(0,1)}]$ respectively. The larger red transparant points are positioned at $\log(\ell)+\pi\img m$, with $\ell=1,2,3$ and $m\in\bZ$. At large imaginary values, some `spurious' poles appear due to numerical inaccuracies.}
\label{fig:BorelPade}
\end{figure}

Unfortunately, this expression cannot be taken at face value, as the
asymptotic expansion of $P^{(0,1)}$ and $P^{(1,0)}$ is not Borel
summable. More precisely, there are singularities on the integration
contours, as we can see in \fig{BorelPade}, where we show the
singularities of the Padé approximants of the Borel transforms of
\eq{largeOrderTransseriesSectors} for diagram $(a)$.  We make a few
observations. First of all, we do not only see singularities at positions
$\log(\ell)$, but also at shifts of $2\pi \img m$ for $m\in\bZ$. This
is related to the fact that logarithms are not  uniquely
defined: for instance one could replace
\begin{equation}
\log(\ell)\to \log(\ell)+2\pi\img m\,.
\end{equation}
The second important observation is that we only observe singularities
at $\log(2)+2\pi\img m$ and $\log(3)+2\pi\img m$ for $\cB[P^{(1,0)}]$
and $\log(1)+(2m+1)\pi\img$ and $\log(3)+\pi\img m$ for
$\cB[P^{(0,1)}]$.  These singularities can be traced back to the
expansion of $\cB[D_a^{(0,0)}](u)$ around $\log(1-u)$ and $\log(1+u)$:
\begin{equation}
  \label{eq:9}
\left(\frac{6}{(3-u)(2-u)^2u} - 1 \right) \cdot \frac{\log(1-u)}{(1-u)}\,,
\qquad\text{and}\qquad
\frac{2\log(1+u)}{(1-u)^2(2-u)u(2+u)}\,,
\end{equation}
where in the first expression we subtracted the $\frac{\log(1-u)}{1-u}$ term from \eq{BorelTransNLOb}
as its contribution, i.e. the large order growth initiated by the
coefficient $\tilde{d}_0^{(1,0)[1]}$, is already subtracted in
\eq{largeOrderWithoutPolyGammaRewritten}.  That is, the expansions of \eqref{eq:9}
have singularities at $u=1,2,3$ and $u=-2,1,2$ respectively
\footnote{The pole around $u=0$ cancels against the first term in the
  expansion of the logarithms.}. Taking the logarithm of these
values and using the fact that the logarithm is multi-valued, one indeed finds the 
observed positions of the singularities in \fig{BorelPade}.  For
this paper, the above observations are sufficient, but
they are indicative of further interesting phenomena which we plan to come back to in a
forthcoming publication \cite{CM-AvS-MV-to-appear}.

For both diagrams $(b)$ and $(c)$, we have simple $(0,1)$-sectors
with only a single coefficient, so we do not need to perform a
Borel-Pad\'e summation on these sectors to be able to subtract them. For the $(1,0)$ sector, we 
obtain a similar plot as in \fig{BorelPade10}, where the positions of the
singularities are traced back to locations of the poles in the
expansions around $\log(1-u)$,
given in \eqs{expansionAroundLog(1-u)b}{expansionAroundLog(1-u)c}. To be
precise, for both diagram $(b)$ and $(c)$, we observe singularities at
$\log(2)+2\pi\img m$ and $\log(3)+2\pi\img m$.  

Since the Borel transforms in particular have singularities on the
real positive axis, we need to deal with the ambiguity of the Laplace
contour. In order to do so, we performed a numerical integration under
a small positive angle just above the real axis in \fig{BorelPade01}. See \app{borelPade}
for details about the numerical integration. This choice for the
`upper contour' implies that we also get an imaginary part. However, as we will see in a moment, this imaginary part does not play a role in determining the perturbative coefficients through large order formulas, because of a similar and corresponding ambiguity in the choice of logarithm branch cuts in the Borel plane.
Of course, we
also could have chosen the `lower contour'; the imaginary contribution will then have opposite sign, but will still be `canceled' leading to the
same large order results.

\begin{figure}
    \centering
    \includegraphics[width=.7\textwidth]{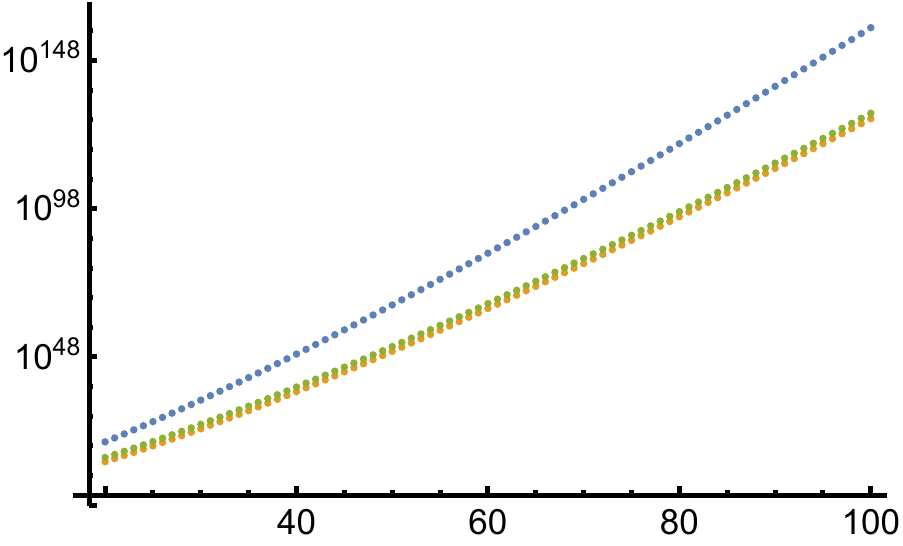}
    \caption{Plot of the coefficients $d_k^{(0,0)}$ (top, blue) and
      Re$[\delta_k^{(1)}]$ (bottom, orange), Im$[\delta_k^{(1)}]$ (middle, green)
      for $20\leq k\leq 100$ for diagram $(a)$. We clearly observe
      that the coefficients $\delta_k^{(1)}$ are `less asymptotic'
      compared to the perturbative coefficients $d_k^{(0,0)}$, a sign
      that this new sequence probes the subleading non-perturbative $(2,0)$ and $(0,2)$ sectors.}
    \label{fig:coeffsPlotExactDiagram}
\end{figure}

With the Borel summed $(1,0)$ and $(0,1)$ sectors at our disposal, we can now turn our attention to the resurgence of the second non-perturbative $(2,0)$ and $(0,2)$ sectors. As in the previous subsection, we start with a discussion of diagram $(a)$. We can now subtract the numerically performed integral
 from the perturbative coefficients. Recalling that $\beta=-1$ and $A_1=-A_2=1$, we
obtain the new sequence
\begin{align}
\delta_k^{(1)}
&= d_k^{(0,0)}
- \tilde{d}_0^{(1,0)[1]}\frac{\Ga(k-\beta)\psi(k-\beta)}{A_1^{k-\beta}}\nn\\
&\hspace{4cm}
- \frac{\Ga(k-\beta)}{A_1^{k-\beta}}\cS[P^{(1,0)}]\left(\frac1k\right)
- \frac{\Ga(k-\beta)}{A_2^{k-\beta}}\cS[P^{(0,1)}]\left(\frac1k\right)\,,
\label{eq:deltak1}
\end{align}
which probes the contribution of the
$(2,0)$ and $(0,2)$ sectors to the large order behaviour of the perturbative coefficients. In \fig{coeffsPlotExactDiagram} we
show $d_k^{(0,0)}$ together with $\delta_k^{(1)}$ for $20\leq k\leq 100$ for
diagram $(a)$. We clearly observe that the (real) perturbative coefficients
$d_k^{(0,0)}$ diverge faster than both Re$[\delta_k^{(1)}]$ and
Im$[\delta_k^{(1)}]$. This is
already a good sign that these coefficients probe the subleading non-perturbative
$(2,0)$ and $(0,2)$ sectors.

Let us now discuss the fact that $\delta_k^{(1)}$ for diagram $(a)$ appears to have an imaginary part. This may seem at odds with the fact that we are describing the large order behaviour of a {\em real} expansion of the Adler function, but in fact it is not. Our numerical resurgence analysis yields
\begin{equation}\label{eq:largeOrderImPartDiagrama}
\img \, \text{Im}\left[\delta_k^{(1)}\right]
\sim \pm \pi\img
\left[
 - \frac{\Ga(k+2)}{2^{k+2}} \cdot 3
 - \frac{\Ga(k+1)}{2^{k+1}} \cdot \frac{3}{2}
 + \frac{\Ga(k+1)}{(-2)^{k+1}} \cdot \frac{1}{36}
 -\frac{\Ga(k+1)}{3^{k+1}}
\right]\,,
\end{equation}
where the overall sign ambiguity comes from a choice of contour in the Borel-Padé evaluation of \eq{deltak1}, either above or below the singularities on the positive real axis in the Borel plane. Note that this imaginary contribution can itself be Borel transformed into
\begin{equation}
\pm \img \pi \left[-3\frac{1}{(2-u)^2} - \frac{3}{2} \frac{1}{2-u} -\frac{1}{36}\frac{1}{2+u} - \frac{1}{3-u}\right]\,.
 \label{eq:imgcontr}
\end{equation}
Now, these same imaginary and ambiguous contributions will also appear in the Borel transform of diagram (a) itself, \eq{BorelTransNLOb}. For example, an ambiguity
is present in the expansion of the logarithm $\log(1-u)$ in that expression around $u=2$. This expansion takes the form
\begin{equation}
\frac{6\log(1-u)}{(3-u)(2-u)^2(1-u)u}\bigg|_{u=2} 
= -  \frac{\pm3\pi\img}{(2-u)^2}-\frac{\pm\frac32\pi\img+3}{2-u}+...
\end{equation}
where we see the exact same imaginary ambiguity as in the first and second term of \eq{imgcontr}. As a result, the coefficients of the $(2,0)$ sector that can be read off from either \eq{largeOrderImPartDiagrama} or \eq{imgcontr} will {\em not} be ambiguous and moreover will be purely real.

A similar reasoning can be applied to the third ambiguous term in \eq{imgcontr} which will reappear in the expansion around $\log(1+u)$ of \eq{BorelTransNLOb}:
\begin{equation}
\frac{2\log(1+u)}{(1-u)^2(2-u)u(2+u)}\bigg|_{u=-2} 
= -\frac{1}{36} \cdot \frac{\pm\pi\img }{2+u} + ...\,.
\end{equation}
For the fourth ambiguous term in \eq{imgcontr}, which will determine a $(3,0)$ sector coefficient, we expand around $u=3$:
\begin{equation}
\frac{6\log(1-u)}{(3-u)(2-u)^2(1-u)u}\bigg|_{u=3}
= -\frac{\pm\pi\img-\log(2)}{3-u}+...\,.
\end{equation}
Taking the same steps as we did for the $(1,0)$ and $(0,1)$ sectors,
we can now determine the coefficients in the $(2,0)$ and $(0,2)$ sectors by doing a large order analysis on the $\delta_k^{(1)}$. Above, we have already analysed the imaginary part of $\delta_k^{(1)}$. The real part does not have a contribution to the $(0,2)$ sector, but is does affect the $(2,0)$ sector. From it, we extract $\beta_{2,0}=-1$ and the $(2,0)$ coefficients
\begin{align}
\tilde{d}_0^{(2,0)[1]}
&= -2\\
\tilde{d}_h^{(2,0)[0]}
&=
\begin{cases}
3-2\ga_E+\frac{1}{4}\log(3)
&\qquad h=0\\
\Ga(h)\Big(\left(\frac23h +\frac{11}{9}\right)(-1)^{h}+ 1
-\frac29\frac{1}{2^{h}}\Big)
&\qquad h>0\,.
\end{cases}\label{eq:diagrama200sectorreal}
\end{align}
As we see, the $(2,0)$ sector contains again a $\log(\al)$ contribution in the transseries, given by the coefficient $d_0^{(2,0)[1]}$. We were able to extract the first 15 coefficients in \eq{diagrama200sectorreal} (as well as those of the simpler imaginary part, \eq{largeOrderImPartDiagrama}) numerically from \eq{deltak1} up to 9 decimal places, after which we inferred the exact expressions. Once again, we know that this is exact to all orders, because we can compare with the exact Borel transform. (Recall how we did this for the $(1,0)$ and $(0,1)$ sectors in \eqs{borelDiagramaAround1}{borelDiagramaAround1LogTerm}.)

Let us now turn to diagram $(b)$. Similar to diagram $(a)$, before we can probe the second non-perturbative $(2,0)$ and $(0,2)$ sectors, we want to subtract the $(1,0)$ and $(0,1)$ sectors. As both the $(0,1)$ sector and the log part of the $(1,0)$ sector (i.e. the $(1,0)[1]$ sector) contain just one coefficient, we can directly subtract these parts. The $(1,0)[0]$ coefficients however grow asymptotically, and therefore we have to perform a Borel-Padé summation on this part. This leads to the new sequence
\begin{align}\label{eq:deltak1bandc}
\delta_k^{(1)}
&= d_k^{(0,0)}
- \tilde{d}_0^{(1,0)[1]}\frac{\Ga(k-\beta)\psi(k-\beta)}{A_1^{k-\beta}}
- \frac{\Ga(k-\beta)}{A_2^{k-\beta}}\tilde{d}_0^{(0,1)[1]}
- \frac{\Ga(k-\beta)}{A_1^{k-\beta}}\cS[P^{(1,0)}]\left(\frac1k\right)\,,
\end{align}
with $\beta=-1$ and $A_1=-A_2=1$ and the coefficients $\tilde{d}_0^{(1,0)[1]}$ and $\tilde{d}_0^{(1,0)[0]}$ given in \eqs{101sectorDiagramb}{010sectorDiagramb} respectively. A numerical analysis on the imaginary part of $\delta_k^{(1)}$ yields a contribution to the $(2,0)$ sector:
\begin{equation}\label{eq:imPartDiagramb20}
\img\,\text{Im}\left[\delta_k^{(1)}\right]
\sim \pm6\pi\img\frac{\Ga(k+1)}{2^{k+1}} + \ord{3^{-k}}\,.
\end{equation}
From the real part of $\delta_k^{(1)}$ we extract for both sectors $\beta_{2,0}=\beta_{0,2}=-1$. Furthermore, we extract a single non-perturbative coefficient for the $(0,2)$ sector,
\begin{equation}
\tilde{d}_0^{(0,2)[0]} = -\frac16+\frac29\log(2)\,, 
\end{equation}
and the first few coefficients of the $(2,0)$ sectors. We list the first four:
\begin{equation}
\tilde{d}_0^{(2,0)[0]} = -2\,, 
\qquad
\tilde{d}_1^{(2,0)[0]}= -\frac{14}{3}\,,
\qquad
\tilde{d}_2^{(2,0)[0]} = -\frac{5}{2}\,,
\qquad
\tilde{d}_3^{(2,0)[0]} = -\frac{7}{6}-8\zeta_3\,.
\end{equation}
We have obtained these coefficients numerically up to at least 9 decimal places, as well as some further ones not displayed here, after which we inferred the analytic expression. Furthermore, notice that this diagram does not have a $\log$ sector, i.e. all the coefficients $\tilde{d}_h^{(2,0)[1]}$ vanish. 

We can improve on these results using again the method of alien derivatives acting on the convolution integral to find all $\tilde{d}_h^{(2,0)[0]}$ exactly. We repeat the convolution integral, \eqs{convoMethodDiagramb}{convoMethodDiagramb2}, for this diagram:
\begin{align}
\cB[D_b^{(0,0)}](u)
=\frac{-6\Ga(1+u)}{(1-u)\Ga(3-u)}\cB[\Psi^{(0)}](u),
\end{align}
with:
\begin{equation}
\cB[\Psi^{(0)}](u)
=\int_0^u du_1\, \cB[F](u_1)\cB[F](u-u_1),
\quad\text{with}\quad
\cB[F](u) 
= \frac{\Ga(1-u)}{(2-u)\Ga(2+u)}\,.
\end{equation}
In order to extract the $(2,0)$ sector, we need the expansion of $\cB[\Psi^{(0)}]$ around $u=2$, i.e. we need the second non-perturbative sector $\Psi^{(2)}$. 
As discussed in \sec{convoInt}, this implies we need
\begin{equation}\label{eq:delta1FandDelta2F}
\De_1F = \frac12(2\pi\img),
\qquad\text{and}\qquad
\De_{2}F = -\frac16(2\pi\img)\left(\frac{1}\al +\frac{17-12\ga_E}{6}\right)\,,
\end{equation}
which is read off from the expansions around the poles of $\cB[F]$ positioned at $u=1$ and $u=2$.
We notice that, similar to the LO Adler function, $\De_1^2F=0$. However, $\De_2F$ is now non-vanishing.  
This yields
\begin{align}
\left(S_1^\Psi\right)^2\Psi^{(2)}
&= 2\cdot \frac12 (\De_1F)^2 + 2 \cdot F\De_2F\nn\\
&=\frac14(2\pi\img)^2
-\frac13(2\pi\img)\left(\frac{1}\al +\frac{17-12\ga_E}{6}\right)  
\sum_{n=0}^\infty F_n\al^{n+1}\nn\\
&= 2\pi\img\left(\frac12 \pi\img - \frac{1}{3}F_0\right)
- \frac{1}{3}(2\pi\img)\sum_{n=0}^\infty \left(
F_{n+1}
+\frac{17-12\ga_E}{6}\sum_{n=0}^\infty F_{n}
\right)\al^{n+1}\,.
\end{align}
Recalling the discussion at the end of \sec{convoDoublePole}, in particular \eqs{doublePoleConvoTrick}{doublePoleConvoTrick2}, we can write the Borel transform of $\Psi^{(2)}$ in terms of $\cB[F](u)$ and its derivative $\cB[F]'(u)$ w.r.t. $u$. Therefore, the Borel transform $\cB[\Psi^{(0)}]$ around $u=2$ reads
\begin{align}
\cB[\Psi^{(0)}](u)\Big|_{u=2}
&= \frac{\frac12\pi\img-\frac13F_0}{u-2}-\left(S_1^\Psi\right)^2\cB[\Psi^{(2)}](u-2)\frac{\log(1-\frac{u}{2})}{2\pi\img}+...\\
&= \frac{\frac12\pi\img-\frac13F_0}{u-2}
+ \frac{1}{3}\left(\cB[F]'(u-2)+\frac{17-12\ga_E}{6}\cB[F](u-2)\right)
\log\left(1-\frac{u}{2}\right)
+...\,.\nn
\end{align}
Adding back in the prefactor to get $\cB[D_b](u)$ and $F_0=\frac12$ yields, after a brief calculation\footnote{Note that in \eq{imPartDiagramb20} the factor $6\pi \img$ appears with an ambiguous sign. As usual, in large order formulas this imaginary ambiguity should not contribute to the perturbative coefficients; it will be `canceled' by the $6\pi i$ terms in the present equation. One may be worried that here, no ambiguous sign appears in front of the $6 \pi \img$, but this is a result of the fact that we have implicitly chosen the `upper contour' by applying a particular form of the Stokes automorphism and the resulting alien derivatives.},
\begin{align}
\cB[D_b^{(0,0)}](u)\Big|_{u=2}
&=\frac{-6\Ga(1+u)}{(1-u)\Ga(3-u)}\cB[\Psi^{(0)}](u)\Big|_{u=2}+...\nn\\
&= \frac{6\pi\img-2}{2-u}
-S_{1,0}^2\cB[D_b^{(2,0)}](u-2)\frac{\log(1-\frac{u}{2})}{2\pi\img}
+...\,,\label{eq:sector2NLO2}
\end{align}
with 
\begin{equation}\label{eq:sector2NLO2BorelPart}
\frac{S_{1,0}^2}{2\pi\img}\cB[D_b^{(2,0)}](u-2)
= \frac{2u}{(1-u)(4-u)}\left(\frac{1}{4-u}+\frac{17-12\ga_E}{6}
-\psi(3-u)-\psi(u)
\right)\,.
\end{equation}
The exact coefficients $\tilde{d}_h^{(2,0)[0]}$ are obtained from the residue of the simple pole around $u=2$ in \eq{sector2NLO2}, and after applying an inverse Borel transform on \eq{sector2NLO2BorelPart}.

At this point, we want to come back to a left over point of the previous subsection: the resurgence of the $\tilde{d}_0^{(1,0)[0]}$ and $\tilde{d}_0^{(0,1)[0]}$ coefficients -- which we were not yet able to determine there -- from the convolution integral. (We shall see momentarily how the same reasoning also leads to the resurgence of the $\tilde{d}_0^{(0,2)[0]}$ coefficient.) 
In principle, as $\cB[F](u)$ only has singularities at positive integers of $u$, also its convolution integral $\cB[\Psi^{(0)}](u)$ only has possible singularities at positive integers of $u$. Therefore, with the techniques of studying the convolution integral with alien derivatives, it seems we can only make statements about the local expansions of $\cB[\Psi^{(0)}](u)$ around these values. However, the prefactor 
\begin{equation}
\frac{-6\Ga(1+u)}{(1-u)\Ga(3-u)}
\end{equation}
in front of this convolution integral has singularities at $u=1$ and the negative integers of $u$. Therefore, these singularities pick up regular terms of $\cB[\Psi^{(0)}](u)$ at these points.
In particular, the prefactor also leads to the pattern of singularities at negative integers in $u$ of $\cB[D_b^{(2,0)}](u-2)$. Furthermore, there is also a singularity at $u=1$ and therefore we notice that this sector contributes to the coefficients $\tilde{d}_0^{(1,0)[0]}$, $\tilde{d}_0^{(0,1)[0]}$ and $\tilde{d}_0^{(0,2)[0]}$. For example, expanding the Borel transform of the $(0,2)$ sector around $u=1$ yields
\begin{equation}
S_{1,0}^2\cB[D_b^{(2,0)}](u-2)\frac{\log(1-\frac{u}{2})}{2\pi\img}\Big|_{u=1} = \frac{13\log(2)}{9(u-1)}+...
\end{equation}
with the ellipsis denoting regular terms around $u=1$. We notice that this indeed corresponds with part of the coefficient $\tilde{d}_0^{(1,0)[0]}$ already obtained numerically in \eq{100sectorDiagramb}. The remaining part of this coefficient follows from the fact that we multiplied a pole at $u=1$ of the prefactor with the expansion of $\cB[\Psi^{(0)}[u]$ around $u=1$. 
Similarly, we can expand the same expression around $u=-1$ and $u=-2$ to find (parts of) the coefficients $\tilde{d}_0^{(0,1)[0]}$ and $\tilde{d}_0^{(0,2)[0]}$ as given above. In conclusion, we see in particular that knowing the subleading $(2,0)$ and $(0,2)$ sectors actually teaches us a bit more about the leading $(1,0)$ and $(0,1)$ sectors that we were not able to deduce from the perturbative $(0,0)$ sector alone.

Finally, we discuss the resurgence of the $(2,0)$ and $(0,2)$ sectors of diagram $(c)$. For this, we can use the same sequence as in \eq{deltak1bandc}, but now using the coefficients of the $(1,0)$ and $(0,1)$ sectors of diagram $(c)$ given in Eqs. \eqref{eq:101diagramc}-\eqref{eq:010diagramc}. A numerical analysis yields
\begin{equation}\label{eq:imdiagramc}
\img\,\text{Im}\left[\delta_k^{(1)}\right]
\sim \pm3\pi\img\frac{\Ga(k+2)}{2^{k+2}}\left(1-\frac{1}{k+1}\right) + \ord{3^{-k}}
\end{equation}
for the imaginary part. 
Similar to diagram $(b)$, there is only a single non-vanishing non-perturbative coefficient in the $(0,2)$ sector:
\begin{equation}
\tilde{d}_0^{(0,2)[0]} = -\frac{17}{72}+\frac13\log(2)\,.
\end{equation}
However, the $(2,0)$ sector of diagram $(c)$ has a $\log$ contributing in the transseries, given by the non-perturbative coefficient
\begin{equation}
\tilde{d}_0^{(1,0)[1]} = -2\,.
\end{equation}
Again we list the first four coefficients of the $(2,0)[0]$ sector and we show in a moment how higher coefficients are obtained using the convolution method:
\begin{equation}\label{eq:200diagramc}
\tilde{d}_0^{(2,0)[0]} = 4-2\ga_E+\frac14\log(3)\,, 
\quad
\tilde{d}_1^{(2,0)[0]}= \frac13\,,
\quad
\tilde{d}_2^{(2,0)[0]} = \frac{11}{12}\,,
\quad
\tilde{d}_3^{(2,0)[0]} = -\frac{43}{12}+4\zeta_3\,.
\end{equation}
The numerical analysis on $\de_k^{(1)}$ agrees with Eqs. \eqref{eq:imdiagramc}-\eqref{eq:200diagramc} up to at least 11 decimal places.

These same results are reproduced using the convolution integral method. We repeat this integral, \eqs{convoMethodDiagramc1}{convoMethodDiagramc2}, 
\begin{align}
\cB[D_c^{(0,0)}](u)
=\frac{-6\Ga(u)}{(1-u)(2-u)\Ga(3-u)}\cB[\Phi^{(0)}](u),
\end{align}
with 
\begin{equation}
\cB[\Phi^{(0)}](u)
=\int_0^u du_1\, \cB[F](u_1)\cB[G](u-u_1),
\qquad\text{with}\qquad
\cB[G](u) 
= \frac{u\Ga(1-u)}{\Ga(2+u)}\,,
\end{equation}
and $\cB[F]$ the same as for diagram $(b)$ given in \eq{convoMethodDiagramb2}.
Again, for the resurgence of the $(2,0)$ and $(0,2)$ sectors of diagram $(c)$, we need access the second non-perturbative sector of $\Phi$. To compute this this, we use $\De_1F$ and $\De_2F$ already given in \eq{delta1FandDelta2F}, and also
\begin{equation}
\De_1 G=\frac12(2\pi\img),
\qquad\text{and}\qquad
\De_2G=-\frac13(2\pi\img)\,.
\end{equation}
As $\De_1^2F=\Delta_1^2G=0$, we get
\begin{align}
\left(S_1^\Phi\right)^2\Phi^{(2)}
&=2\cdot\frac12\De_1F\De_1G+F\De_2G+G\De_2F\\
&=\frac14(2\pi\img)^2
-\frac13(2\pi\img) \sum_{n=0}^\infty F_n\al^{n+1}
-\frac16(2\pi\img)\left(\frac{1}\al +\frac{17-12\ga_E}{6}\right)\sum_{n=0}^\infty G_n\al^{n+1}\nn\\
&= (2\pi\img)\left(\frac12\pi\img - \frac{1}{6}G_0\right)
-\frac16(2\pi\img) \sum_{n=0}^\infty \left(
2F_n
+ G_{n+1}
+\frac{17-12\ga_E}{6} G_{n}
\right)\al^{n+1}.\nn
\end{align}
We can write the Borel transform $\cB[\Phi^{(2)}](u)$ in terms of $\cB[F](u)$, $\cB[G](u)$, and $\cB[G]'(u)$. This yields
\begin{align}
\cB[\Phi^{(0)}](u)\Big|_{u=2}
&= \frac{\frac12\pi\img-\frac16G_0}{u-2}-\left(S_1^\Phi\right)^2\cB[\Phi^{(2)}](u-2)\frac{\log(1-\frac{u}{2})}{2\pi\img}+...\\
&= \frac{\frac12\pi\img-\frac16G_0}{u-2}
+ \frac16\bigg(2\cB[G](u-2)+\cB[G]'(u-2)\nn\\
&\hspace{4cm}
+\frac{17-12\ga_E}{6}\cB[G](u-2)\bigg)
\log\left(1-\frac{u}{2}\right)
+...\,.\nn
\end{align}
Adding back in the prefactor to get $\cB[D_c](u)$ and $G_0=0$ yields
\begin{align}
\cB[D_c^{(0,0)}](u)\Big|_{u=2}
&=\frac{-6\Ga(u)}{(1-u)(2-u)\Ga(3-u)}\left[\cB[\Phi^{(0)}](u)\Big|_{u=2}+...\right]\nn\\
&= -\frac{3\pi\img}{(2-u)^2}-\frac{3\pi\img}{2-u}
-S_{1,0}^2\cB[D_c^{(2,0)}](u-2)\frac{\log(1-\frac{u}{2})}{2\pi\img}+...\,,
\label{eq:expansionAroundLog(2-u)c}
\end{align}
with 
\begin{equation}
\frac{S_{1,0}^2}{2\pi\img}\cB[D_c^{(2,0)}](u-2)
= -\frac{1}{1-u}\left(\frac{1}{4-u}
-\frac{2}{2-u}+\frac{17-12\ga_E}{6}
-\psi(3-u)-\psi(u)
\right)\,.
\end{equation}
An inverse Borel transform indeed yields the coefficients $\tilde{d}_h^{(2,0)[0]}$ given above. Note that the pole in the prefactor in the first line of \eq{expansionAroundLog(2-u)c} has a pole at $u=2$, and therefore a constant term from the regular part between square brackets also contributes to the singular terms in the last line. This corresponds to the coefficient $\tilde{d}_0^{(2,0)[0]}$. With the same reasoning as for diagram $(b)$, we can get parts of the coefficients of $\tilde{d}_0^{(1,0)[0]}$, $\tilde{d}_0^{(0,1)[0]}$ and $\tilde{d}_0^{(0,2)[0]}$ by expanding the sector $\cB[D_c^{(2,0)}]$ around the singular points $u=1$, $u=-1$ and $u=-2$ respectively.

This finishes the analysis and discussion of the resurgence of the $(2,0)$ and $(0,2)$ sectors. In our analysis, we saw that for diagram $(b)$ and $(c)$ the $(2,0)$ sector was more interesting from a resurgence point of view, with resurgence relations that involve multiple non-perturbative sectors. In the next two subsections, we study some further non-perturbative sectors and draw the alien lattices for these diagrams.

\subsection{Furter non-perturbative sectors and alien derivative structure}
In the previous two subsections, we have focused on the resurgence of the $(1,0)$, $(0,1)$, $(2,0)$ and $(0,2)$ sectors from the perturbative $(0,0)$ sector. As already observed in these sections, the coefficients of these non-perturbative sectors show asymptotic growth, so these sectors themselves have resurgence relations to other non-perturbative sectors. We will see momentarily that the resurgence relations know more about the full transseries due to the fact that at order $1/N_f^2$ also non-perturbative sectors like $(n,m)$ with $n\neq0$ and $m\neq0$ will appear, something we have not observed before.  

Let us once again start our discussion with diagram $(a)$. In \eqs{borelDiagramaAround1}{borelDiagramaAround1LogTerm}, we saw how in the Borel plane of $\cB[D_a^{(0,0)}](u)$, the $(1,0)$ sector resurges in the form 
\begin{equation}
\frac{S_{1,0}}{2\pi\img}\cB[D^{(1,0)[0]}](u-1)
= \frac{3}{(2-u)^2}+\frac{3}{2(2-u)} - \frac{1}{2u} + \frac{1}{3-u}\,.
\end{equation}
Likewise, one finds the Borel transform of the $(0,1)$ sector:
\begin{equation}\label{eq:BorelDiagrama01Sector}
\frac{S_{0,1}}{2\pi\img}\cB[D^{(0,1)[0]}](u+1)
= -\frac{1}{4(2-u)}-\frac{2}{3(1-u)^2}+\frac{2}{9(1-u)} - \frac{1}{2u} + \frac{1}{36(2+u)}\,.
\end{equation}
In particular, we notice that these expressions have a singularity at $u=0$. One can wonder why the perturbative $(0,0)$ sector does not contain this singularity in its own Borel transform. The reason is that, to get the $(0,0)$ sector, one has to include the logarithms:
\begin{equation}
\frac{S_{1,0}}{2\pi\img}\cB[D^{(1,0)[0]}](u-1)\log(1-u)\Big|_{u=0}
= \frac12+\ord{u}
\end{equation}
and
\begin{equation}
\frac{S_{0,1}}{2\pi\img}\cB[D^{(0,1)[0]}](u+1)\log(1+u)\Big|_{u=0}
= -\frac12+\ord{u}\,.
\end{equation}
That is, as seen from the point of view of the $(0,0)$ sector, the two constant terms that would contribute to a singularity at $u=0$ cancel. However, from the perspective of the $(1,0)$ and $(0,1)$ sectors, the individual singularities contribute to the large order growth of these sectors. Note that what these sectors see is {\em not} the $(0,0)$ sector again, as that sector has a vanishing constant term (see \eq{diagramaPerturbativeCoeffs}) and moreover, as in examples with bridge equations, one would not expect a sector like $(1,0)$ to be able to detect the $(0,0)$ sector. Thus, we conclude that what we find here is the contribution of a $(1,1)$ sector, i.e. we get
\begin{equation}
D^{(1,1)[0]} = \frac12\,,
\end{equation}
which is a sector containing only a single non-perturbative coefficient. 
This relation is best viewed in the alien lattice, for which we now learn that it must contain the following arrows:
\begin{equation*}
\vcenter{\hbox{\includegraphics[width=.45\textwidth]{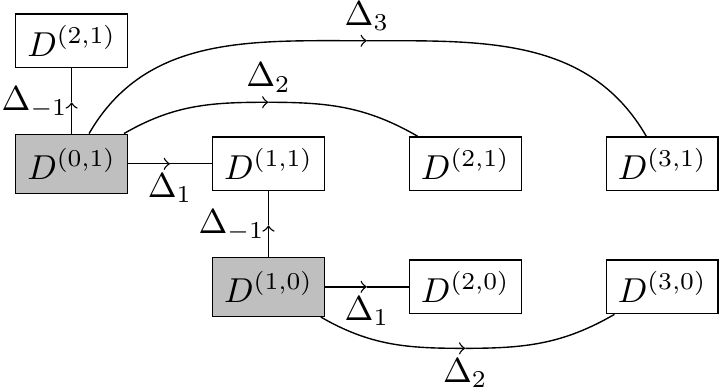}}}
\end{equation*}
Here we also included the motions towards other non-perturbative sectors. Let us make two more remarks about this diagram.

First of all, observe that from the $(0,1)$ sector we have also drawn alien derivative arrows towards a $(2,1)$ and a $(3,1)$ sector. The existence of a $(2,1)$ sector can be argued in a similar way as above and we will show this explicitly in a moment, by looking at the singularities in the relevant Borel transforms and observing that these do not correspond to sectors that are already known. The same argument for a $(3,1)$ sector is on less solid ground, as the coefficient we find for that sector is also contained in the relevant perturbative coefficient in the $(2,0)$ sector, and so we cannot be certain that it is indeed the former sector that the $(1,0)$ sector detects. Thus, for now the existence of a $(3,1)$ sector is a conjecture, which we hope to come back to in future work.

Secondly, notice the alien derivative $\De_1$ pointing from the $(1,0)$ sector to the $(2,0)$ sector; this alien derivative was discussed and explained around \eq{imgcontr}. In this diagram, we drew the $(2,0)$ sector with a white, solid box, which would imply that it is not an asymptotic series, contrary to the fact that we already know from the previous subsection that this {\em is} an asymptotic sector. The reason for this representation here is that the $(1,0)$ sector only sees a small, non-asymptotic part of the $(2,0)$ sector. 

\bigskip
\noindent
Similarly, we observe that the Borel transform of the $(0,1)$ sector has a double pole at $u=1$, whereas this does not resurge from the $(0,0)$ sector directly, i.e.\ in our analysis we only found a single pole given by the coefficient $\tilde{d}_0^{(1,0)[0]}=\frac13-3\gamma_E+\frac{13}{9}\log(2)$. A closer inspection of the $(2,0)$ sector reveals the same double pole at $u=1$, i.e. we have
\begin{equation}
\frac{S_{1,0}^2}{2\pi\img}\cB[D^{(2,0)[0]}](u-1)
= -\frac{2}{3(1-u)^2}+\frac{11}{9(1-u)} +\frac{1}{3-u}- \frac{2}{9(4-u)}\,,
\end{equation}
with the same coefficient $\frac23$ as in \eq{BorelDiagrama01Sector}.
Again, from the point of view of the $(0,0)$ sector, one has to include the logarithms:
\begin{equation}
\frac{S_{1,0}^2}{2\pi\img}\cB[D^{(2,0)[0]}](u-1)\log\left(1-\frac{u}{2}\right)\Big|_{u=1}
= \frac{2\log(2)}{3(1-u)^2} +\frac{\frac23+\frac{11}{9}\log(2)}{1-u}+...
\end{equation}
and 
\begin{equation}
\frac{S_{0,1}}{2\pi\img}\cB[D^{(0,1)[0]}](u+1)\log(1+u)\Big|_{u=1}
= -\frac{2\log(2)}{3(1-u)^2} +\frac{-\frac13+\frac{2}{9}\log(2)}{1-u}+...\,.
\end{equation}
This means that the $(0,0)$ sector does not see the double pole, but the $(2,0)$ and $(0,1)$ sectors independently do. Furthermore, we observe that the $(2,0)$ and the $(0,1)$ sectors see the same double pole, and we thus conclude that this is part of the $(2,1)$ sector. 

We notice however that the single pole, as seen from the point of view of the $(2,0)$ and $(0,1)$ sectors has a different residue in the two cases, i.e.\ $11/9$ and $2/9$ respectively. This is a result of the fact that both sectors have resurgence relations towards both the $(1,0)$ sector and the $(2,1)$ sector. As the $(2,0)$ sector sees the $(2,1)$ and $(1,0)$ sectors with different weights compared to the $(0,1)$ sector (which depend on different Stokes constants), we conclude that it is hard to distinguish which part of the single pole at $u=1$ is part of the $(2,1)$ sector and which part corresponds to the $(1,0)$ sector. Being able to distinguish between the two would allow one to extract an interesting relation between the Stokes constants of the Adler function. This is another point that we hope to come back to in future work.

Again, these relations are best viewed in terms of the alien lattice, which as we now have learned also contains the following ingredients:
\begin{equation*}
\vcenter{\hbox{\includegraphics[width=.6 \textwidth]{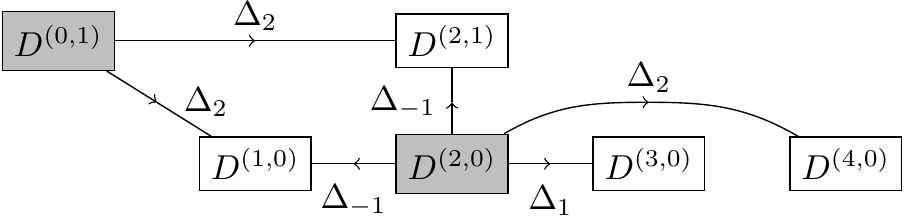}}}
\end{equation*}
The full alien lattice for diagram $(a)$ can now be obtained by combining the previous two diagrams with the similar diagram for the $(0,0)$ sector:
\begin{equation*}
\vcenter{\hbox{\includegraphics[width=.6\textwidth]{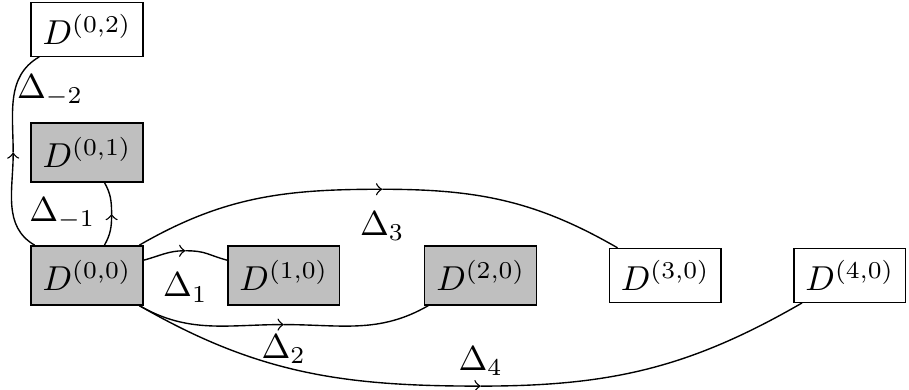}}}
\end{equation*}
where we also inlcuded motions from the perturbative $(0,0)$ sector towards a $(3,0)$ and $(4,0)$ sector (see also the discussion in the next subsection).

For diagrams $(b)$ and $(c)$, we have seen that only the $(1,0)$ and $(2,0)$ sectors are asymptotic. To be precise, we derived for diagram $(b)$ the following Borel transform of the $(1,0)$ sector:
\begin{equation}
\frac{S_{1,0}}{2\pi\img}\cB[D_b^{(1,0)[0]}](u-1) 
= \frac{6}{2-u}-\frac{3}{3-u}\,.
\end{equation}
The singularities at $u=2$ and $u=3$ can be seen as part of the $(2,0)$ and $(3,0)$ sectors, i.e.\ in terms of alien lattices:
\begin{equation}
\vcenter{\hbox{\includegraphics[width=.3\textwidth]{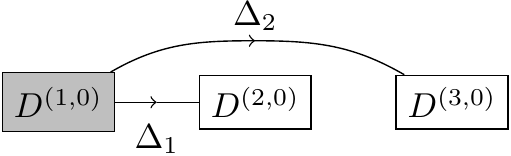}}}
\end{equation}
Diagram $(c)$ on the other hand is slightly different. The Borel transform of its $(1,0)$ sector reads
\begin{equation}
\frac{S_{1,0}}{2\pi\img}\cB[D_c^{(1,0)[0]}](u-1) 
= \frac{1}{2u}+\frac{3}{(2-u)^2}+\frac{3}{2(2-u)}+\frac{1}{2(3-u)}\,.
\end{equation}
Again we notice a singularity at $u=0$ which is not seen by the perturbative $(0,0)$ sector. We therefore conclude that this is the effect of a non-zero $(1,1)$ sector. As before, this new sector is not an asymptotic sector, but it only contains one non-perturbative coefficient. This yields the following part of the alien lattice:
\begin{equation*}
\vcenter{\hbox{\includegraphics[width=.3\textwidth]{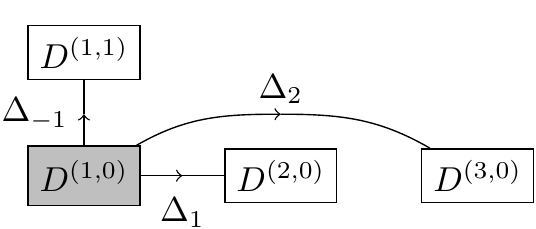}}}
\end{equation*}
We already observed earlier that the $(2,0)$ sector of both diagram $(b)$ and $(c)$ has an infinite number of singularities in $u$ at both positive and negative integers. 
Coming back to our discussion in \sec{prefactor}, we want to mention here that from the convolution integral alone, \eq{convoMethodDiagramc2}, we would not have expected singularities at $u=0$ or negative values of $u$. However the prefactor in front of the convolution integral, \eq{convoMethodDiagramc2} adds singularities at non-positive $u$. It is therefore hard to distinguish which part of the pole at $u=1$ for diagram $(c)$ is related to the $(0,1)$ sector or the $(1,2)$ sector. 
First of all, the singularities at positive integers of $u$ follow mainly from the convolution integral, but they also arise from the prefactors in front of the convolution integral since these add a pole at $u=1$ for diagram $(b)$ and a pole at $u=1,2$ for diagram $(c)$. We therefore expect that this contributes to both a $(1,0)$ and a $(2,1)$ sector as seen from the $(2,0)$ sector. Likewise, the prefactor adds an infinite number of singularities at negative $u$, leading to both $(0,n)$ as well as $(2,2+n)$  sectors, for $n\geq1$, or in terms of motions on the alien lattice:
\begin{equation*}
\vcenter{\hbox{\includegraphics[width=.6\textwidth]{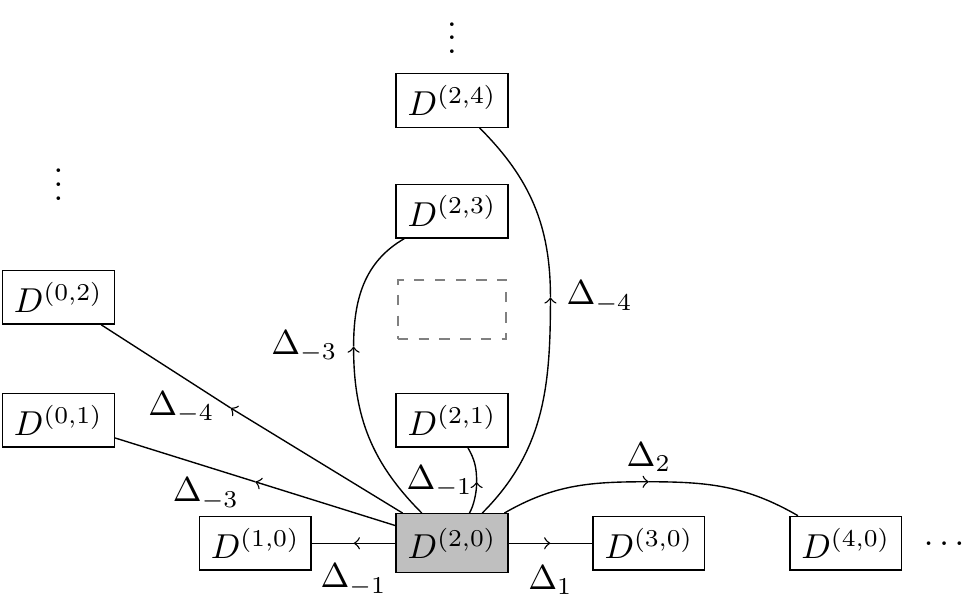}}}
\end{equation*}
All in all, we see that at order $1/N_f^2$ in the number of flavours, the Adler function displays many more interesting resurgence features than were present at order $1/N_f$.

\subsection{Overview: transseries sectors and relations}
We have discussed the resurgence of the first non-perturbative sectors ($(1,0)$ and $(0,1)$) as well as the second non-perturbative sectors ($(2,0)$ and $(0,2)$) of diagrams $(a)$-$(c)$, plus their own resurgence towards other sectors. We now briefly sketch the resurgence structure of higher sectors for these diagrams.  

Looking at the closed form expression for diagram $(a)$ in \eq{BorelTransNLOb}, we notice from the singularity structure that the only remaining non-vanishing sectors are the $(3,0)$ and $(4,0)$ sectors. Furthermore, as these singularities are not branch cuts but just simple poles, we conclude that they describe non-asymptotic sectors containing just a single non-vanishing non-perturbative coefficient.

From our discussion in \sec{convoEquidistant}, it might seem that we need to act with many alien derivatives to get access to the higher sectors of diagrams $(b)$ and $(c)$. However, the constituents of the convolution integral $F$ and $G$ of diagram $(b)$ and $(c)$ (recall \eqs{convoMethodDiagramb}{convoMethodDiagramc1} respectively), have $\De_\w^nF=\De_\w^nG=0$ for $n>1$. This implies that we only get non-vanishing contributions by acting with at most two alien derivatives -- one for each factor in the convolution product. Still, that leaves us with quite a few options to consider, but as we see momentarily, we can categorise all remaining options into a few classes. We discuss each of these classes case by case. Furthermore, we will see that not all cases contribute to the resurgence of higher sectors.

Let us first discuss the higher non-perturbative asymptotic sectors of diagram $(b)$. The non-vanishing alien derivatives acting on $F$ of that diagram are given by
\begin{equation}
\De_1F = \frac12(2\pi\img)\,,
\qquad
\De_{2}F = -\frac16(2\pi\img)\left(\frac{1}\al +\frac{17-12\ga_E}{6}\right)\,,
\end{equation}
and
\begin{equation}
\De_nF = \frac{(-1)^n}{(2-n)(n+1)!(n-1)!}(2\pi\img)\,,
\qquad n\geq3\,.
\end{equation}
First of all, using the convolution method we get an asymptotic contribution to the $(n,0)$, $n\geq3$, sector via:
\begin{equation}
F\De_n F\,.
\end{equation}
One can show that this gives a contribution to $\cB[D_b^{(0,0)}](u)$ around $u=n$ proportional to
\begin{align}
&\frac{-6\Ga(1+u)}{(1-u)\Ga(3-u)}\cB[F](u-n)\log\Big(1-\frac{u}{n}\Big)\nn\\
&\hspace{2cm}
=\frac{-6\Ga(1+u)}{(1-u)\Ga(3-u)}\frac{\Ga(n+1-u)}{(n+2-u)\Ga(2+u-n)}\log\Big(1-\frac{u}{n}\Big)\,.
\end{align}
The ratio of gamma functions, is regular for $n\geq3$ and furthermore it cancels the pole at $u=1$. We therefore see that this asymptotic part of the $(n,0)$ sector is completely determined by the singularity at $u=n+2$, i.e. by the sector $(n+2,0)$.

Secondly, one gets contributions from acting with two alien derivatives:
\begin{equation}\label{eq:diagrambTwoAlienDerivatives}
\De_k\,F\De_{(n-k)}F\,.
\end{equation}
However, many of these cases do not lead to a contribution to the resurgence of diagram $(b)$ due to the prefactor in front of the convolution integral. That is, the contributions of \eq{diagrambTwoAlienDerivatives} that lead to a simple pole do not contribute as the 
prefactor contains a term $1/\Ga(3-u)$ which has zeros at $u=n$ for $n\geq3$. As a result, the only non-vanishing contribution follows from the case $k=2$ as this leads to a double pole at $u=n$. In particular, this results in a contribution to $\cB[D_b^{(0,0)}](u)$ at $u=n$ proportional to:
\begin{align}
\frac{-6\Ga(1+u)}{(1-u)\Ga(3-u)}\frac{1}{(n-u)^2}\,.
\end{align}
Combining the above observations leads to the contributions to the alien lattices shown in the second column of \tab{NLOsummary}.

\begin{table}[]
\renewcommand{\arraystretch}{2}
\centering
\vspace{-.7cm}
\makebox[\textwidth][c]{
\begin{tabular}{||c|c|c|c||}
\hline\hline
 & \includegraphics[width=.18\textwidth]{figures/adlerNLO1.pdf} & \includegraphics[width=.18\textwidth]{figures/adlerNLO2.pdf} & \includegraphics[width=.18\textwidth]{figures/adlerNLO3.pdf}
 \\
\hline\hline
\raisebox{1.5cm}{\rotatebox[]{90}{$(0,0)$}}
& \raisebox{.09\totalheight}{\includegraphics[width=.45\textwidth]{figures/alien_lattice_NLO_a-0-0.pdf}}
& \multicolumn{2}{|c||}{\includegraphics[width=.4\textwidth]{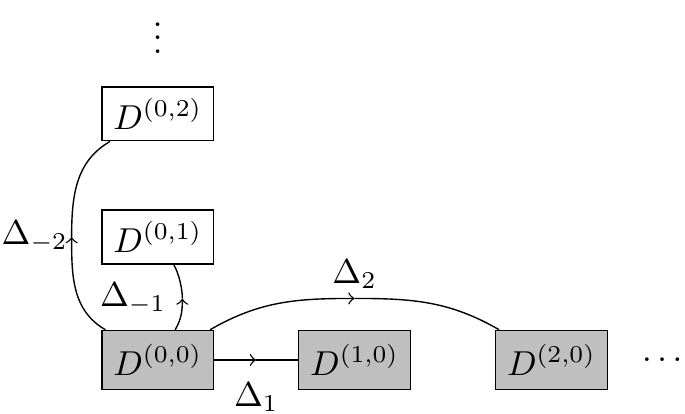}} \\
\hline
\raisebox{.8cm}{\rotatebox[]{90}{$(1,0)$}} 
& \includegraphics[width=.3\textwidth]{figures/alien_lattice_NLO_a-1-0.pdf}
& \includegraphics[width=.3\textwidth]{figures/alien_lattice_NLO_b-1-0.pdf}
& \includegraphics[width=.3\textwidth]{figures/alien_lattice_NLO_a-1-0.pdf} \\
\hline
\raisebox{1cm}{\rotatebox[]{90}{$(0,1)$}} 
& \includegraphics[width=.38\textwidth]{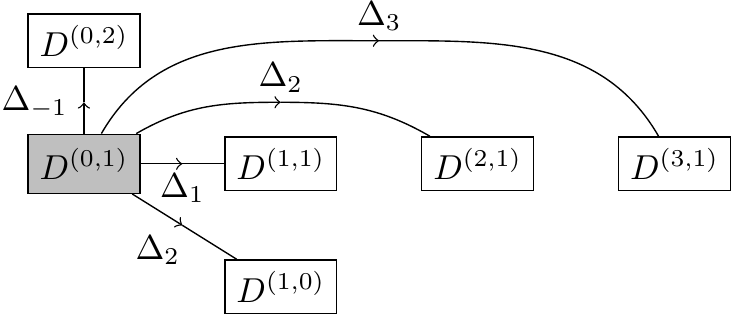}
& \multicolumn{2}{|c||}{\raisebox{1.7\totalheight}{\includegraphics[width=.07\textwidth]{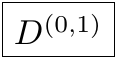}}} \\
\hline
\raisebox{2.4\totalheight}{\rotatebox[]{90}{$(2,0)$}} 
& \raisebox{.6\totalheight}{\includegraphics[width=.4\textwidth]{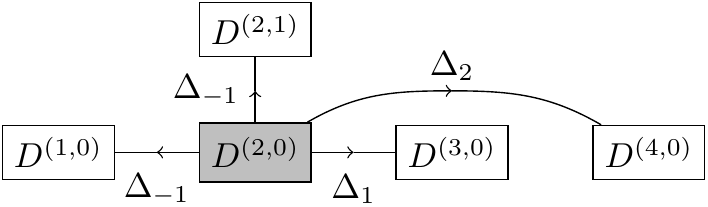}}
& \multicolumn{2}{|c||}{\includegraphics[width=.5\textwidth]{figures/alien_lattice_NLO_b-2-0.pdf}} \\
\hline
\raisebox{.15\totalheight}{\rotatebox[]{90}{$(0,2)$}}
& \includegraphics[width=.07\textwidth]{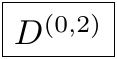} 
& \multicolumn{2}{|c||}{\includegraphics[width=.07\textwidth]{figures/alien_lattice_NLO_b-0-2.pdf}}  \\
\hline
\raisebox{.9\totalheight}{\rotatebox[]{90}{$(3,0)$}} 
& \raisebox{1.1\totalheight}{\includegraphics[width=.07\textwidth]{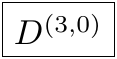}} 
& \raisebox{.5\totalheight}{\includegraphics[width=.2\textwidth]{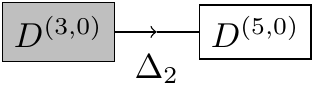}} 
& \includegraphics[width=.3\textwidth]{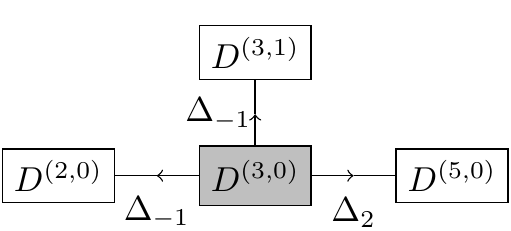} \\
\hline
\raisebox{.15\totalheight}{\rotatebox[]{90}{$(0,3)$}}
& - 
& \multicolumn{2}{|c||}{\includegraphics[width=.07\textwidth]{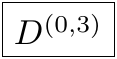}} \\
\hline
\raisebox{.22\totalheight}{\rotatebox[]{90}{$(n,0)$}}
& \includegraphics[width=.07\textwidth]{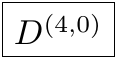} 
& \multicolumn{2}{|c||}{\includegraphics[width=.2\textwidth]{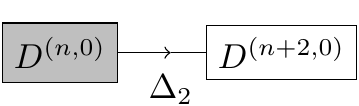} }\\
\hline
\raisebox{.15\totalheight}{\rotatebox[]{90}{$(0,n)$}}
& - 
& \multicolumn{2}{|c||}{\includegraphics[width=.07\textwidth]{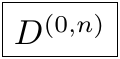}} \\
\hline\hline
\end{tabular}
}
\caption{Final summary of the relations between non-perturbative sectors at order $1/N_f^2$.}
\label{tab:NLOsummary}
\end{table}

The structure of higher sectors of diagram $(c)$ is similar to that of diagram $(b)$. The non-vanishing alien derivatives acting on $G$ are
\begin{equation}
\De_nG = \frac{n(-1)^n(2\pi\img)}{(n+1)!(n-1)!}\,,
\qquad
n\geq1\,.
\end{equation}
Acting with a single alien derivative on the product $FG$, we get asymptotic contributions to the $(n,0)$ sectors ($n\geq3$) from
\begin{equation}
F\De_nG+G\De_nF\,.
\end{equation}
First, we consider $n=3$, where $G\De_3F$ yields a contribution to $\cB[D_c^{(0,0)}](u)$ proportional to
\begin{equation}
\frac{\log(1-\frac{u}{3})}{8(2-u)}\,.
\end{equation}
Likewise, $F\De_3G$ yields a contribution proportional to
\begin{equation}
\left(\frac{1}{8(2-u)}+\frac{1}{4(5-u)}\right)\log(1-\frac{u}{3})\,.
\end{equation}
Notice in particular that the expansion of this around $u=2$ leads to part of the coefficient $\tilde{d}_0^{(2,0)[0]}$. 

For $n>3$, the only non-vanishing contribution of the convolution integral to the resurgence of diagram $(c)$ coming from a single alien derivative is given by $F\De_nG$. This yields a contribution proportional to 
\begin{equation}
\frac{\log(1-\frac{u}{n})}{(2+n-u)}\,.
\end{equation}
Similar to diagram $(b)$, when acting with two alien derivatives, one can show that only a double pole coming from the convolution integral contributes to the resurgence of diagram $(c)$. This means that the contribution has to come from $\De_2F\De_{n-2}G$, and we get a term proportional to
\begin{equation}
\frac{-6\Ga(u)}{(1-u)(2-u)\Ga(3-u)}\frac{1}{(n-u)^2}
\end{equation}
We have added the resulting motions on the alien lattice in \tab{NLOsummary}.

\bigskip
\noindent
This brings us to the end of our resurgence analysis of diagrams $(a)$, $(b)$ and $(c)$. The results are summarized in \tab{NLOsummary}. Compared to the LO Adler function, at order $1/N_f^2$ the $(n,0)$ sectors (and also the $(0,1)$ sector for diagram $(a)$) are asymptotic sectors with their own resurgence relations towards further non-perturbative sectors. In particular we find sectors like $(n,m)$ with $n\neq0$ and $m\neq0$. We notice however that the asymptotic non-perturbative sectors only have resurgence relations to sectors that contain only one or two coefficients, quite similar to asymptotic perturbative series of the LO Adler function. 
We expect that at higher order in the flavour expansion, the asymptotic non-perturbative sectors will have resurgence towards full asymptotic sectors, thus yielding an alien lattice with more and more motions between sectors.

\section{Conclusion and outlook}
\label{sec:conclusion}
In this paper we have analyzed the transseries of the Adler function, which includes non-perturbative effects,  to order $1/N_f^2$. In our description of the analysis we aimed to connect two communities:  we wished to introduce particle physicists to the techniques of resurgence, but also wanted to show the resurgence community that in particle physics and phenomenology problems, many of the techniques and results known from toy models still appear, in addition to new phenomena.

Let us summarize our main findings.
First, our resurgence analysis of renormalon singularities in the Borel transform of the Adler function led us  to construct a minimal, non-trivial {\em two-parameter} transseries, which -- in contrast to the $1/N_f$ Adler function -- has asymptotic coefficient growth in many of its non-perturbative sectors. 

Second, we found that resurgent properties of the Adler function are best expressed using the calculus and lattice structure  based on the alien derivative (as advocated in \cite{Aniceto:2018bis}). In particular, we introduced in \sec{convoInt} a way to study the alien structure of transseries including renormalons using a convolution technique. Applying this technique in \sec{adlerNLO} gave us a way to extract resurgence properties without having immediate access to the exact Borel transform of a given asymptotic expansion. Furthermore, from this technique we were able to extract locally the form of the Borel transformed result.

Third, while the exact Borel transform of the $\cO(1/N_f)$ Adler function and its singularities has been known for a long time \cite{Broadhurst:1992si,Beneke:1992ch} (reviewed and rederived in \sec{adlerLO}), we constructed a complete resummed transseries expression \eqs{DLOIR2}{DLOUV2} for the Adler function at order $1/N_f$ in the original $\al$ variable.
At this order in $1/N_f$ there are only poles in the Borel plane: UV double poles at negative integer values of the (rescaled) Borel plane parameter $u$, and IR double poles at positive integer values with the exceptions of an absent pole at $u=+1$ and the occurence of a single pole at $u=+2$.

Our fourth finding is a more intricate singularity structure in the Borel plan at order $1/N_f^{2}$, involving logarithmic branch cuts.
  In particular, using the Borel-Padé technique and a large order analysis of the coefficients that we computed, we found many of these branch cuts at IR singularities and a few at UV singularities. Of course, since we have only computed a limited number of diagrams, we cannot state with full certainty which singularities have branch cuts (and therefore which sectors have asymptotic expansions) in the full Adler function at all orders in $1/N_f$. However, we conjecture that all singularities in fact become of this sort - except of course a potential IR singularity at $u=+1$ which on physical grounds (due to the absence of a corresponding operator in the OPE \cite{Parisi:1978az,Mueller:1984vh}) we expect to remain absent.

A fifth and final finding for the order $1/N_f^2$ transseries is that overlapping poles and branch point singularities in the Borel plane lead to $\log(\al)$ transmonomials. These factors were already anticipated in \cite{Beneke:1995qq}; we indeed find that they occur. These transmonomials indicate a new type of power correction that we also expect to persist at higher orders in $1/N_f$.

The picture that emerges from these findings is that the structure of our transseries, displayed in detail for order $1/N_f^2$ in \tab{NLOsummary}, is interestingly different from that occurring in many models studied in the literature. In particular, we have seen that for the Adler function, `multiple forward alien derivatives' such as $\De_2$ give nonzero results, whereas in most simple models that have a bridge equation, only $\De_1$ provides a nonvanishing `forward motion'. Furthermore, even though opposite non-perturbative exponentials $\pm A$ as well as logarithmic transmonomials appear in the Adler function transseries, the transseries turns out {\em not} to be resonant.

\bigskip
\noindent With our analysis and aim to start bridging the gap between the particle physics and resurgence communities, many open questions and opportunities for further research suggest themselves. Let us mention a few interesting directions.

To start verifying further the conjecture that for the full Adler function all non-perturbative sectors are asymptotic, one could take a closer look at specific diagrams at higher orders in $1/N_f$ and study their resurgence properties using the techniques developed in this paper. Ultimately, it would be interesting to also be able to perform a sum over all orders in $1/N_f$. For example, this may allow one to see if the interesting shift in renormalon pole locations of \cite{Marino:2021dzn} also occurs for the Adler function.

A second interesting area for further research concerns the structure of the Adler transseries. We have seen that for our purposes, two parameters for this transseries suffice, but it is conceivable that one would in fact require  more parameters. A related problem is that our Adler transseries has a nonvanishing `two steps forward' alien derivative $\De_2$; a `three-steps forward' $\De_3$ etc. This may already hint at a situation where sectors like the $e^{-2A/\al}$ sector come with their own transseries paramaters. Whether such parameters should be viewed as integration constants that need to be fixed, or as true non-perturbative expansion parameters is an interesting open question (cf.\ \cite{Gu:2022fss}, where a similar interplay was observed in a different model). One way to answer such questions would be using a bridge equation, but unfortunately we do not have one at our disposal. Nevertheless, the case of two parameters being sufficient seems supported by our finding that the Adler function has only two Stokes lines and therefore two Stokes automorphisms.

In conclusion, the Adler function remains an  interesting object to study using resurgence techniques. It is sufficiently complicated to show features not displayed by  many simpler models, but is also amenable to further perturbative studies that might uncover some of the subtleties mentioned. We expect many more lessons and surprises to lie ahead.

\acknowledgments
\noindent 
We thank Jos Vermaseren for assistance with \texttt{FORM} and \texttt{SUMMER} and Alexander van Spaendonck for useful discussions.

\newpage
\appendix
\section{Master integrals}
\label{app:masters}
In this appendix we discuss the master scalar integrals we used in the calculation of the diagrams considered in this paper. The diagrams given in Figs. \ref{fig:adlerLO1}, \ref{fig:adlerNLO1}, \ref{fig:adlerNLO2} and \ref{fig:adlerNLO3} can be computed using standard one-loop techniques and the one-loop massless master integral
\begin{align}
  \label{eq:6}
\hat{I}^{(1,2)}(n_1,n_2) 
&\equiv (-q^2)^{n_{12}-d/2} \int \frac{d^dk}{{\img \pi^{d/2}}}\frac{1}{(-k^2)^{n_1}(k-q)^{2n_2}}\nn\\
&= \frac{\Ga(n_{12}-\frac{d}{2})}{\Ga(d-n_{12})}
\frac{\Ga(\frac{d}{2} - n_1)}{\Ga(n_1)} 
\frac{\Ga(\frac{d}{2} - n_2)}{\Ga(n_2)}\,,
\end{align}
with $d$ the number of space-time dimensions. We have adopted the following notation: the superscripts for an integral $\hat{I}^{(l,n)}$ indicate the number of loops and the number of propagators. The prefactor ensures that the result is independent of $(-q^2)$ and we do not get factors of $\pi$. Furthermore, we introduced the shorthand notation $n_{12} = n_1+n_2$ with its natural extension to the cases with more indices. 
\bigskip

\noindent
For the calculation of the diagrams in \figs{adlerLO2}{adlerNLO4}, we need to consider the massless two loop two-point function
\begin{align}
\hat{I}^{(2,5)}&(n_1,n_2,n_3,n_4,n_5) \nn\\
&\equiv (-q^2)^{n_{12345}-d}\int \frac{d^dk_1}{\img\pi^{d/2}}
\frac{d^dk_2}{\img\pi^{d/2}}
\frac{1}{(-k_1^2)^{n_1}(-k_2^2)^{n_2}(-k_3^2)^{n_3}(-k_4^2)^{n_4}(-k_5^2)^{n_5}},
\end{align}
where $k_3=k_2-q$, $k_4=k_1-q$ and $k_5=k_1-k_2$. 
A general representation for the massless two-point function with arbitrary powers $n_i$ of the propagators can be found in~\cite{Bierenbaum:2003ud}. For a historical overview of this integral, see \cite{Grozin:2012xi}. 
For the diagram with one bubble chain, given in \fig{adlerLO2}, we need the result for
\begin{equation}
\hat{I}^{(2,5)}(1,1,1,1,u+1). 
\end{equation}
We should mention here that we can work in $d=4$ dimensions, as the UV and IR divergences are regulated by the power $u$ in the propagator.  
This integral was first solved using IBP relations and the method of uniqueness in \cite{Kazakov:1983pk}, and the result in $d=4$ dimensions reads
\begin{equation}\label{eq:resultZakarov}
\hat{I}^{(2,5)}(1,1,1,1,u+1)
= -\frac{1}{2u}\Big[
\psi^{(1)}\Big(\tfrac{1+u}{2}\Big)
-\psi^{(1)}\Big(\tfrac{1-u}{2}\Big)
+\psi^{(1)}\Big(\tfrac{2-u}{2}\Big)
-\psi^{(1)}\Big(\tfrac{2+u}{2}\Big)
\Big],
\end{equation}
with $\psi^{(1)}(z)=\td{^2}{z^2}\log\Ga(z)$ the trigamma function.
\bigskip

\noindent
For the diagram in \fig{adlerNLO4} with two bubble chains, we need the two-loop master integral
\begin{equation}
\hat{I}^{(2,5)}(1,u_2+1,1,1,u_1+1). 
\end{equation}
This master was computed in~\cite{Kotikov:1995cw, Broadhurst:1996ur} for three arbitrary powers using IBP relations. The result reads\footnote{In~\cite{Broadhurst:1996ur}, the integral is denoted by $I_4(\al,\beta,\ga,\de)$.}
\begin{align}\label{eq:I4}
&\frac{\hat{I}^{(2,5)}(\al,\beta,1,1,\ga)}{\ga\de \hat{I}^{(1,2)}(1,\de+1)}\nn\\
&\hspace{1cm}=
\frac{\hat{I}^{(1,2)}(\al,\ga+1)}{d-3} S\Big(\tfrac{d}{2}-\al-1,\beta-1,\tfrac{d}{2}+\al-\delta-2,\delta-\beta\Big) + (\al\leftrightarrow\beta)
\end{align}
with $\de=\al+\beta+\ga-\frac{d}{2}$. The function $S$ encapsulates all symmetries of this integral and is given by:
\begin{align}
S(a,b,c,d) 
&= \frac{\psi(1-c)-\psi(c)}{H(a,b,c,d)} - \frac{1}{c} - \frac{b+c}{bc}F(a+c,-b,-c,b+d)
\end{align}
where $\psi(z)=\td{}{z}\log\Ga(z)$ is the digamma function and
\begin{align}
H(a,b,c,d)&= \frac{\Ga(1+a)\Ga(1+b)\Ga(1+c)\Ga(1+d)\Ga(1+a+b+c+d)}{\Ga(1+a+c)\Ga(1+a+d)\Ga(1+b+c)\Ga(1+b+d)},\\
F(a,b,c,d)&= \sum_{n=1}^\infty \frac{(-a)_n(-b)_n}{(1+c)_n(1+d)_n} = {}_3F_2(-a,-b,1;1+c,1+d;1)-1\,.\label{eq:hypergeometricfunction}
\end{align}
We have used this result to calculate the expansion of $\hat{I}^{(2,5)}(1,u_2+1,1,1,u_1+1)$ in $u_1$ and $u_2$. After the expansion, one still needs to carry out the infinite sum in \eq{hypergeometricfunction}. In order to perform the sums, we made use of the package \texttt{Summer} \cite{Vermaseren:1998uu}, which itself relies on \texttt{FORM} \cite{Vermaseren:2000nd,Ruijl:2017dtg}. 
The first few terms read
\begin{align}
\hat{I}^{(2,5)}(1,u_2+1,1,1,u_1+1)
&= 6\zeta_3
+5\zeta_5\left(3 u_1^2 + 3 u_1 u_2 + 2 u_2^2\right)\nn\\
&\hspace{1cm}
-6\zeta_3^2\left(u_1^2 u_2 + u_1 u_2^2\right)
+ \frac{525}{8}\zeta_7u_1^2u_2^2
+ \ORd{u_1^3,u_2^3}
\end{align}
Notice that in the limit $u_2\to0$ and $u_1\to u$, one recovers the expansion of \eq{resultZakarov} around $u=0$. For the calculation of the full diagram, we want to take the convolution integral, \eq{flavourExpansionConvoInt}, to get the Borel transform in terms of $u=u_1+u_2$. As
\begin{equation}
\int_0^udu_1\ u_1^m(u-u_1)^n = \frac{m!n!}{(m+n+1)!}u^{m+n+1},
\end{equation}
we see that if we want the Borel transform up to order $m+n+1$, we need all terms where the powers of $u_1$ and $u_2$ add up to $m+n$. 
We have added the expansion in terms of $u_1$ and $u_2$ of $\hat{I}^{(2,5)}(1,u_2+1,1,1,u_1+1)$ up to combined order $m+n=18$ in a separate file that is available from the authors upon request.
The result can be expressed in terms of regular zeta values and {\em multiple-zeta values} (MZVs) defined as
\begin{equation}
\zeta_{m_1,m_2,...,m_k} 
\equiv \sum_{i_1>i_2>...>i_k>0}\, \frac{1}{i_1^{m_1}i_2^{m_2}...i_k^{m_k}}\,,
\end{equation}
for positive integers $m_j$. The number of indices of the sums is called the {\em depth} and the sum of the powers, i.e. $m_1+...+m_k$, the {\em weight}. Of course, MZVs of depth one are just the regular zeta values.  
The \texttt{Summer} package contains all sums up to weight nine. In order to express the sums of higher weight in terms of MZVs, we made use of the tables in the multiple zeta value data-mine up to the maximal available weight 22 \cite{Blumlein:2009cf}. 
To give an example, at order $u_1^4 u_2^4$ one finds
\begin{equation}
\zeta_{5,3,3}
= \sum_{i_1=1}^\infty\frac{1}{i_1^5}
\sum_{i_2=1}^{i_1-1}\frac{1}{i_2^3}
\sum_{i_3=1}^{i_2-1}\frac{1}{i_3^3}\,,
\end{equation}
which in the data file are denoted by \texttt{z5z3z3}, with similar notations for sums of higher depth. 
In general, at higher orders in $u_1$ and $u_2$, the weight increases and one also finds MZVs of larger depth. 
For numerical values of these MZVs, up to 100 decimal places, we have used the EZ-Face calculator that can still be found at \href{http://wayback.cecm.sfu.ca/cgi-bin/EZFace/zetaform.cgi}{http://wayback.cecm.sfu.ca/cgi-bin/EZFace/zetaform.cgi} and checked its output up to several decimal places.

\section{Diagram momentum integrals}
\label{app:momentumIntegrals}
Recall the definition of the Adler function \eqref{eq:adlerintro}
\begin{equation}
D(Q^2) = 4\pi^2Q^2\td{\Pi(Q^2)}{Q^2}\,,
\end{equation}
where $\Pi(Q^2)$ is related to the correlation function of two vector currents $j_\mu=\bar \psi\ga_\mu \psi$ of massless quarks,
\begin{equation}
(-i)\int d^4xe^{-iqx} \mae{0}{T\{j_\mu(x)j_\nu(0)\}}{0} = (q_\mu q_\nu - \eta_{\mu\nu}q^2)\Pi(Q^2)\,,
\end{equation}
and where $Q^2=-q^2$. In order to extract $\Pi(Q^2)$, one can use 
\begin{equation}
\frac{\eta^{\mu\nu}}{(d-1)Q^2}\,,
\end{equation}
as the projector for $\Pi(Q^2)$. 
At order $1/N_f$, this yields the following integral expression for the diagram of \fig{adlerLO1},
\begin{align}
\cB[\Pi_a(Q^2)](u)
&= -\img\frac{(4\pi)^2(\mu^{2\eps})^2}{(d-1)Q^2} 
\int \frac{d^dk_1}{(2\pi)^d}\frac{d^dk_2}{(2\pi)^d} 
\cB\big[\al D^{\rho\si}(k_2)\big](u)\nn\\
&\hspace{2cm}\times
\tr\bigg[
\ga^\mu
\frac{\slashed{k}_1}{k_1^2}
\ga_\rho
\frac{(\slashed{k}_1-\slashed{k}_2)}{(k_1-k_2)^2}
\ga_\si
\frac{\slashed{k}_1}{k_1^2}
\ga_\mu
\frac{(\slashed{k}_1-\slashed{q})]}{(k_1-q)^2}
\bigg],\label{eq:LoopIntLO1}
\end{align}
where the Borel transform of the bubble chain, $\cB\big[\al
D^{\rho\si}(k_2)\big](u)$, is given in \eq{BorelChain}.
After performing the spinor trace, \eq{LoopIntLO1} can be written in terms of the master integral of \eq{6}.
Likewise, for the diagram of \fig{adlerLO2} we get
\begin{align}
\cB[\Pi_b(Q^2)](u)
&= -\img\frac{(4\pi)^2(\mu^{2\eps})^2}{(d-1)Q^2} 
\int \frac{d^dk_1}{(2\pi)^d}\frac{d^dk_2}{(2\pi)^d} 
\cB\big[\al D^{\rho\si}(k_1-k_2 )\big](u)\nn\\
&\hspace{2cm}\times
\tr\bigg[
\ga^\mu
\frac{\slashed{k}_1}{k_1^2}
\ga_\rho
\frac{\slashed{k}_2}{k_2^2}
\ga_\mu
\frac{(\slashed{k}_2-\slashed{q})}{(k_2-q)^2}
\ga_\si
\frac{(\slashed{k}_1-\slashed{q})]}{(k_1-q)^2}
\bigg]\,,
\end{align}
which after the fermion trace has been performed, reduces to the master integrals of \eqs{6}{resultZakarov}.

Let us consider the diagrams of \fig{AdlerNLO} studied in \sec{adlerNLO} at order $1/N_f^2$. For the diagram of \fig{adlerNLO1}, the loop integral reads
\begin{align}
\cB[\Pi_a(Q^2)](u_1,u_2)
&= -\img\frac{(4\pi)^4(\mu^{2\eps})^3}{(d-1)Q^2} 
\int \frac{d^dk_1}{(2\pi)^d}\frac{d^dk_2}{(2\pi)^d}\frac{d^dk_3}{(2\pi)^d}  
\cB\big[\al D^{\rho\si}(k_1-k_3)\big](u_1)\\
&\hspace{0cm}\times
\cB\big[\al D^{\al\beta}(k_1-k_2)\big](u_2)
\tr\bigg[
\ga^\mu
\frac{\slashed{k}_1}{k_1^2}
\ga_\rho
\frac{\slashed{k}_3}{k_3^2}
\ga_\si
\frac{\img\slashed{k}_1}{k_1^2}
\ga_\al
\frac{\slashed{k}_2}{k_2^2}
\ga_\beta
\frac{\slashed{k}_1}{k_1^2}
\ga_\mu
\frac{(\slashed{k}_1-\slashed{q})]}{(k_1-q)^2}
\bigg]\nn\,.
\end{align}
After the spinor trace has been performed, the remaining integrals reduce to the master of \eq{6}. Likewise, the diagrams of \figs{adlerNLO2}{adlerNLO3} read 
\begin{align}
\cB[\Pi_b&(Q^2)](u_1,u_2)\nn\\
&= -\img\frac{(4\pi)^4(\mu^{2\eps})^3}{(d-1)Q^2} 
\int \frac{d^dk_1}{(2\pi)^d}\frac{d^dk_2}{(2\pi)^d}\frac{d^dk_3}{(2\pi)^d}  
\cB\big[\al D^{\rho\si}(k_1-k_3)\big](u_1)
\cB\big[\al D^{\al\beta}(k_1-k_2)\big](u_2)\nn\\
&\hspace{2cm}\times
\tr\bigg[
\ga^\mu
\frac{\slashed{k}_1}{k_1^2}
\ga_\rho
\frac{\slashed{k}_3}{k_3^2}
\ga_\si
\frac{\slashed{k}_1}{k_1^2}
\ga_\mu
\frac{(\slashed{k}_1-\slashed{q})}{(k_1-q)^2}
\ga_\al
\frac{(\slashed{k}_2-\slashed{q})}{(k_2-q)^2}
\ga_\beta
\frac{(\slashed{k}_1-\slashed{q})}{(k_1-q)^2}
\bigg]\,,
\end{align}
and
\begin{align}
\cB[\Pi_c&(Q^2)](u_1,u_2)\nn\\
&= -\img\frac{(4\pi)^4(\mu^{2\eps})^4}{(d-1)Q^2} 
\int \frac{d^dk_1}{(2\pi)^d}\frac{d^dk_2}{(2\pi)^d}\frac{d^dk_3}{(2\pi)^d}  
\cB\big[\al D^{\rho\si}(k_1-k_2)\big](u_1)
\cB\big[\al D^{\al\beta}(k_2-k_3)\big](u_2)\nn\\
&\hspace{2cm}\times
\tr\bigg[
\ga^\mu
\frac{\slashed{k}_1}{k_1^2}
\ga_\rho
\frac{\slashed{k}_2}{k_2^2}
\ga_\al
\frac{\slashed{k}_3}{k_3^2}
\ga_\beta
\frac{\slashed{k}_2}{k_2^2}
\ga_\si
\frac{\slashed{k}_1}{k_1^2}
\ga_\mu
\frac{(\slashed{k}_1-\slashed{q})]}{(k_1-q)^2}\,,
\bigg]
\end{align}
which reduce to the same master integral of \eq{6}. Finally, we studied the diagram of \fig{adlerNLO4} which reads
\begin{align}
\cB[\Pi_d&(Q^2)](u_1,u_2)\nn\\
&= -\img\frac{(4\pi)^4(\mu^{2\eps})^3}{(d-1)Q^2} 
\int \frac{d^dk_1}{(2\pi)^d}\frac{d^dk_2}{(2\pi)^d}\frac{d^dk_3}{(2\pi)^d}  
\cB\big[\al D^{\rho\si}(k_1-k_2)\big](u_1)
\cB\big[\al D^{\al\beta}(k_3)\big](u_2)\nn\\
&\hspace{2cm}\times
\tr\bigg[
\ga^\mu
\frac{\slashed{k}_2}{k_2^2}
\ga_\al
\frac{(\slashed{k}_2-\slashed{k}_3)}{(k_2-k_3)^2}
\ga_\beta
\frac{\slashed{k}_2}{k_2^2}
\ga_\rho
\frac{\slashed{k}_1}{k_1^2}
\ga_\mu
\frac{(\slashed{k}_1-\slashed{q})}{(k_1-q)^2}
\ga_\si
\frac{(\slashed{k}_2-\slashed{q})}{(k_2-q)^2}
\bigg]\,.
\end{align}
This diagram can be written in terms of the master integrals of \eqs{6}{I4}
where the latter master integral requires the most computational effort. See
\app{masters} for the details of this computation.  We refer to
\app{allDiagrams} for a brief discussion on the remaining diagrams at
order $1/N_f^2$.

\section{Remaining \texorpdfstring{$\ord{1/N_f^2}$}{ } diagrams}
\label{app:allDiagrams}
%
\begin{figure}
\centering
\begin{subfigure}{3.5cm}
    \includegraphics[width=\textwidth]{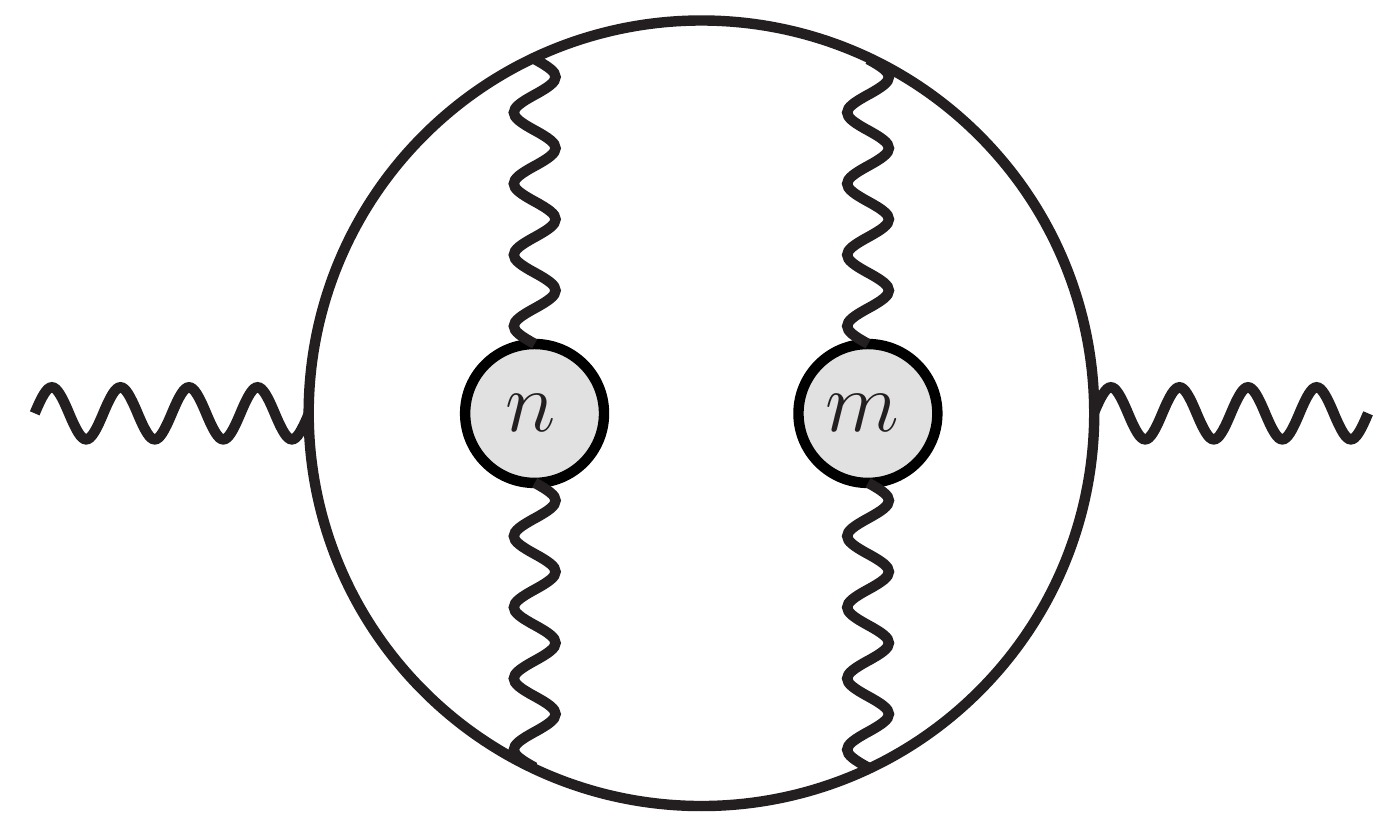}
    \caption{}
    \label{fig:adlerNLO5}
\end{subfigure}
\begin{subfigure}{3.5cm}
    \includegraphics[width=\textwidth]{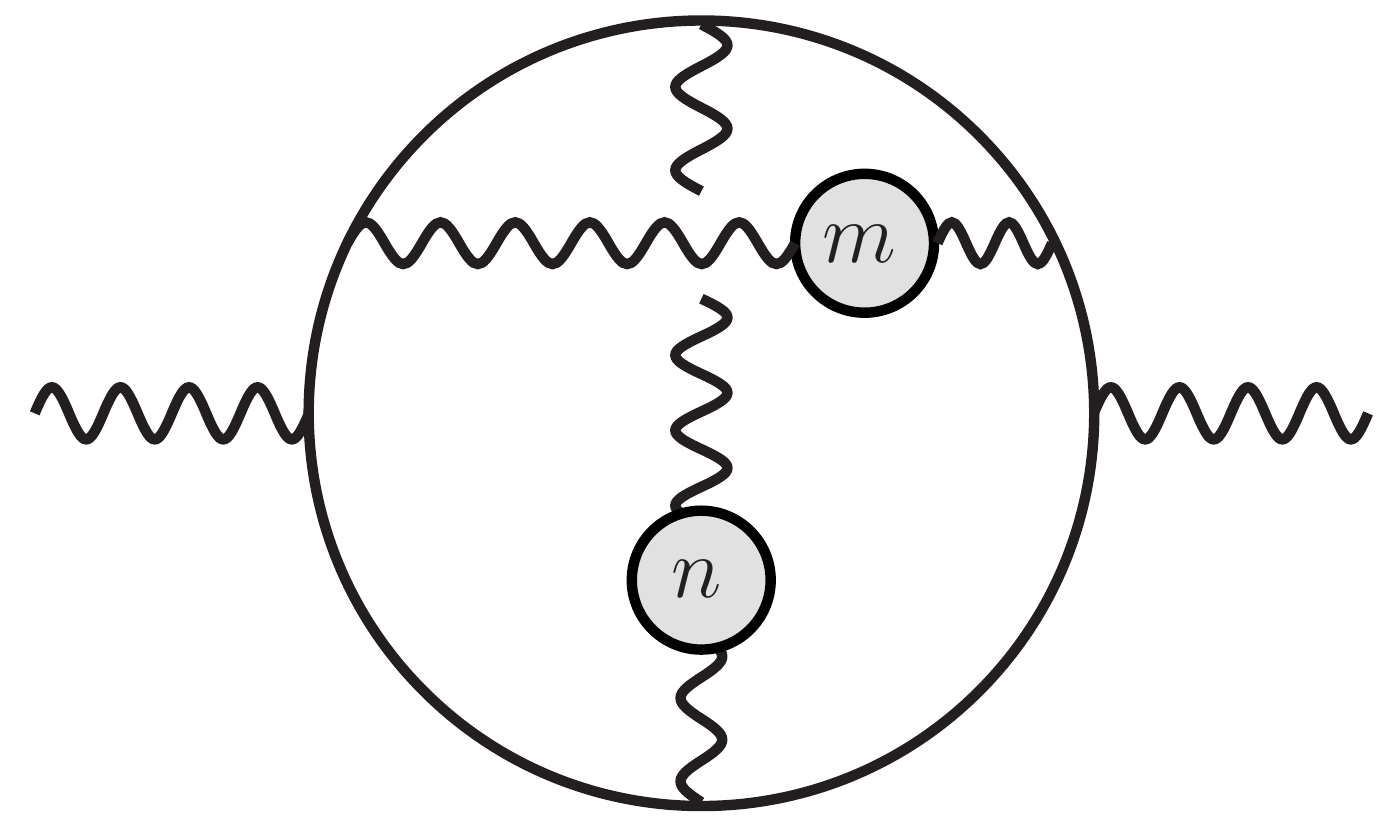}
    \caption{}
    \label{fig:adlerNLO7}
\end{subfigure}
\begin{subfigure}{3.5cm}
    \includegraphics[width=\textwidth]{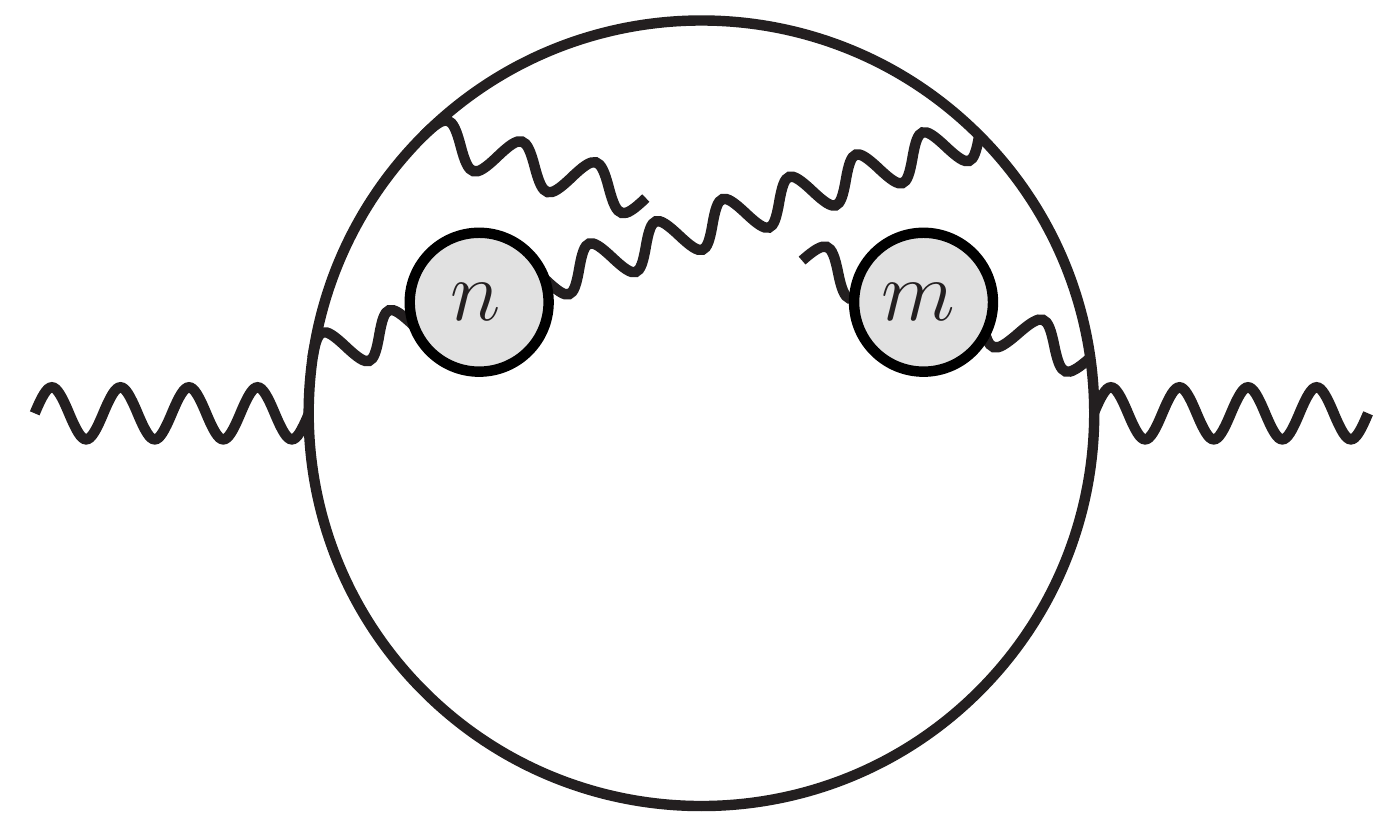}
    \caption{}
    \label{fig:adlerNLO8}
\end{subfigure}
\begin{subfigure}{3.5cm}
    \includegraphics[width=\textwidth]{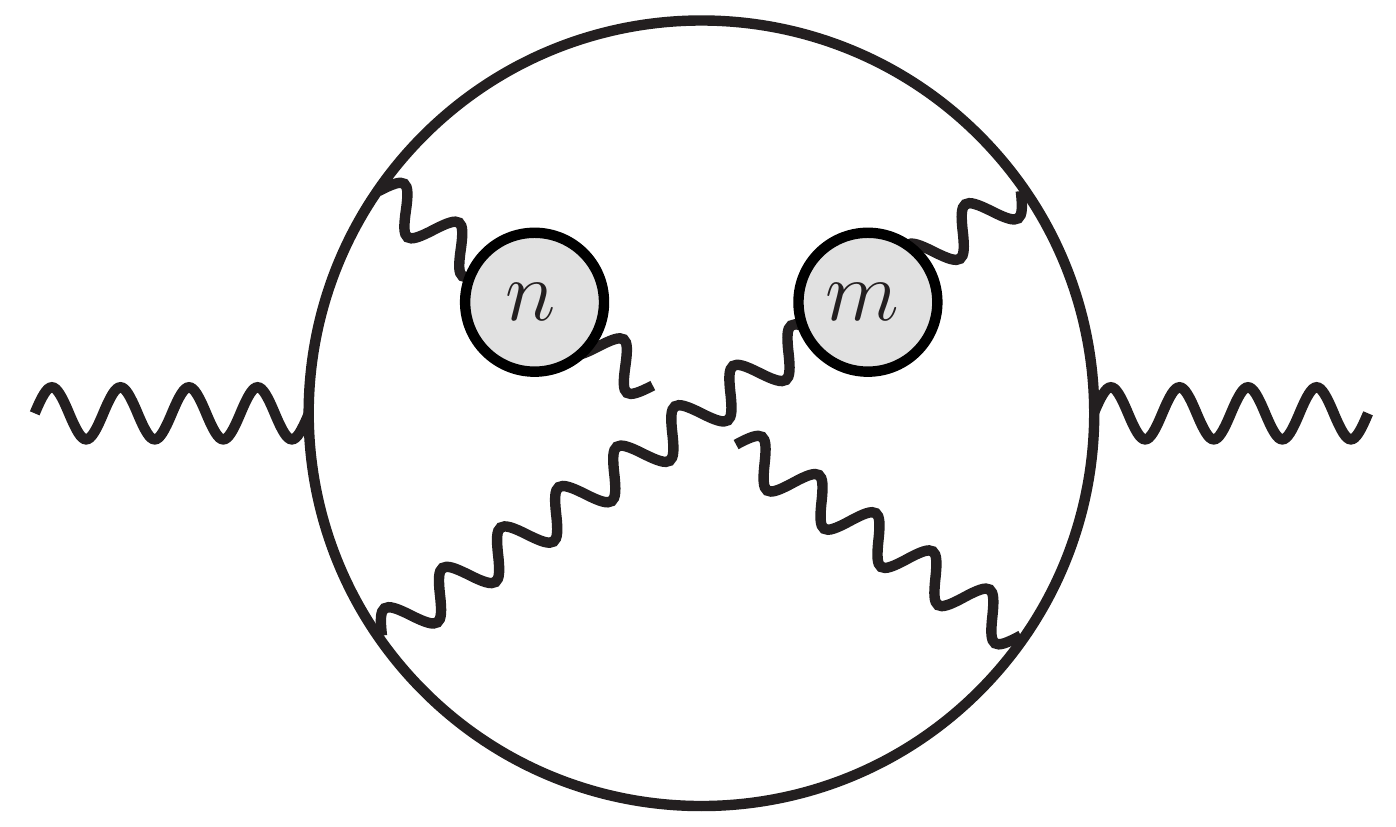}
    \caption{}
    \label{fig:adlerNLO9}
\end{subfigure}
\begin{subfigure}{3.5cm}
    \includegraphics[width=\textwidth]{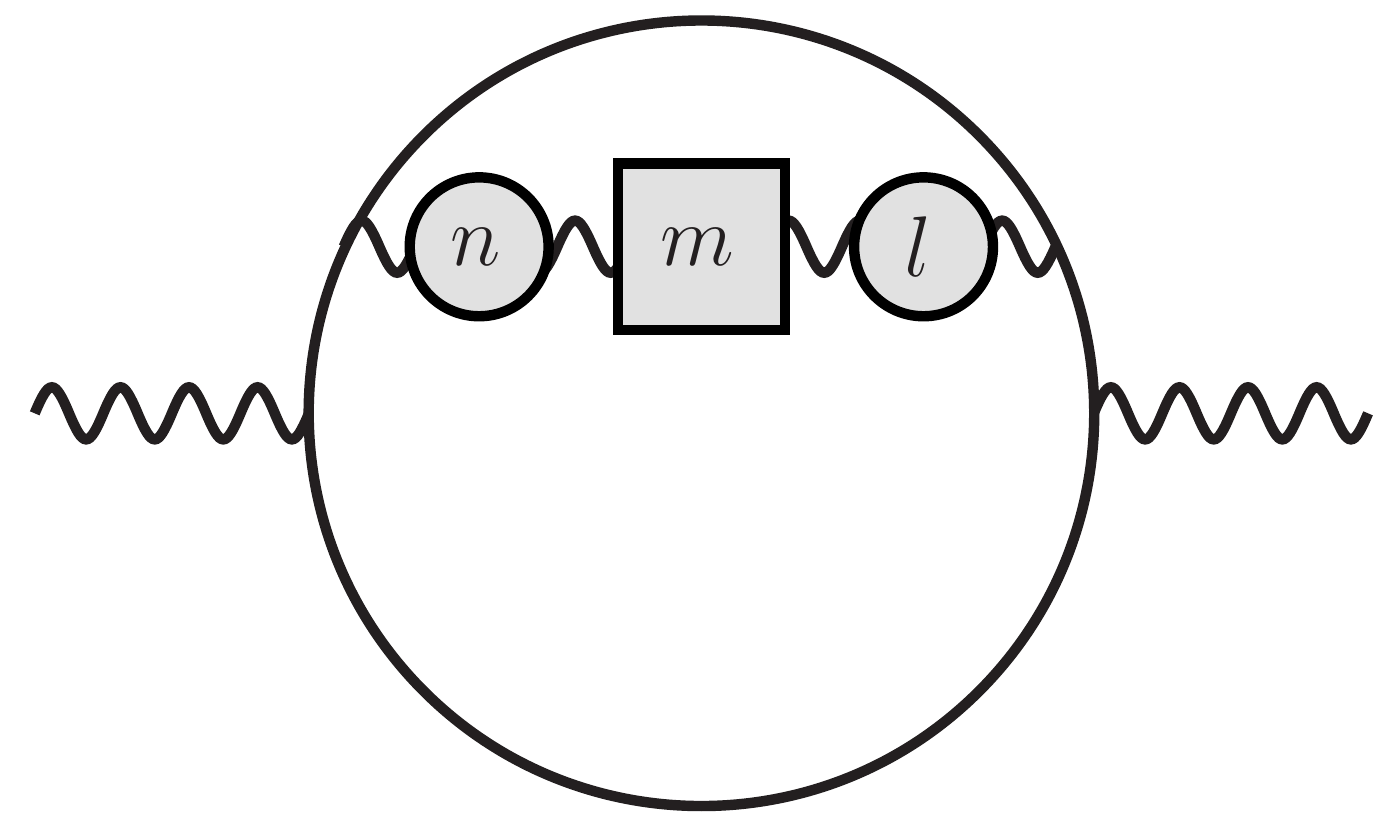}
    \caption{}
    \label{fig:adlerNLO10}
\end{subfigure}
\begin{subfigure}{3.5cm}
    \includegraphics[width=\textwidth]{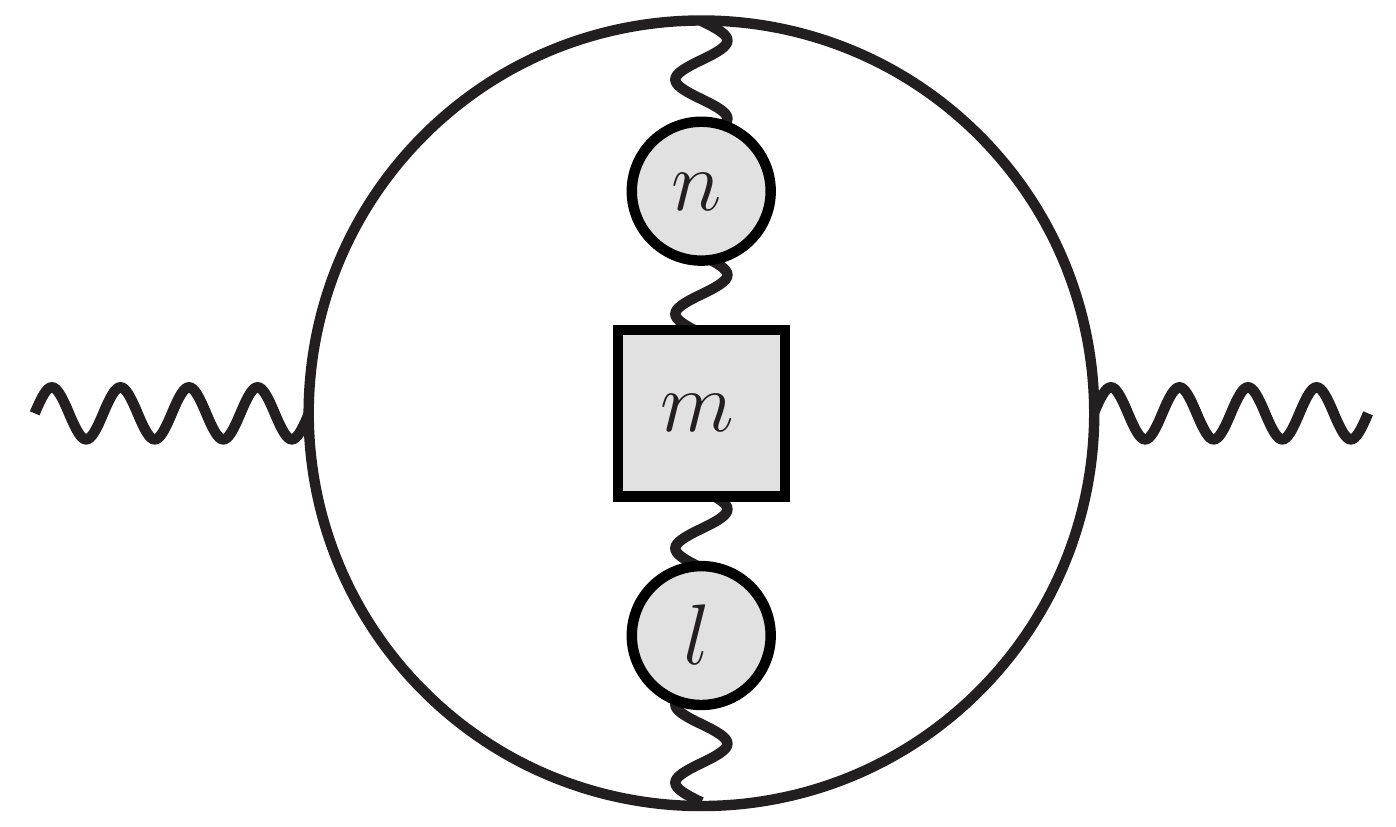}
    \caption{}
    \label{fig:adlerNLO11}
\end{subfigure}
\begin{subfigure}{6.3cm}
    \includegraphics[width=\textwidth]{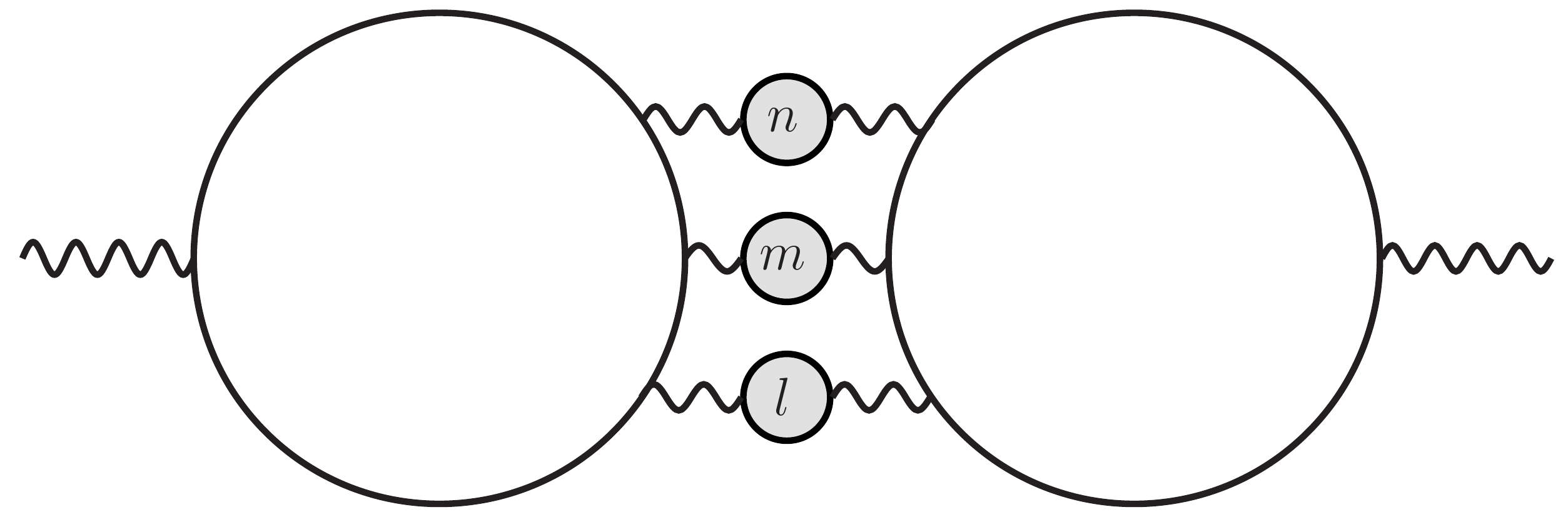}
    \caption{}
    \label{fig:adlerNLO6}
\end{subfigure}
\caption{Diagrams at order $1/N_f^2$ in the flavour expansion that are not computed in this paper.}
\label{fig:allAdlerNLOdiagrams}
\end{figure}
In \sec{adlerNLO} of this paper, we considered a subset of the diagrams at order $1/N_f^2$ in the flavour expansion. In this appendix, we briefly discuss the current status of the remaining diagrams at order $1/N_f^2$, given in \fig{allAdlerNLOdiagrams}.  
Notice that these are the abelian diagrams and we did not present the diagrams with non-abelian self interactions of the gauge field. As explained at the end of \sec{flavour}, one uses the process of naive non-abelianization, i.e. replace $\beta_{0f}$ by $\beta_{0}$, to model the non-abelian effects.

The two bubble chain diagrams $(a)$-$(d)$ of \fig{allAdlerNLOdiagrams} were considered in \cite{Beneke:1995qq}. In particular, these authors focused on the leading UV singularity because a full calculation of these diagrams is not possible yet, as the corresponding master integrals are not known. Notice that the results given are for the vacuum polarization (taking the derivative with respect to $Q^2$, \eq{pitoadler}, yields the results for the Adler function). The leading singularity found is\footnote{The parameter $u$ in \cite{Beneke:1995qq} is defined with an opposite sign compared to the definition used in this paper. Here, we present the result in our conventions.}
\begin{equation}
\frac{\log(1+u)}{(1+u)^2}\,,
\end{equation}
which would imply the leading order growth of the perturbative coefficients to be
\begin{equation}\label{eq:LOgrowthRemainingDiagrams}
d_n\sim (-1)^n\Ga(n+2)\psi(n+2)\,.
\end{equation}
where $\psi$ is the digamma function. At the level of the transseries, this translates to a non-perturbative contribution of the form 
\begin{equation}
\frac1\al\log(\al)\,e^{\frac1\al}\,.
\end{equation}
Recall that an exponential like this is typical for UV renormalons and grows exponentially large when $\al\to0$. Possible solutions were already discussed in \sec{discussionLO}: either the transseries parameter in front of such a momomial vanishes, or as long as $\al$ has a definite value, an expression like this still makes sense. 
Furthermore, we notice that the leading order growth of \eq{LOgrowthRemainingDiagrams} is enhanced by a factor $n$ compared to leading order growth $\Ga(n+1)\Psi(n+1)$ of the diagrams considered in \sec{adlerNLO}. 
In order to construct the transseries, and the relations between the non-perturbative sectors, one either needs access to sufficient perturbative data of these diagrams, or one needs the expansion of the Borel transform around the singularity at $u=-1$ as well as expansions around other singular points in $u$. In particular, we recall from diagram $(d)$ discussed in \sec{adlerNLO}, that for both procedures one has to compute a number of coefficients that is at least in the double digits, which is a challenging task to perform.

In \figs{adlerNLO10}{adlerNLO11}, we show the so called nested diagrams, where the square box denotes the LO ($1/N_f$) diagrams considered in \sec{adlerLO}, i.e. 
\begin{equation}
\vcenter{\hbox{\includegraphics[width=.18\textwidth]{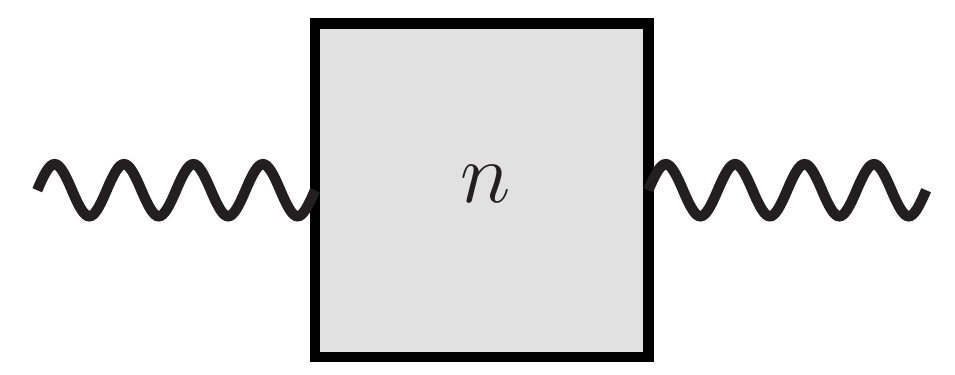}}}
\equiv 
\vcenter{\hbox{\includegraphics[width=.18\textwidth]{figures/adlerLO2.pdf}}}
+2\vcenter{\hbox{\includegraphics[width=.18\textwidth]{figures/adlerLO1.pdf}}}\,.
\end{equation}
A discussion of the singularity structure in the Borel plane of these diagrams can be found in \cite{Dondi:2020qfj} (where again, the results given are for the vacuum polarization). Using Borel Padé techniques, these authors showed that the singularities for the sum of these two diagrams at $u=1$ drops out (as expected by OPE arguments). Furthermore, they were able to deduce that the singularities at $u=2$ and $u=-1$ are branch points, suggesting the presence of further asymptotic sectors. However, is was not possible to compute sufficient perturbative data to determine the precise nature of these branch points. 

As far as we know, the diagram of \fig{adlerNLO6} has not been considered so far in the literature, as it is a difficult diagram with three bubble chains and its corresponding master integral is not known.

In order to compute the diagrams of \fig{allAdlerNLOdiagrams}, one needs to compute difficult skeleton diagrams (i.e. the diagrams where the bubble chains are replaced by the analytic regularized propagators of \eq{BorelChain}). In particular, this means that  diagrams $(e)$ and $(f)$ are two loop skeleton diagrams, diagram $(a)$-$(d)$ are three loop skeleton diagrams and diagram $(g)$ is a four loop skeleton diagram. We expect that diagrams $(d)$ and $(g)$ are the most challenging diagrams to compute as these, respectively, correspond to three loop, non-planar master integrals with two analytic regularized propagators and a four loop master integral with three analytic regularized propagators.

\section{Numerical methods}
\label{app:numerical}
Richardson transforms and Pad\'e approximants 
are often-used numerical methods in the resurgence literature because
they are well-suited to extract as much information as possible from a limited set of
perturbative data. Since we make considerable use of both methods in this paper,
we discuss them here in some detail.

\subsection{Richardson Transform}
\label{app:richardson}
\noindent 
Consider a sequence $\bS(k)$ which we know how to compute numerically
for arbitrary $k$, and which we also know can be expressed in the form
\begin{equation}\label{eq:SequenceForRichardsonTransform}
\bS(k)
= s_0 + \frac{s_1}{k} + \frac{s_2}{k^2}+\hdots\,.
\end{equation}
Our first aim is to calculate $s_0$, which we can do by
computing $\lim_{k\to\infty}\mathbb{S}(k)$. However, it might well be that one
needs to compute $\bS(k)$ for many and large values of $k$ to confidently
judge the limit. The {\em Richardson  transform}~\cite{1911RSPTA.210..307R,ri27} (see \cite{bender78:AMM} for a clear exposition) is a method to accelerate this process and enable one to
reach the desired limit with (significantly) fewer values of $k$.
It is defined recursively by
\begin{equation}\label{eq:recursiveFormulaRT}
\begin{aligned}
\text{RT}[\bS](k,0)
&= \mathbb{S}(k),\\
\text{RT}[\bS](k,N) 
&= \text{RT}[\bS](k+1,N-1) 
+ \frac{k}{N}\Big(\text{RT}[\bS](k+1,N-1)-\text{RT}[\bS](k,N-1) \Big) 
\end{aligned}
\end{equation}
where we denote the $N^\textsuperscript{th}$ Richardson transform of
$\bS(k)$ by $\text{RT}[\bS](k,N)$. To see how it works, consider the
case for $N=1$, which corresponds to
\begin{equation}
\text{RT}[\bS](k,1)
= \mathbb{S}(k+1) + k\Big(\mathbb{S}(k+1)-\mathbb{S}(k)\Big)
= s_0+\frac{s_1-s_2}{k^2} + \ORd{\frac1{k^3}}\,.
\end{equation}
Clearly the limit $\lim_{k\to\infty}\text{RT}[\bS](k,1)=s_0$ has a
faster rate of convergence than $\bS(k)$, as the subleading part
proportional to $s_1$  now scales as $\ORd{\frac1{k^2}}$. The estimate
of $s_0$ thus becomes better when both
$k$ and $N$ increase, since  the Richardson transform $\text{RT}[\bS](k,N)$ cancels the
subleading terms in $\bS(k)$ up to order $\frac{1}{k^N}$.  

\begin{figure}
\centering
\begin{subfigure}{7.5cm}
    \includegraphics[width=\textwidth]{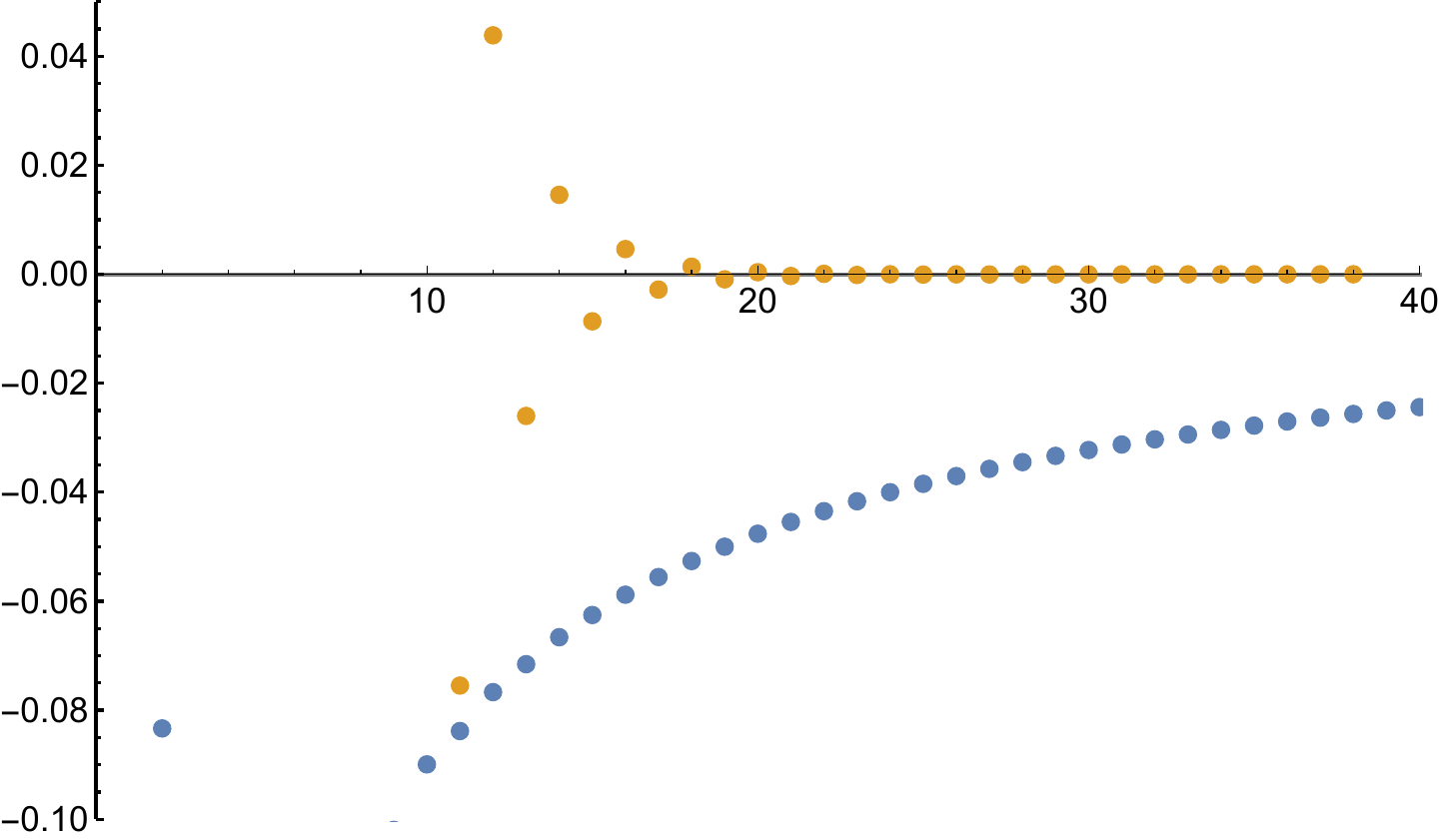}
    \caption{}
    \label{fig:RTexample1}
\end{subfigure}
\begin{subfigure}{7.5cm}
    \includegraphics[width=\textwidth]{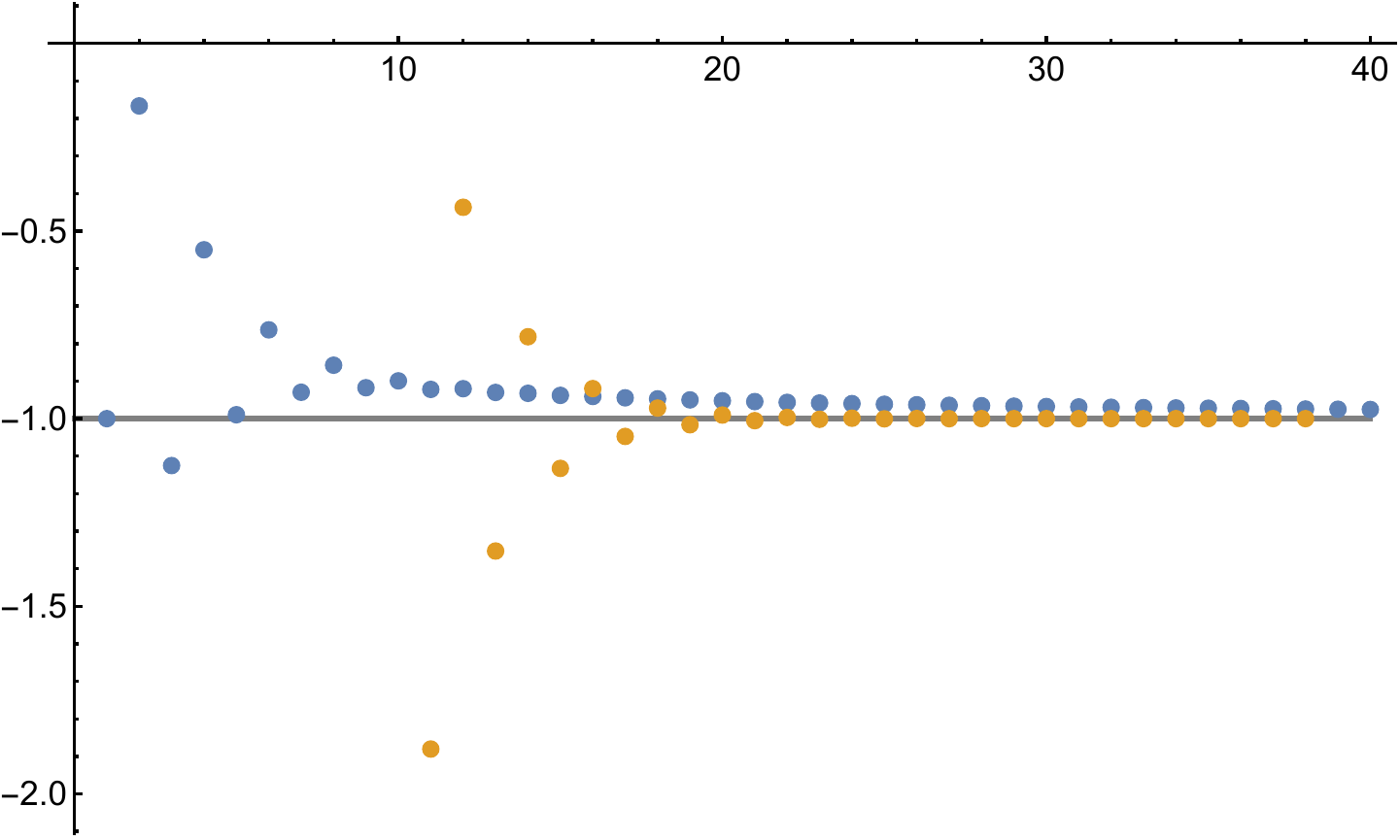}
    \caption{}
    \label{fig:RTexample2}
\end{subfigure}
\caption{In (a), we show the coefficients $c_k$ (bottom curve, blue), \eq{RTexample1}, together with their second Richardson transform (top curve, orange). In (b), we show the sequence $k\big(\bT(k)-0\big)$ (top curve, blue) of \eq{RTexample2}, again with their second Richardson transform (bottom curve, orange).}
\label{fig:RTexample}
\end{figure}

We can also find a closed form, in terms of
$\mathbb{S}(k),\mathbb{S}(k+1), \hdots,\mathbb{S}(k+N)$, for the
$N^\textsuperscript{th}$ Richardson transform by solving the following
set of equations,
%
\begin{equation}
\begin{aligned}
\mathbb{S}(k) &= s_0 + \frac{s_1}{k} + \hdots +\frac{s_N}{k^N}\\
\mathbb{S}(k+1) &= s_0 + \frac{s_1}{k+1} + \hdots +\frac{s_N}{(k+1)^N}\\
&\vdots\\
\mathbb{S}(k+N) &= s_0 + \frac{s_1}{k+N} + \hdots +\frac{s_N}{(k+N)^N}
\end{aligned}
\end{equation}
where we truncated the series in \eq{SequenceForRichardsonTransform}
after the $N^\textsuperscript{th}$ term. The solution for $s_0$, up to
order $\ord{\frac{1}{k^{N+1}}}$, is given by
\begin{equation}\label{eq:directFormulaRT}
s_0
\approx \text{RT}[\bS](k,N) 
= \sum_{n=0}^N(-1)^{n+N} \frac{(k+n)^N}{n!(N-n)!}\mathbb{S}(k+n) + \ORd{\frac{1}{k^{N+1}}}.
\end{equation}
Let us see how the Richardson transform works in practice by considering the function
\begin{equation}
\label{eq:borelexample}
f(u) 
= \frac{1}{1+\frac{u}{2}} + \frac{\log(1-u)}{u}
= \sum_{k=0}^\infty \bigg(\bigg(-\frac{1}{2}\bigg)^k-\frac{1}{k+1}\bigg)u^k. 
\end{equation}
If one is only given numerical values for the first, say 40,
coefficients $\bT(k)$: 
\begin{equation}\label{eq:RTexample1}
\bT(k)= \bigg(-\frac{1}{2}\bigg)^k-\frac{1}{k+1}\,,
\end{equation}
one can use the Richardson transform to extract the leading growth of these coefficients:
\begin{equation}
\bT(k) \approx -\frac{1}{k} + \frac{1}{k^2} + ... 
\end{equation}
where the ellipsis contains terms for both the subleading $\frac1k$ behaviour as
well as the $(-\frac12)^k$ term. In \fig{RTexample1} we show the
coefficients $\bT(k)$ in blue for $1\leq k\leq 40$, together with their
second Richardson transform. Here we used
\eq{directFormulaRT} to calculate the Richardson transform as the
direct formula has a faster implementation than the recursive formula
given in \eq{recursiveFormulaRT}. We see that the Richardson transform
gives a clear acceleration of the series to the value $s_0=0$ with
\begin{equation}
\text{RT}[\bT(k)](38,2) = -0.00001562865083298977613
\end{equation}
Higher order Richardson transform give even better estimates of
$s_0$.

We may go even further and extract the $s_1$, by constructing the sequence
\begin{equation}\label{eq:RTexample2}
k\big(\bT(k)-0\big)
\approx -1 + \frac{1}{k} + ... 
\end{equation}
This converges to $-1$ in the $k\to\infty$ limit. In \fig{RTexample2}
we see this sequence together with its second Richardson transform,
showing again good convergence towards $-1$:
\begin{equation}
\text{RT}\big[k\big(\bT(k)-0\big)\big](38,2) = -0.9999841284122182302
\end{equation}
One can naturally continue this process and also extract the
coefficients $s_i$, $i>1$.
%

\subsection{Pad\'e approximants and Borel-Padé summation}
\label{app:borelPade}
\noindent 
Consider an analytic function $f(u)$, or even a formal power series
\begin{equation}
  \label{eq:13}
 f(u) = \sum_{k=0}^\infty a_k u^k\,. 
\end{equation}
One can define the {\em Pad\'e approximant} of $f(u)$ ~\cite{ASENS_1892_3_9__S3_0,Frobenius1881} as (see \cite{bender78:AMM} for a more extensive exposition)
\begin{equation}
 P_{N,M}[f](u) \equiv \frac{\sum_{n=0}^Nb_nu^n}{\sum_{m=0}^Mc_mu^n}\,.
\end{equation}
This rational function should be thought of as an approximation to
$f(u)$ itself: the coefficients $b_n$ and $c_n$ are precisely chosen
such that the first $N+M+1$ coefficients of the Taylor expansions of
$f(u)$ and $P_{N,M}[f](u)$ agree. (Note that the Padé approximant has
$N+M+2$ coefficients in total, but an overall scaling of these
coefficients has no effect on the rational function.) Given a power
series $f(u)$, software packages like Mathematica can compute Padé
approximants up to around 100 coefficients exactly, and even further
numerically. In practice, one usually computes the {\em diagonal} Padé
approximant of order $N$, $P_{N}[f](u)$ (defined as $P_{N,M}[f](u) $ with $M=N$), and we shall always do so in this article.

The main use of Padé approximants lies in the fact that they can
encode singularities. If for example $f(u)$ describes a function with
a pole at some point $u_0 \neq 0$, a truncated Taylor expansion of
$f(u)$ will not `see' such a pole easily, simply because a Taylor
expansion is a polynomial which has a finite value at $u_0$. The Padé
approximant, on the other hand, by construction must have a number of
poles, and essentially always\footnote{In rare cases where a singularity does not show up, it can be made visible by changing the order of the Padé approximant. As a check, we have always compared Padé approximants at slightly different orders for the computations in this paper.} one of these will turn out to be near $u_0$.

This ability of Padé approximants to encode singularities of $f(u)$ goes even further: a Padé approximant can also spot more complicated singularities in $f(u)$ such as logarithmic branch cuts. Of course the rational function $P_{N}[f](u)$ does not have such branch cuts, but for large $N$ it does have a large number of poles, and these poles can accumulate into half-lines that start at some point $u_0$, thus mimicking a branch cut\footnote{This process is similar to what happens in e.g.\ matrix models, where at finite matrix size $N$ one deals with a resolvent which has a pole at every eigenvalue location, but when $N \to \infty$ this resolvent turns into a function with branch cuts.}, not only in a pictorial way but also in the sense that integrals around branch cuts are well-approximated by integrals around the poles.

For a real power series, the coefficients in the Padé approximant will also be real. In general, this leads to Padé poles that are also at real locations, but occasionally `spurious' poles occur when the denominator of the Padé approximant has two non-real complex conjugate zeroes. The occurrence of such spurious poles is usually not very problematic, as they occur sporadically and can often be removed by taking a slightly higher or lower order Padé approximant.

\begin{figure}
    \centering
    \includegraphics[width=.7\textwidth]{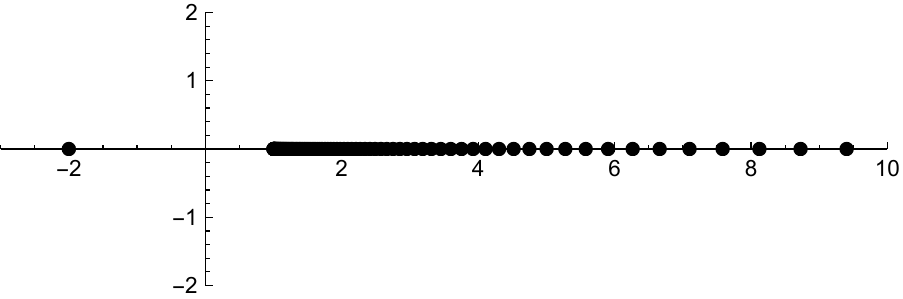}
    \caption{Poles of the order 120 diagonal Padé approximant to the function in \eq{borelexample}. One clearly sees the (single) pole at $u=-2$ as well as the mimicked `branch cut' starting at $u=1$.}
    \label{fig:padeplot}
\end{figure}

To illustrate the process of Padé approximantion, let us once again look at the example
\eq{borelexample}. The function $f(u)$ has a simple pole at $u=-2$ and
a logarithmic branch cut starting at $u=1$. Using the exact Taylor
expansion in \eq{borelexample}, and from that result computing $P_{120}[f](u)$
using Mathematica, we find a rational function whose poles we plot in
\fig{padeplot}. Both the single pole at $u=-2$ and the accumulation of
all other poles to mimick the branch cut at $u=1$ are clearly visible
in the figure.

In this article, we use Padé approximants in two different ways. First of all, we use these approximants to spot singularities for functions (usually in the Borel plane) for which we only have a finite number of expansion coefficients available -- see e.g.\ \figs{PadeNLO}{BorelPade}. A second application of Padé approximants is that they can be used to
replace an actual Borel transform $\cB[O](t)$ of an asymptotic series
before doing the inverse Laplace transform. That is, instead of
computing \be \cS[O](g) \approx \int_0^\infty dt \, \cB[O]_N(t) \,
e^{-\frac{t}{g}} \ee using an $N$th order Taylor expansion of
$\cB[O](t)$, one instead computes 
\be
 \label{eq:borelpadelaplace}
 \cS[O](g) \approx \int_0^\infty dt \, P_N[\cB[O]](t) \, e^{-\frac{t}{g}}.  
\ee 
The reason why this is useful is that it does not
make much sense to Laplace transform a Taylor series of a Borel
transform directly: such a Taylor series is simply a polynomial, and
its Laplace transform does nothing but reinsert factors of $n!$ in
front of the $n$th coefficient of $\cB[O](t)$ -- that is, it simply
returns a cut-off version of the original asymptotic series, which
when we make $N$ larger and larger will still diverge for any value of
the parameter.

If instead of directly Laplace transforming a Borel plane power series
one takes a diagonal Padé approximant first, the resulting functions
are much better behaved. There are two reasons for this: first of all
in a Laplace transform of the form \eq{borelpadelaplace} the function
$P_N[\cB[O]](t)$ behaves very nicely as $t\to\infty$: it simply
becomes a constant, so the total integral including the decaying
exponential converges very fast. More importantly, however, as we have
seen $P_N[\cB[O]](t)$ encodes the singularities of $\cB[O](t)$ very
well, and these singularities contain information about the
non-perturbative contributions to the function $O(g)$ that we are
trying to approximate -- non-perturbative contributions that `cure'
the asymptotic growth of the power series expansion.

As a result, whenever we want to numerically compute an inverse
Laplace transform of a Borel transform whose analytical form is
unknown, we use a Padé approximant. This procedure is known as Borel-Padé summation or sometimes
Borel-Écalle-Padé summation; further details can be found in e.g.\
\cite{Aniceto:2011nu,Aniceto:2018bis}.

\bibliographystyle{JHEP}

\providecommand{\href}[2]{#2}\begingroup\raggedright
\endgroup

\end{document}